\def\Tex{$T_{\rm ex}$}
\newcommand{\Msun}{\ifmmode {M_{\odot}} \else ${M_{\odot}}$ \fi} 
\def\Msunpyr{$M_{\rm \odot}~\rm yr^{-1}$} 
\def\Macc{$\dot{M}_{\rm acc}$} 
\newcommand{\Rsun}{\ifmmode {R_{\odot}} \else ${R_{\odot}}$ \fi} 
\newcommand{\Lsun}{\ifmmode {L_{\odot}} \else ${L_{\odot}}$ \fi} 
\newcommand{\Lbol}{\ifmmode {L_{\rm bol}} \else ${L_{\rm bol}}$ \fi} 
\def\Lacc{$L_{\rm acc}$} 
\def\Lline{$L_{\rm line}$} 
\def\Fline{$F_{\rm line}$} 

\newcommand{\FeI}{Fe\,{\footnotesize I}}
\newcommand{\FeII}{Fe\,{\footnotesize II}} 
\newcommand{\SII}{S\,{\footnotesize II}} 
\newcommand{\NII}{N\,{\footnotesize II}} 
\newcommand{\OI}{O\,{\footnotesize I}}
\newcommand{\CaI}{Ca\,{\footnotesize I}}
\newcommand{\CaII}{Ca\,{\footnotesize II}}
\newcommand{\CI}{C\,{\footnotesize I}}
\newcommand{\MgI}{Mg\,{\footnotesize I}}
\newcommand{\NaI}{Na\,{\footnotesize I}}
\newcommand{\HeI}{He\,{\footnotesize I}}
\newcommand{\KI}{K\,{\footnotesize I}}
\newcommand{\kms}{\ifmmode \,\rm km\,s^{-1} \else $\,\rm km\,s^{-1}$ \fi}
\newcommand{\halp}{H$\alpha$} 
\newcommand{\hbet}{H$\beta$} 
\newcommand{\brgam}{Br$\gamma$} 

\newcommand{\um}{$\mu$m}
\newcommand{\hmol}{H$_{2}$}
\newcommand{\yr}{\ifmmode {\rm yr^{-1}} \else yr${^{-1}}$ \fi} 
\newcommand{\Av}{$A_V$}

\documentclass[twocolumn]{aastex63}
\shorttitle{V899 Mon}
\shortauthors{Park et al.}
\graphicspath{{./}{figures/}}

\begin{document}
\title{V899 Mon: a peculiar eruptive young star close to the end of its outburst}

\correspondingauthor{Sunkyung Park}
\email{sunkyung.park@csfk.org}

\author[0000-0003-4099-1171]{Sunkyung Park}
\affiliation{Konkoly Observatory, Research Centre for Astronomy and Earth Sciences, E\"otv\"os Lor\'and Research Network (ELKH), Konkoly-Thege Mikl\'os \'ut 15-17, 1121 Budapest, Hungary}

\author[0000-0001-7157-6275]{\'Agnes K\'osp\'al}
\affiliation{Konkoly Observatory, Research Centre for Astronomy and Earth Sciences, E\"otv\"os Lor\'and Research Network (ELKH), Konkoly-Thege Mikl\'os \'ut 15-17, 1121 Budapest, Hungary}
\affiliation{Max Planck Institute for Astronomy, K\"onigstuhl 17, 69117 Heidelberg, Germany}
\affiliation{ELTE E\"otv\"os Lor\'and University, Institute of Physics, P\'azm\'any P\'eter s\'et\'any 1/A, 1117 Budapest, Hungary}

\author[0000-0002-4283-2185]{Fernando Cruz-S\'aenz de Miera}
\affiliation{Konkoly Observatory, Research Centre for Astronomy and Earth Sciences, E\"otv\"os Lor\'and Research Network (ELKH), Konkoly-Thege Mikl\'os \'ut 15-17, 1121 Budapest, Hungary}

\author[0000-0001-5018-3560]{Micha{\l} Siwak}
\affiliation{Konkoly Observatory, Research Centre for Astronomy and Earth Sciences, E\"otv\"os Lor\'and Research Network (ELKH), Konkoly-Thege Mikl\'os \'ut 15-17, 1121 Budapest, Hungary}

\author{Marek Dr{\'o}{\.z}d{\.z}}
\affiliation{Mount Suhora Astronomical Observatory, Cracow Pedagogical University, ul. Podchorazych 2, 30-084 Krak{\'o}w, Poland}

\author{Bernadett Ign\'acz}
\affiliation{Konkoly Observatory, Research Centre for Astronomy and Earth Sciences, E\"otv\"os Lor\'and Research Network (ELKH), Konkoly-Thege Mikl\'os \'ut 15-17, 1121 Budapest, Hungary}

\author{Daniel T. Jaffe}
\affiliation{Department of Astronomy, University of Texas at Austin, 2515 Speedway, Austin, TX, USA}

\author{R\'eka K\"onyves-T\'oth}
\affiliation{Konkoly Observatory, Research Centre for Astronomy and Earth Sciences, E\"otv\"os Lor\'and Research Network (ELKH), Konkoly-Thege Mikl\'os \'ut 15-17, 1121 Budapest, Hungary}

\author{Levente Kriskovics}
\affiliation{Konkoly Observatory, Research Centre for Astronomy and Earth Sciences, E\"otv\"os Lor\'and Research Network (ELKH), Konkoly-Thege Mikl\'os \'ut 15-17, 1121 Budapest, Hungary}

\author{Jae-Joon Lee}
\affiliation{Korea Astronomy and Space Science Institute 776, Daedeok-daero, Yuseong-gu, Daejeon, 34055, Republic of Korea}

\author{Jeong-Eun Lee}
\affiliation{School of Space Research, Kyung Hee University 1732, Deogyeong-daero, Giheung-gu, Yongin-si, Gyeonggi-do, 17104, Republic of Korea}

\author{Gregory N. Mace}
\affiliation{Department of Astronomy, University of Texas at Austin, 2515 Speedway, Austin, TX, USA}

\author{Waldemar Og{\l}oza}
\affiliation{Mount Suhora Astronomical Observatory, Cracow Pedagogical University, ul. Podchorazych 2, 30-084 Krak{\'o}w, Poland}

\author[0000-0001-5449-2467]{Andr\'as P\'al}
\affiliation{Konkoly Observatory, Research Centre for Astronomy and Earth Sciences, E\"otv\"os Lor\'and Research Network (ELKH), Konkoly-Thege Mikl\'os \'ut 15-17, 1121 Budapest, Hungary}

\author{Stephen B. Potter} 
\affiliation{South African Astronomical Observatory, PO Box 9, Observatory 7935, South Africa}
\affiliation{Department of Physics, University of Johannesburg, PO Box 524, Auckland Park 2006, South Africa}

\author[0000-0001-9830-3509]{Zs\'ofia Marianna Szab\'o}
\affiliation{Konkoly Observatory, Research Centre for Astronomy and Earth Sciences, E\"otv\"os Lor\'and Research Network (ELKH), Konkoly-Thege Mikl\'os \'ut 15-17, 1121 Budapest, Hungary}
\affiliation{Max-Planck-Institute f\"{u}r Radioastronomie, Auf dem H\"{u}gel 69, 53121 Bonn, Germany}

\author[0000-0003-3904-6754]{Ramotholo Sefako}
\affiliation{South African Astronomical Observatory, PO Box 9, Observatory 7935, South Africa}

\author{Hannah L. Worters}
\affiliation{South African Astronomical Observatory, PO Box 9, Observatory 7935, South Africa}


\begin{abstract}
V899~Mon is an eruptive young star showing characteristics of both FUors and EXors. It reached a peak brightness in 2010, then briefly faded in 2011, followed by a second outburst. We conducted multi-filter optical photometric monitoring, as well as optical and near-infrared spectroscopic observations of V899~Mon. The light curves and color-magnitude diagrams show that V899~Mon has been gradually fading after its second outburst peak in 2018, but smaller accretion bursts are still happening. Our spectroscopic observations taken with Gemini/IGRINS and VLT/MUSE show a number of emission lines, unlike during the outbursting stage. We used the emission line fluxes to estimate the accretion rate and found that it has significantly decreased compared to the outbursting stage. The mass loss rate is also weakening. 
Our 2D spectro-astrometric analysis of emission lines recovered jet and disk emission of V899~Mon.
We found the emission from permitted metallic lines and the CO bandheads can be modeled well with a disk in Keplerian rotation, which also gives a tight constraint for the dynamical stellar mass of 2\,${M_{\odot}}$. 
After a discussion of the physical changes that led to the changes in the observed properties of V899~Mon, we suggest this object is finishing its second outburst.
\end{abstract}

\section{Introduction \label{sec_intro}}

Throughout low-mass star formation, the material is transferred from the surrounding envelope to the protostellar disk via infall motion. Then, the material is accreted onto a central protostar until the final stages of the star-forming process. A constant replenishing of material from the envelope onto the disk is needed to sustain the accretion rate for the growth of the protostar. The mass accretion rate was originally thought to be constant \citep[$\sim$ 2 $\times$ 10$^{-6}$\Msun{}yr$^{-1}$;][]{shu1977}, however, observations show that the luminosities of young stellar objects (YSOs) are about ten times lower than required by the standard accretion model. This discrepancy of the luminosity is known as the luminosity problem \citep{kenyon1990, dunham2010}, and episodic accretion is suggested as a possible solution \citep[][and references therein]{audard2014}. In this scenario, protostars stay mostly in their quiescent-accretion phases (which explains the observed low luminosities), while their mass accretion is dominated by a relatively brief phase of episodic bursts.

Eruptive YSOs are observable evidence of enhanced accretion rates undergoing large amplitude accretion outbursts. The outbursts, attributed to highly enhanced accretion from the disk to the central protostar, play an important role not only in the growing mass of the central protostar but also in the evolution of the circumstellar disk. Based on their light curves and spectroscopic properties, eruptive YSOs are classified as FU Orionis-type objects (FUors) or EX Lupi-type objects (EXors).

FUors brighten by more than 4 magnitudes in the optical domain and remain in their brightened state for several decades. Spectroscopic characteristics are P~Cygni profiles in \halp\ 6563\,\AA, broad blue-shifted absorption profiles of \hbet\ 4861\,\AA, \FeII\ 5018\,\AA, \MgI\ 5183\,\AA, and \NaI\ D 5890/5896\,\AA{} doublet, broad and strong CO overtone bandhead absorption features, triangular H band continuum shape, and wavelength-dependent spectral types ranging from F/G supergiants in visual bands to K/M supergiants in near-infrared (NIR). In FUors spectra, emission lines are rarely present except for the P~Cygni profiles \citep{hartmann1996, herbig2003, audard2014, connelley2018}.

EXors also brighten by a few magnitudes, but unlike FUors, they remain in the high brightness stage for a few months to a few years. Observations of repetitive outbursts are almost the rule for the majority of EXors, e.g., EX~Lup \citep{rigliaco2020} and V1647~Ori \citep{ninan2013}. Another major difference between the two classifications is the rich emission line spectrum found in EXors \citep{herbig2007, aspin2010, kospal2011, audard2014, hodapp2019, hodapp2020}.

V899~Mon (also known as IRAS~06068-0641; $\alpha_{\rm J2000}$=06$^{\rm h}$\,09$^{\rm m}$\,19$\fs$24, $\delta_{\rm J2000}$=$-$6$^{\circ}$\,41$'$\,55$\farcs$89), is an eruptive young star which shows photometric and spectroscopic characteristics of both FUors and EXors.
It is located near the Monoceros R2 molecular cloud, at a distance of $809\pm17$~pc \citep{gaiacollaboration2021}. 
V899~Mon was classified as flat-spectrum or an early Class~II object \citep{greene1994}. The outburst of this star was first noticed on 2009 November 10 by \citet{wils2009_ATel}, and follow-up spectroscopic and photometric monitoring observations were conducted by \citet{ninan2015}. The brightness peak of the first outburst was observed in 2010, and then it went back to quiescence for a relatively short period ($<$ 1~yr) in 2011. Afterwards, it brightened again in 2012 (see Figure~7 in \citet{ninan2015} and \autoref{fig_light}). 
In their spectroscopic monitoring observations, a number of emission lines and several P~Cygni profiles are observed in V899~Mon. The high-velocity absorption components of the P~Cygni profiles are varied with time.

The physical parameters of V899 Mon (Table~\ref{tbl_info}) were constrained by \citet{ninan2015}. The peak accretion rate of 10$^{-6}$~\Msun{}\yr{}is typical for EXors \citep[][and references therein]{audard2014}, however, FUors can have comparable values \citep[e.g.,][]{kospal2011_hbc722}. A relatively short period of the brightened stage, repetitive outbursts, and rich emission lines are characteristics of EXors. Overall, the derived photometric and spectroscopic properties are in between the FUors and EXors but show more similarity to EXors. 
\citet{ninan2016} studied the variation of the high-velocity component of outflowing wind and concluded that the outflow is driven by magnetospheric accretion.
\citet{kospal2020b} analyzed the mid-infrared 10\,\um\ silicate emission feature of V899~Mon, and found this target is in the Class~II stage with signs of significant grain growth. Recently, \citet{kospal2021} performed ALMA 1.3\,mm continuum observation and radiative transfer modeling to investigate the disk properties.
The derived disk mass of 0.03\,\Msun is the lowest among the 10 FUors analyzed by the authors.
Furthermore, by comparing the disk mass with the stellar mass (Table~\ref{tbl_info}), the authors found the disk around V899~Mon does not fall within the region of gravitational instability. This mechanism has been suggested as the trigger behind FUor-type outbursts \citep{audard2014}, thus it is likely the episodic accretion of V899~Mon experiences a different triggering mechanism.

In this study, we characterize the current physical properties of V899~Mon by analyzing new and archival photometric and spectroscopic data. New observations and the data reduction, as well as used archival data, are described in Section~\ref{sec_observation}. In Section~\ref{sec_results}, we present the period and color magnitude analyses of photometric observations, as well as spectral analysis and disk modeling results. In Section~\ref{sec_discussion}, we discuss our results and compare them with those from previous studies. Finally, we summarize our results and findings in Section~\ref{sec_conclusion}.

\begin{deluxetable}{ccc}
\tabletypesize{\scriptsize} 
\tablecaption{Target Information \label{tbl_info}}
\tablewidth{0pt}
\tablehead{\colhead{} & \colhead{Parameters} & \colhead{Reference}}
\startdata
$V_{\rm LSR}$ & 9.63 \kms~& 1 \\ 
Distance & 809.20 $\pm$ 17.08 pc & 2\\
$A_V$ & 2.6 mag & 3 \\
$M_*$ & $\sim$ 2 ${M_{\odot}}$ & 1, 3 \\
$M_{disk}$ & 0.03 ${M_{\odot}}$ & 4 \\
$R_*$ & 4 ${R_{\odot}}$, 6 ${R_{\odot}}$ & 3, 4 \\
$R_{disk}$ & 19 $\pm$ 5 ${R_{\odot}}$ & 4 \\
age & 1 $\sim$ 5 Myr & 3 \\
$i$ & 41 $\sim$ 58 $^{\circ}$ & 1, 4 \\ 
\Lbol{} (outburst) & $\sim$ 150 \Lsun{} & 3 \\
\Lbol{} (2020 Sep.) & $\sim$ 21 \Lsun{} & 1 \\
$\dot{M}_{\rm acc}$ (outburst) & 10$^{-8}$ $\sim$ 10$^{-6}$ \Msun{}\yr{} & 3 \\
$\dot{M}_{\rm acc}$ (2020 Sep.) & $\sim$ 2 $\times$ 10$^{-7}$ \Msun{}\yr{} & 1 \\
$\dot{M}_{\rm out}$ (outburst) & $\sim$ 2.6 $\times$ 10$^{-7}$ \Msun{}\yr{} & 3 \\
$\dot{M}_{\rm out}$ (2020 Sep.) & $\sim$ 2.1 $\times$ 10$^{-8}$ \Msun{}\yr{} & 1 \\
\enddata
\tablerefs{
(1) This Work; (2) \citet{gaiacollaboration2021}; (3) \citet{ninan2015}; (4) \citet{kospal2021}
}
\end{deluxetable}

\section{Observations and Data Reduction \label{sec_observation}}

\subsection{Optical Photometry} \label{sec_photometric_observations}
We obtained ground-based optical photometric observations of V899~Mon between 2017 November and 2021 May with different telescopes around the world (\autoref{fig_light}). At the Piszk\'estet\H{o} Mountain Station of Konkoly Observatory (Hungary), we used two 
telescopes. The first one is a 60/90/180\,cm Schmidt telescope, equipped with an Apogee Alta U16 CCD camera, 1$\farcs$027 pixel scale, $70\farcm0\times70\farcm0$ field of view, Johnson-Cousins $BVR_CI_C$ filters. The second one is a 80\,cm Ritchey-Chretien (RC80) telescope equipped with an FLI PL230 CCD camera, 0$\farcs$55 pixel scale, $18\farcm8\times18\farcm8$ field of view, and Johnson $BV$ and Sloan $g'r'i'$ filters. At the Mount Suhora Observatory (MSO) of the Cracow Pedagogical University (Poland), we used the 60\,cm Carl-Zeiss telescope equipped with an Apogee Aspen-47 camera, 1$\farcs$116 pixel scale, $19\farcm0\times19\farcm0$ field of view, and Sloan $g'r'i'$ filters. At the South African Astronomical Observatory (SAAO), we used Lesedi, the new SAAO 1-m telescope, equipped with a Sutherland High Speed Optical Camera (SHOC), 2x2 binning, 0$\farcs$666 pixel scale, $5\farcm7\times5\farcm7$ field of view, and Bessel $UBVR_CI_C$ filters. At Adiyaman University Observatory (Turkey), we used ADYU60, a PlaneWave 60\,cm f/6.5 corrected Dall-Kirkham Astrograph telescope, equipped with an Andor iKon-M934 camera, 0$\farcs$673 pixel scale, $11\farcm5\times11\farcm5$ field of view, Sloan $g'r'i'$ filters. 
During December, 2020 and January 2021 we also observed V899~Mon by means of OMEGACAM installed on VST, as a part of our ESO programme ID 106.21LL (PI: Siwak). The camera covers 1$\times$1~deg field in $u'g'r'i'z'$ filters.

During the observing nights, typically 3 to 10 images of V899 Mon were taken in each filter. After standard reduction on bias, flat-field, and dark current, we calculated aperture photometry for the science target and several comparison stars in the field of view using an aperture radius of 5$''$. We selected the comparison stars from the APASS9 catalog \citep{henden2015}, which provides Bessel $BV$ and Sloan $g'r'i'$ magnitudes for the potential comparison stars. We calculated the $R_C$ and $I_C$ magnitudes of the comparison stars by plotting their broad-band SED using their APASS9 and 2MASS magnitudes \citep{cutri2003} and spline interpolating for the effective wavelengths of the $R_C$ and $I_C$ filters. We took $U$ band magnitudes for the comparison stars from the Swift/UVOT Serendipitous Source Catalog \citep{yershov2014}. We used the comparison stars for the photometric calibration by fitting a linear color term. Magnitudes taken with the same filter on the same night were averaged. The final uncertainties are the quadratic sum of the formal uncertainties of the aperture photometry, the photometric calibration, and the scatter of the individual magnitudes that were averaged per night. The results can be found in Table~\ref{tab:photometry} in the Appendix~\ref{sec_appendix_photometry}.

In addition to our data, we utilize the public-domain ASAS-SN $V$ and $g$ \citep{shappee2014,kochanek2017}, and ZTF $g'r'$ \citep{masci2018} light curves extracted and calibrated to standard photometric systems by means of dedicated pipelines. These surveys provide data obtained with the typical cadence of 1~day, though one can identify periods with two or more visits per night. The ASAS-SN data utilized in this work were obtained between 2012 February~5 (although the major time-series has started on 2014 December 16) and 2021 May~17, while the ZTF data between 2018 January~9 -- 2021 January~19. 

Finally, to enable the first insights into the small-scale variability unavailable for detailed investigation from the ground, we extracted the light curve from the calibrated full-field images (FFI) obtained with 10~min (0.00694436~d) cadence by TESS \citep{ricker2015} during 2020 December~18 -- 2021 January~13 (Sector~33). Aperture size of 2~pixels was used to extract the stellar flux (\autoref{fig_lightTESS}). The total monitoring time was 25.83304~d, with only 1.57642~d break in the middle of the run, necessary for the data transfer to the ground.

\subsection{Infrared WISE Photometry}
In order to construct a mid-infrared light curve of V899~Mon, we downloaded single exposure data taken by the WISE mission \citep{wright2010} from the AllWISE Multiepoch Photometry Table and the NEOWISE-R Single Exposure (L1b) Source Table. We used only the best quality 
data points and discarded low quality measurements (qi\_fact$<$1). Following Section~2.3 of the Explanatory Supplement \citep{wise_expl_suppl}, we filtered out data that could be contaminated by higher levels of charged particle hits due to the satellite being close to the South Atlantic Anomaly (saa\_sep$<$5). We checked that none of the data points come from regions contaminated by scattered light from the moon (moon\_masked=1). Finally, we rejected data points where the reduced $\chi^2$ of the profile-fit photometry measurement (w1rchi2 or w2rchi2) was higher than 100. Being in the 5--7\,mag mid-infrared brightness range, the WISE photometry of V899~Mon was affected by saturation. We corrected the individual data points for saturation using the correction curves given in Section~2.1 of the Explanatory Supplement. We then calculated the average and standard deviation of the data points for each observing season and found that the scatter was less than or similar to the uncertainty of the individual magnitudes. Therefore, in the following, we use the seasonal averages.

\subsection{Optical Spectroscopy}
Optical integral field spectroscopic observations were conducted for V899~Mon using the Multi Unit Spectroscopic Explorer \citep[MUSE;][]{Bacon2010} on the Very Large Telescope at Paranal Observatory (Chile).
The data were taken on 2021 January~22, as part of our ESO programme ID 106.21KL (PI: Cruz-Sáenz de Miera).
The observations were carried out with the WFM-AO-N mode.
The Wide Field Mode (WFM) provides a 60$''$$\times$60$''$ field of view (FOV) with a pixel size of 0.2$''$$\times$0.2$''$.
The MUSE nominal mode (N) provides a wavelength coverage between 4800\,\AA{} and 9300\,\AA{} with a resolving power between 1770 and 3590.
Due to the Na Notch filter of the adaptive optics (AO) system \citep{Arsenault2008,Strobele2012}, there is a gap in spectral coverage between 5820\,\AA{} and 5970\,\AA{}.

The observations were executed in six 126 second exposures each with a total integration time of 12.6 minutes on V899~Mon.
Each exposure was dithered by $\sim$1$''$, and rotated by 90$^\circ$.
The individual exposures were performed with a clear sky transparency and a seeing between 0.89$''$ and 1.10$''$.
The observations were reduced using the MUSE pipeline \citep[version 2.8.4;][]{Weilbacher2020} by ESO, resulting in a fully calibrated and combined MUSE data cube with the instrumental signature and the sky background removed. The S/N around \halp\ 6563\,\AA{} is about 50. 

The spectrum was extracted using a 5\arcsec\ circular aperture positioned on the peak of the stellar emission.
Then, the the spectrum was shifted by the barycentric velocity of $-12.96$~km s$^{-1}$ calculated by barycorrpy \citep{wright2014} and by the systemic velocity of 9.63~km s$^{-1}$ ($V_{\rm LSR}$) obtained from the ALMA C$^{18}$O line data (Appendix~\ref{sec_Vsys}).
The systemic velocity in the heliocentric system ($V_{\rm helio}$) converted from $V_{\rm LSR}$ is 27.92~km s$^{-1}$. 

\subsection{Near-infrared Spectroscopy}
The NIR spectrum of V899~Mon was obtained with Immersion GRating INfrared Spectrograph (IGRINS) installed on the 8.1m Gemini South telescope on 2020 November~6. IGRINS provides high-resolution (R~$\sim$~45,000; corresponding to a velocity resolution of $\Delta v$~$\sim$ 7~km s$^{-1}$) NIR spectra covering the full H (1.49--1.80 \um) and K (1.96--2.46 \um) bands with a single exposure \citep{yuk2010, park2014, mace2016}.
The spectrum was obtained with a slit scale of 0.34\arcsec~$\times$~5\arcsec. The S/N of H and K bands are about 244 and 313, respectively. The S/N for each band is the median value per resolution element for the wavelength range of 1.58--1.61~\um~and 2.21--2.24~\um.
V899 Mon was observed with two series of ABBA nodding observations at different positions on the slit to better subtract the sky background. The exposure time of each nod observation was 300 sec, and the total exposure time was 2,400 sec. A nearby A0 telluric standard star (HD~42133) was observed immediately after the observation of V899~Mon for telluric correction.

The IGRINS pipeline \citep{lee2017} was used to reduce the spectra for flat-fielding, sky subtraction, correcting the distortion of the dispersion direction, wavelength calibration, and combining the spectra. 
Then, telluric correction and flux calibration were performed in the same way as described in \citet{park2018}. 
Interpolated H and K band magnitudes between our $BVr'i'$ and NEOWISE data observed in 2020 September were adopted for the flux calibration since there is no recent NIR photometry during the fading state.
The barycentric velocity was calculated by the same method as done in the optical spectrum, which is 19.57~km s$^{-1}$. 
Finally, the barycentric and systemic velocity ($V_{\rm sys} = 27.92\,\rm km~s^{-1}$) correction was applied.

\section{Results and Analysis}\label{sec_results}

\subsection{Photometry}
\label{sec_photometry} 

\begin{figure*}
    \centering
    \includegraphics[width=0.9\textwidth, trim=10 70 0 60,clip]{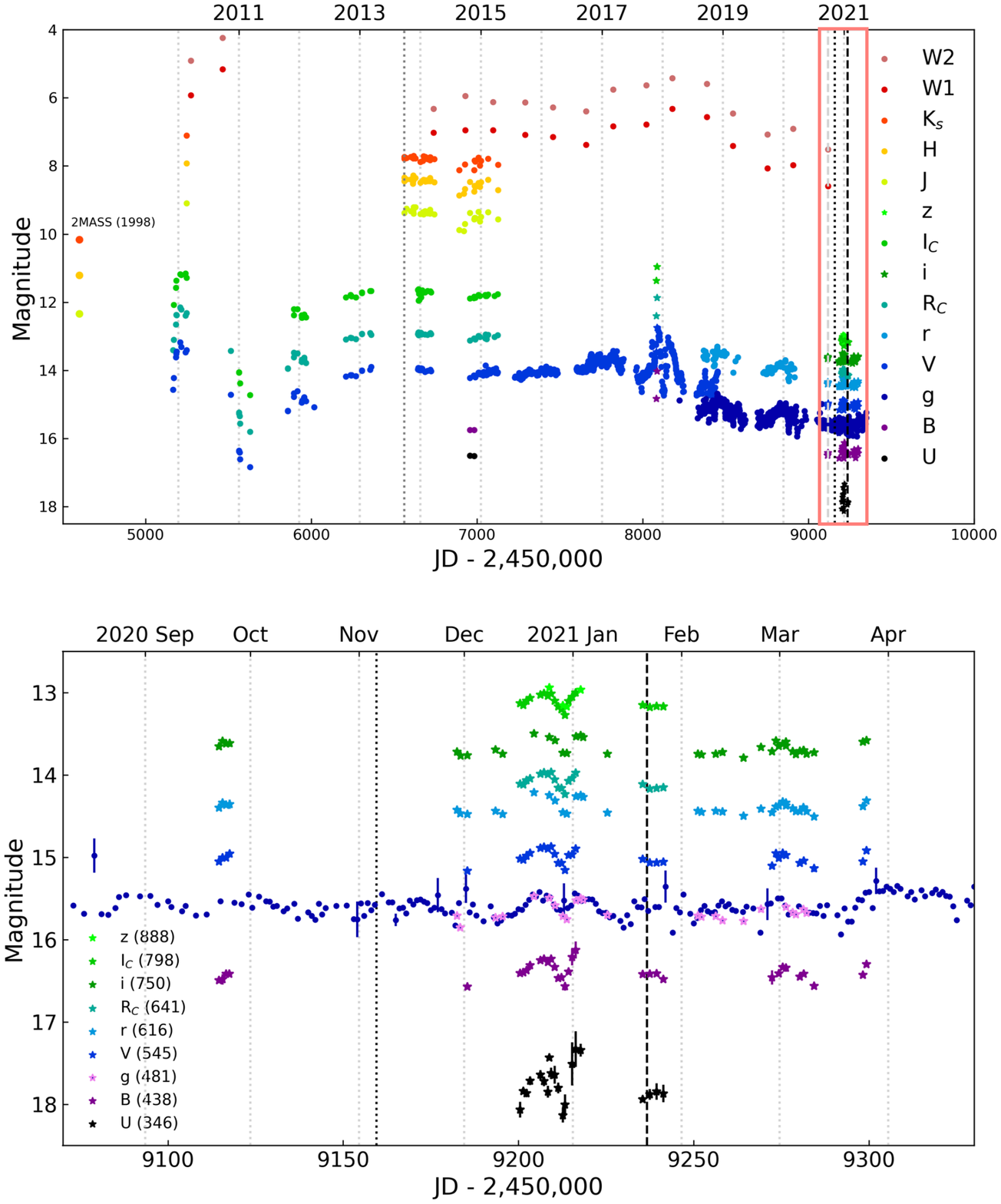}
    \caption{Top: light curve of V899~Mon. Bottom: light curve of V899 Mon for the 2020 September -- 2021 April time period (pink box in the top panel). The bottom figure shows our own observations from Table~\ref{tab:photometry} (Appendix~\ref{sec_appendix_photometry}), complemented with data from the ASAS-SN survey. The photometric uncertainty smaller than the symbol size is not presented. The numbers in brackets indicate the central wavelength of each filter in nm.
    The photometric data from the ASAS-SN \citep[V and g;][]{shappee2014, kochanek2017}, ZTF \citep[g and r;][]{masci2018}, WISE \citep[W1 and W1;][]{mainzer2011}, 2MASS \citep[J, H, K$_{s}$;][]{cutri2003}, and U, B, V, R, I, J, H, and K$_{s}$ from \citet{ninan2015} were used. Circle and star symbols indicate archival data and our data, respectively. Different colors indicate different bandpasses. 
    The black dotted and dashed lines indicate the spectroscopic observation date of IGRINS and MUSE. The dark gray dotted line in the top panel indicates the observation date of \citet{ninan2015}. Dashed gray line in the top panel indicates the data point used in the SED in \autoref{fig_SED}. \label{fig_light}} 
\end{figure*}

In order to get a general view on the long term evolution of V899~Mon, in addition to the multi-band light curves shown in \autoref{fig_light} we also prepared the ''single-band'' light curve composed of $V$-band data collected from 2009 by \citet{ninan2015} as well as our own and public-domain $V$- and $g$-band data. The new data were processed as follows: in order to improve the photon statistic, we averaged typically 3--8 ASAS-SN data points gathered each night in 1-day bins. Outliers were carefully removed prior to this operation. The Schmidt and the most recent RC80 and SAAO $V$-band data were included to this dataset without any shifts. In 2018--2019 the ASAS-SN survey gradually switched to the $g$-band. We supplemented ASAS-SN light curve with the ZTF, VST, MSO and Adiyaman $g$-band data. As the MSO, ASAS-SN, VST and the first two seasons of ZTF $g$-band observations do perfectly overlap, we use them as the reference light curves. Only the third season ZTF data required a $+0.060$~mag shift to match the entire dataset. For the same reason, a $-0.075$~mag shift was applied to all Adiyaman $g$-band points.
We then aligned (with the accuracy of 0.02~mag) the $g$-band data to the $V$-band data  by a constant 0.58~mag shift. As the variability amplitudes are very similar in $V$- and $g$-bands, no correction on this effect was necessary. The part of the combined light curve, important for the description of the most recent light changes, is shown in \autoref{fig_light_events}.

According to \autoref{fig_light}, V899~Mon is currently fading but it is still brighter than during the 2011 quiescent phase. 
After the second outburst starting in 2012, the brightness remained at a relatively high level until 2017, reaching the maximum brightness in 2018. After that, it started to gradually fade at the rate of $0.30\pm0.02$~mag~yr$^{-1}$, as obtained from linear fit to the recently best sampled $g$- and $V$-band combined 2016--2021 data. The less numerous ZTF $r$-band 2018--2021 data indicate a faster decrease, at a rate of $0.36\pm0.01$~mag~yr$^{-1}$.

The evolution of the optical light curve of V899~Mon, however, is far from being smooth. During the early outburst stages, the only feature visible on top of the major light plateau was the major dip in 2011 (\autoref{fig_light}), which was interpreted as an interruption of the enhanced accretion \citep{ninan2015}. The rather sparse sampling during 2009--2014 probably hid brightness changes occurring in the time scales of days and weeks, which became visible later, primarily thanks to the almost daily ASAS-SN observations (see the bottom panel in \autoref{fig_light} and \autoref{fig_light_events}). According to these figures, before 2016, on top of the light plateau caused solely by enhanced accretion we observed rather irregular $\pm0.1$~mag light changes. The amplitude of these background oscillations became higher ($\pm0.2$~mag) in the 2016/2017 season, when they started to resemble accretion bursts, common for classical T Tauri-type stars (CTTS). The most pronounced $\Delta V \approx1$~mag event took place in 2017/2018, and it is marked in red in \autoref{fig_light_events}. The entire accretion burst, defined as the rise from and return back to the virtual continuum, lasted for about 200~days. Assuming that the peak duration is in a way related to Keplerian rotation of the disk, the disk warp from which the accretion flow could originate formed roughly at about 1~au from the 2\Msun star. 
In spite of the large distance, as compared to $\sim$0.05--0.1~au in ordinary CTTS, this view seems to be supported by the presence of the central dip (marked in blue in \autoref{fig_light_events}). It is similar to observed in other CTTS, in which during favorable viewing geometry, the optically thick part of the accretion column is causing total or partial occultations of hot spots created on the stellar photosphere \citep{siwak2018}. In the case of V899~Mon, the eclipse of the associated hot spot appears to be total.

In 2020/2021 another well-defined accretion burst with accompanying central dip appeared. It lasted for about 30 days, which suggests that it originated from a disk warp that formed at $\sim$0.2~au from the star. This value is closer to low-massive CTTS, where the accretion flow originates near the disk co-rotation radius. 

Accretion bursts of smaller amplitudes are visible at other epochs as well, in addition to apparently random $V\approx0.5$~mag dips. These dips are not obviously connected to accretion bursts but may result from dusty clouds passing in front of the bright inner disk and/or the star itself.

\subsubsection{Period analysis of ground-based data} 
\label{sec:period_g-b}

In order to investigate variability occurring on timescales from days to months, we utilize the $V$- and $g$-band data providing the longest and most uniform temporal coverage (\autoref{fig_light_events}, and see the description in the previous paragraph). 

Using the generalised Lomb-Scargle periodogram \citep{zechmeister2009}, we started from unsuccessful attempt to search for periodic variations in the entire data set. This failure, however, is mostly caused by changing morphology of the light curve and the yearly breaks in the data acquisition. Therefore in the second attempt, we decided to analyze each season separately. The seasonal trends visible in the data were removed prior to the analysis by simple 2--3 order polynomial fits. We found that there is no quasi-periodicity persistent for one full observing season. 
This is better illustrated by the spectrum obtained by means of the weighted wavelet Z-transform (WWZ, \citealt{foster1996}), designed for analysis of unevenly sampled light curves and available within the {\sc Vartools} package \citep[][\autoref{fig_freqGround}]{hartman2016}. We show the WWZ spectrum only for one season, but those obtained for other seasons look similar. They do contain information about spacing between dips and accretion bursts, and a mixture of other small-scale variabilities, which should be treated with caution and always accompanied by simultaneous viewing of light curves.
The spectra do reveal that both the accretion burst- and the dip-related quasi-periodic oscillation last for 2-3 oscillatory cycles and then disappear. In the case of accretion burst, this behavior is typical for unstable or moderately-stable accretion \citep[e.g.,][]{blinova2016}. 

\begin{figure}
    \centering
    \includegraphics[width=0.5\textwidth]{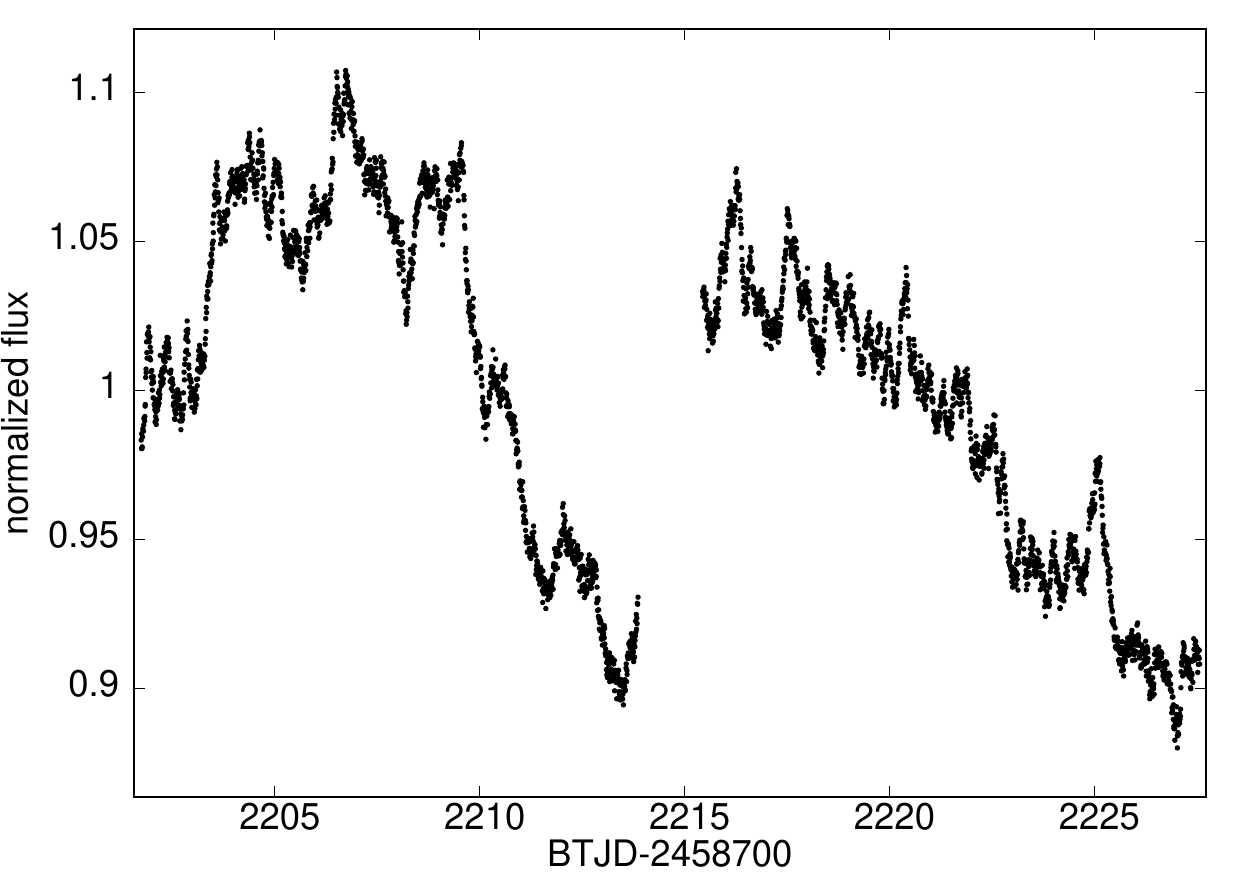}
    \caption{TESS light curve of V899~Mon (Sector~33). \label{fig_lightTESS}}
\end{figure}

\begin{figure*}
    \centering
    \includegraphics[width=\textwidth]{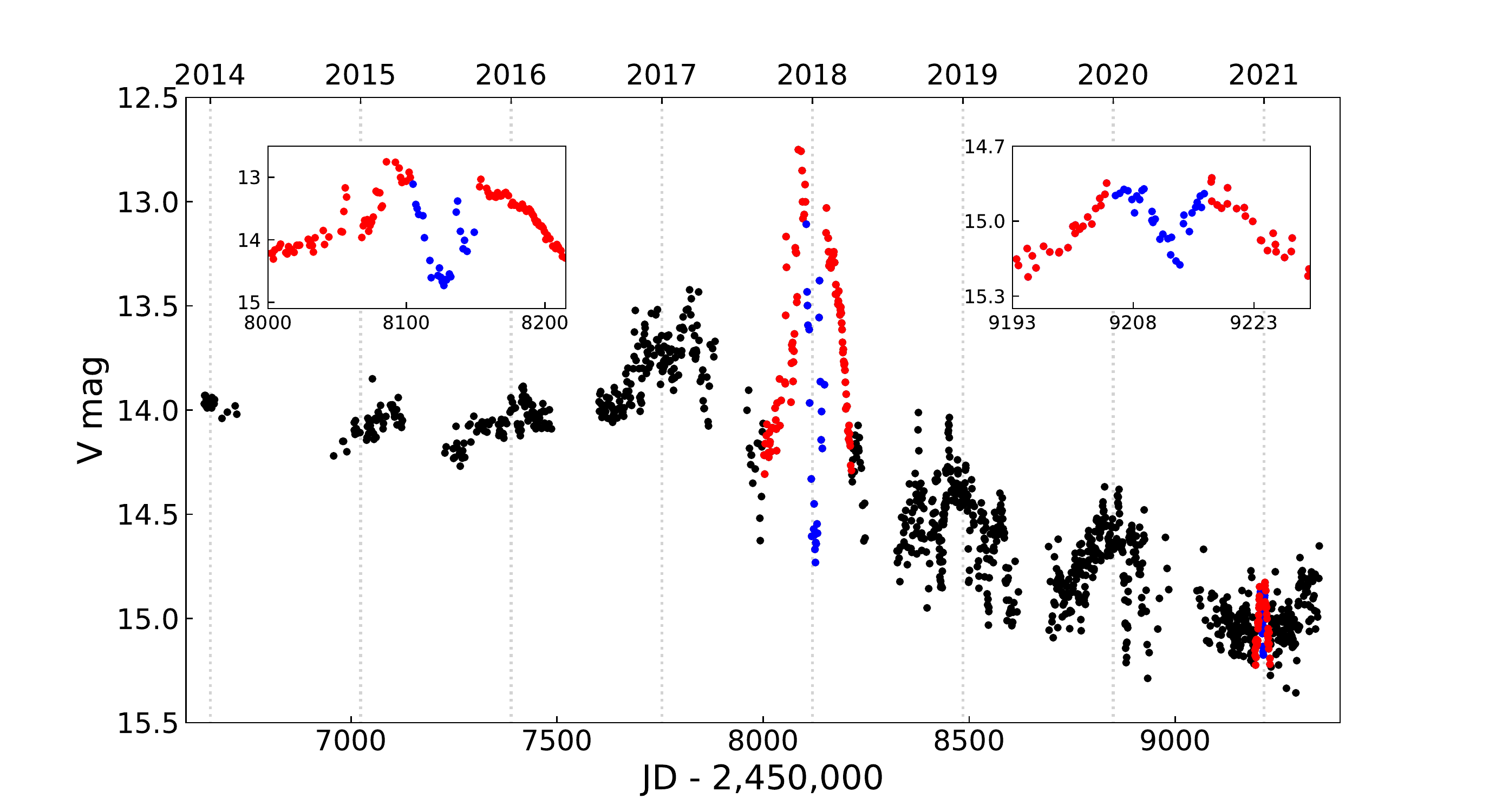}
    \caption{$V$- and $g$-band combined light curve of V899 Mon for the 2014--2021 time period. The two, more widely studied in this paper accretion bursts and accompanying central dips, are indicated by red and blue colors, respectively. Left and right subplots show zoom-in accretion bursts and central dips in 2018 and 2021, respectively. \label{fig_light_events}}
\end{figure*}

\begin{figure}
   \centering
    \includegraphics[width=0.5\textwidth]{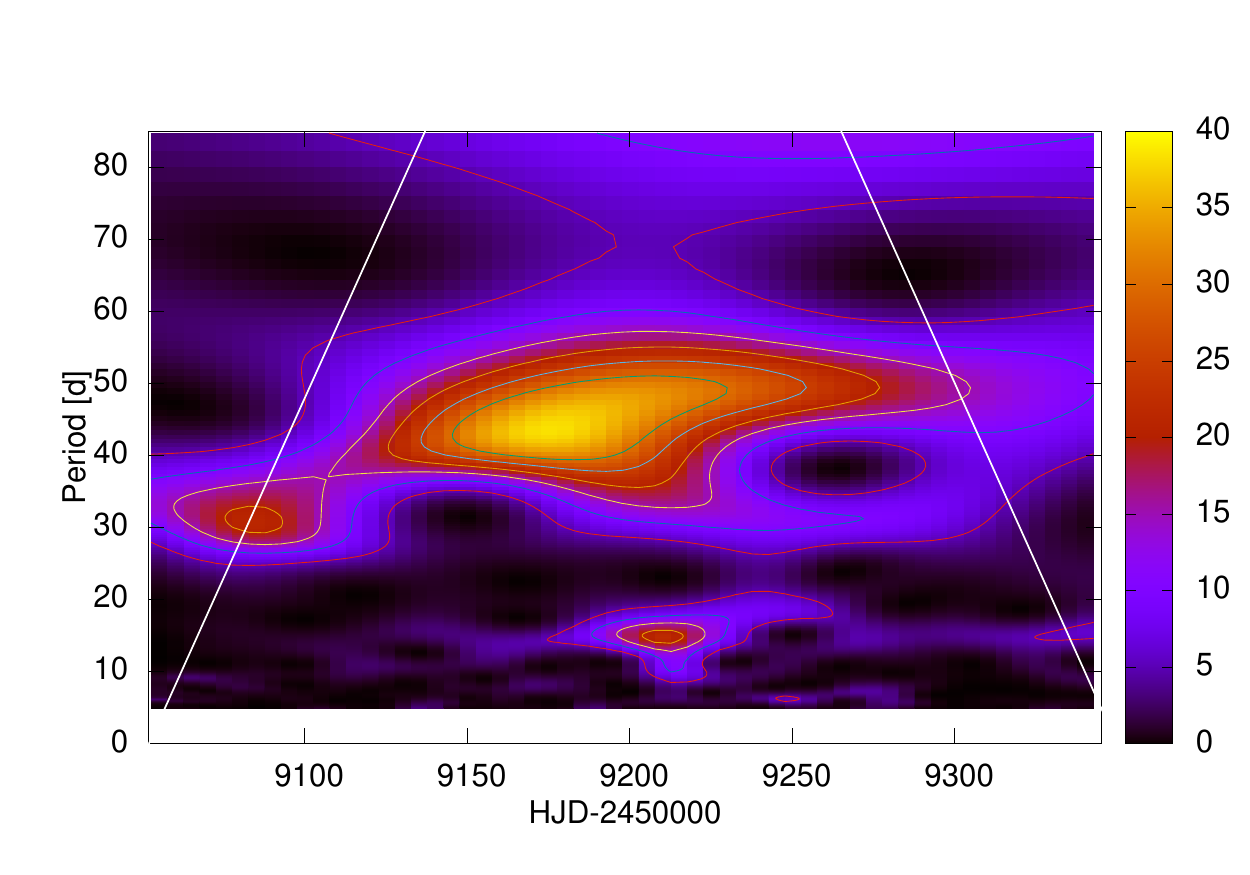}
    \caption{WWZ spectrum calculated for 2020/2021 season. The colors represent the Z-statistic values. Edge effects are contained beyond the white lines. \label{fig_freqGround}} 
\end{figure}

\subsubsection{Period analysis of TESS data}

\begin{figure}
    \centering
    \includegraphics[width=0.45\textwidth]{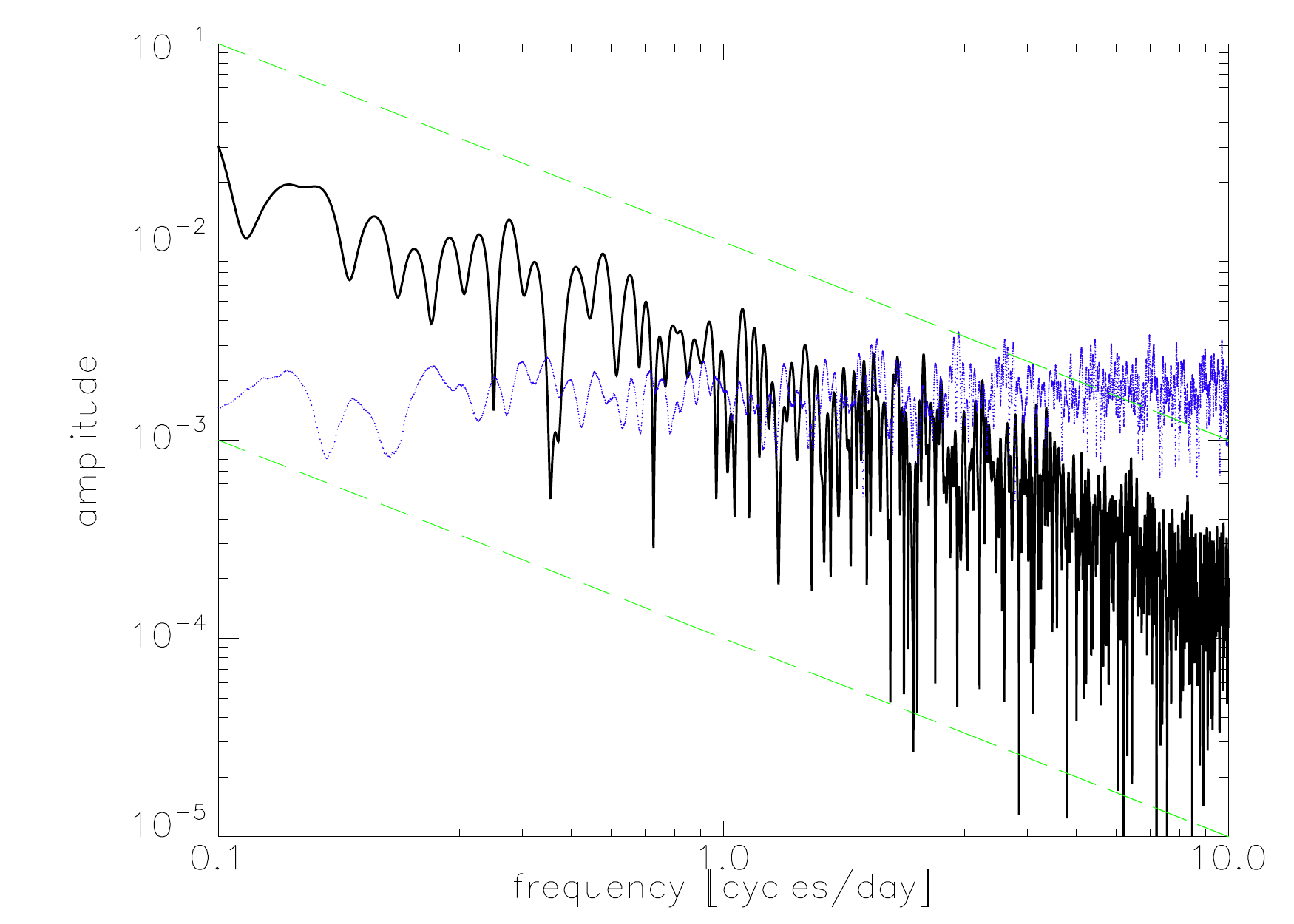}
    \caption{Fourier spectrum (black line) of TESS light curve presented in log-log scale. The stochastic nature ($a_f \sim f^{-1}$) of these oscillations is indicated by the two parallel green dashed lines. The amplitude errors are represented by blue dots. \label{fig_freq_tess}}
\end{figure}

In order to get a detailed look into the shortest variability time scales, we performed a frequency analysis of the 26~day-long TESS light curve obtained during 2020 December 16 -- 2021 January 11 (\autoref{fig_lightTESS}). The amplitude-frequency spectrum does not reveal any well-defined periods, nor quasi-periods (\autoref{fig_freq_tess}). The lack of the outstanding frequencies and the spectrum slope $a_f \sim f^{-1}$ indicative of the random-walk nature of these light changes \citep{press1978} strongly suggest that these stochastic processes are driven by the magnetically-controlled accretion, as in CTTS and some Herbig~Ae stars \citep{rucinski2010, stauffer2014, siwak2018}. 
The wavelet analysis also does not reveal any persistent oscillations: there is some indication of a 2.5~d quasi-periodic oscillation in the first half of the light curve, 
but it disappears after 4--5 cycles and is apparently absent in the second half. Except for the above, both the careful look into the light curve and zoom into the short periodic part of the wavelet spectrum reveals that the shortest accretion bursts have a typical duration time of 0.17~d (4~hr). This enables us to gain even deeper insight into the physical properties of the accretion flow: the disk plasma is not transferred smoothly within the stream but in the form of clumps, a phenomenon first observed in BP Tau \citep{gullbring1996}, but not yet well understood.

\begin{figure*}
    \centering
    \includegraphics[width=0.45\textwidth]{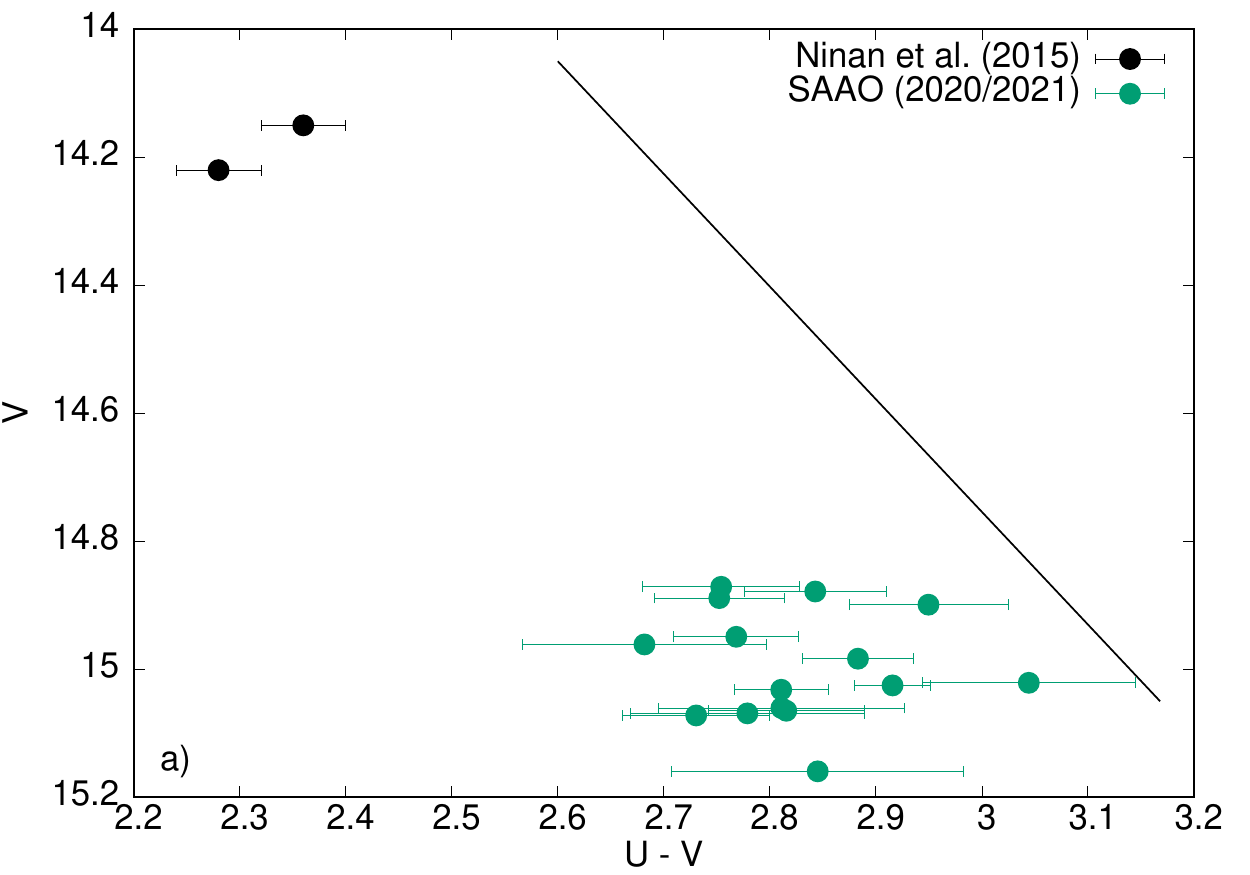}
    \includegraphics[width=0.45\textwidth]{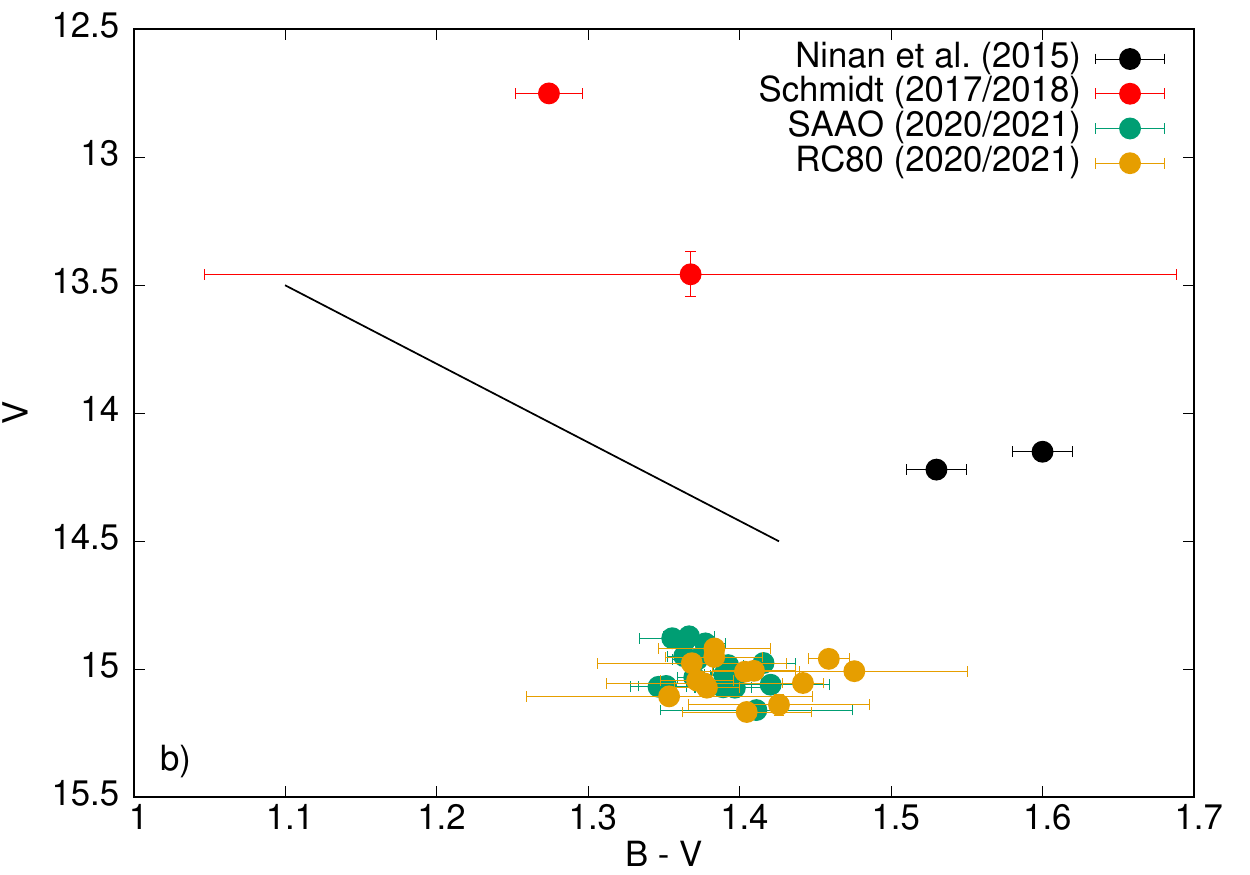}
    \includegraphics[width=0.45\textwidth]{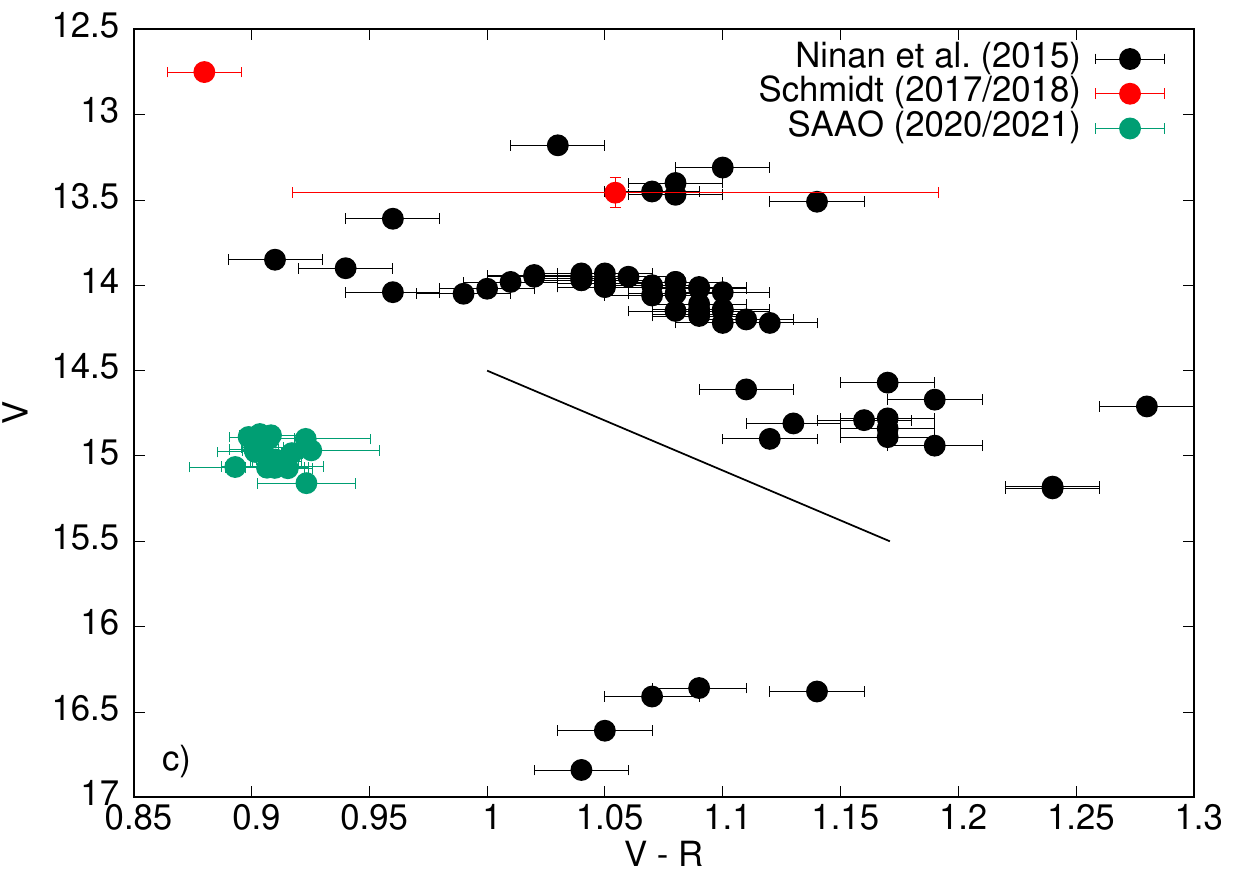}
    \includegraphics[width=0.45\textwidth]{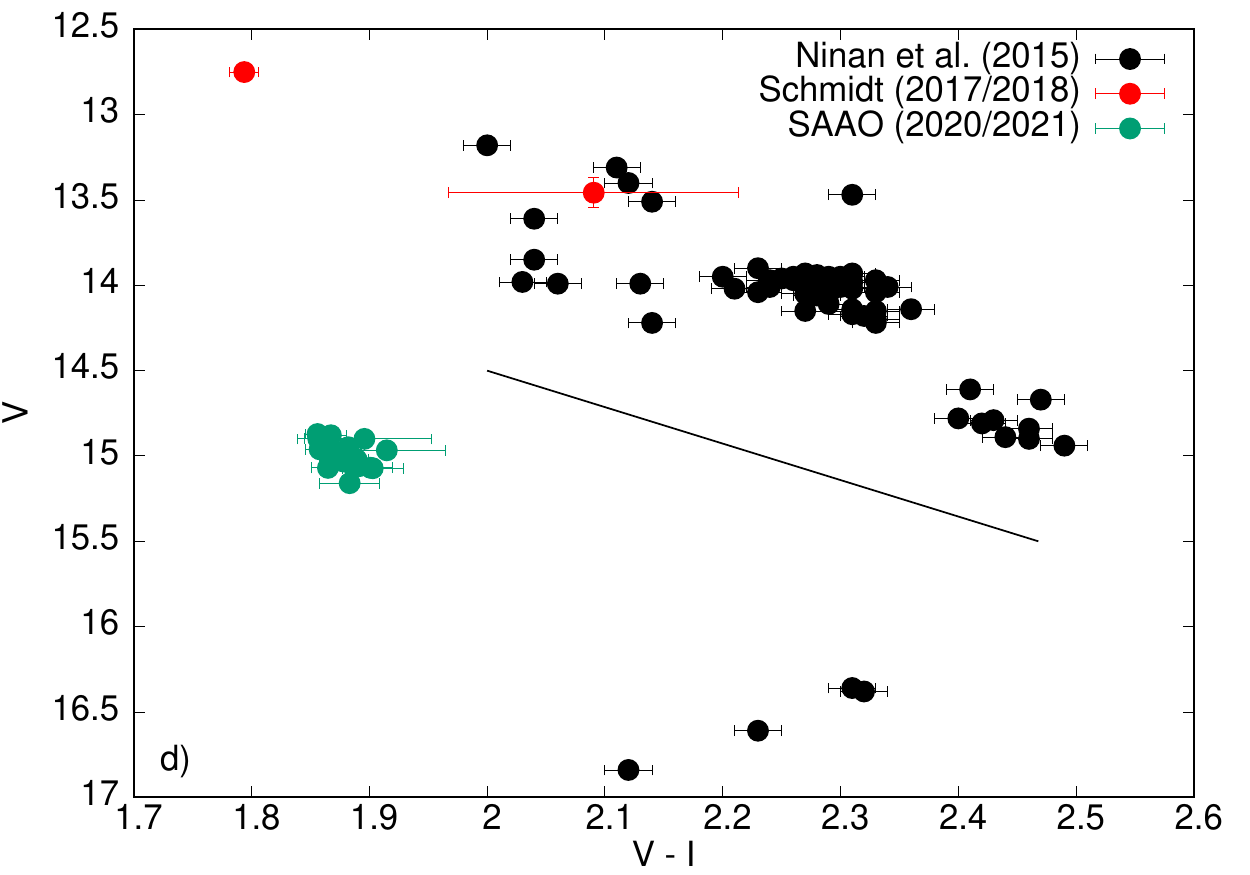}
    \includegraphics[width=0.45\textwidth]{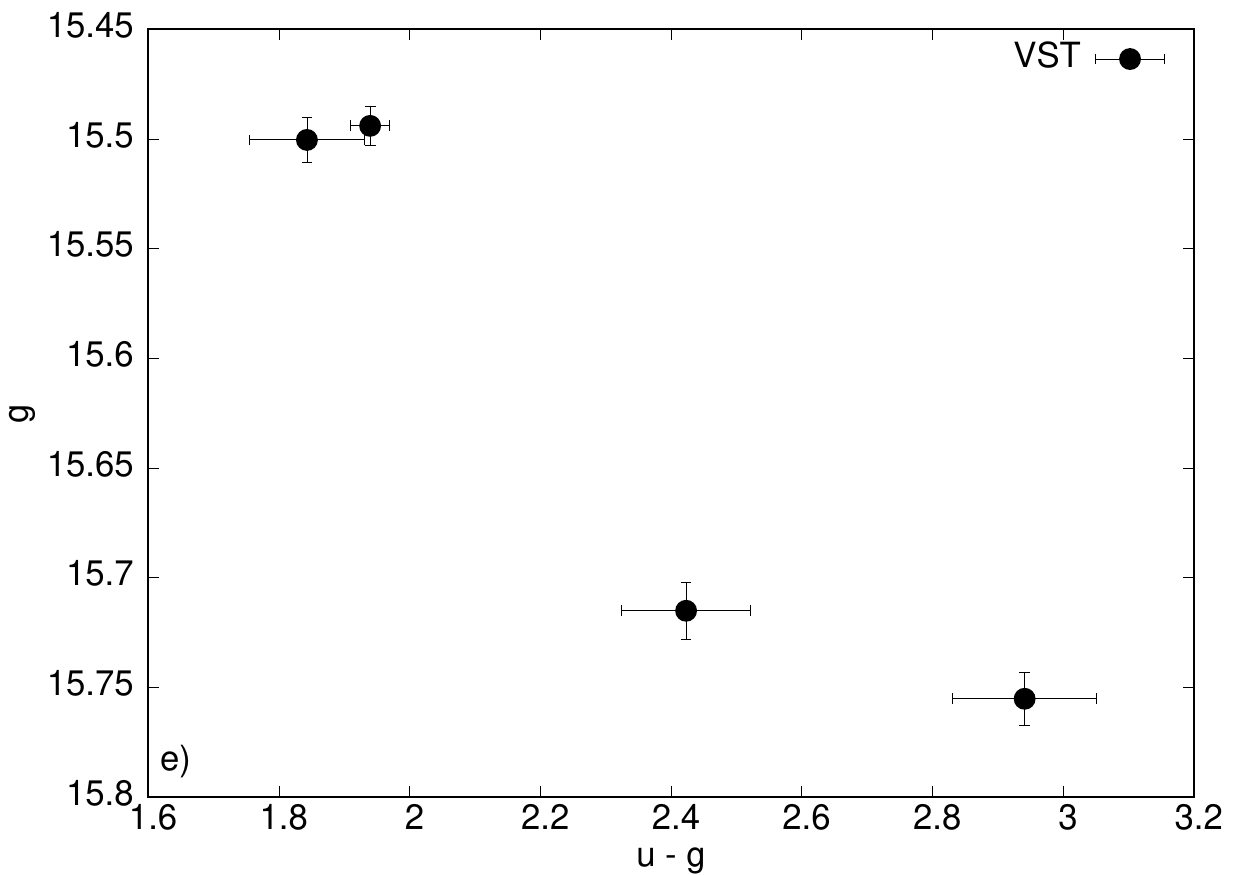}
    \includegraphics[width=0.45\textwidth]{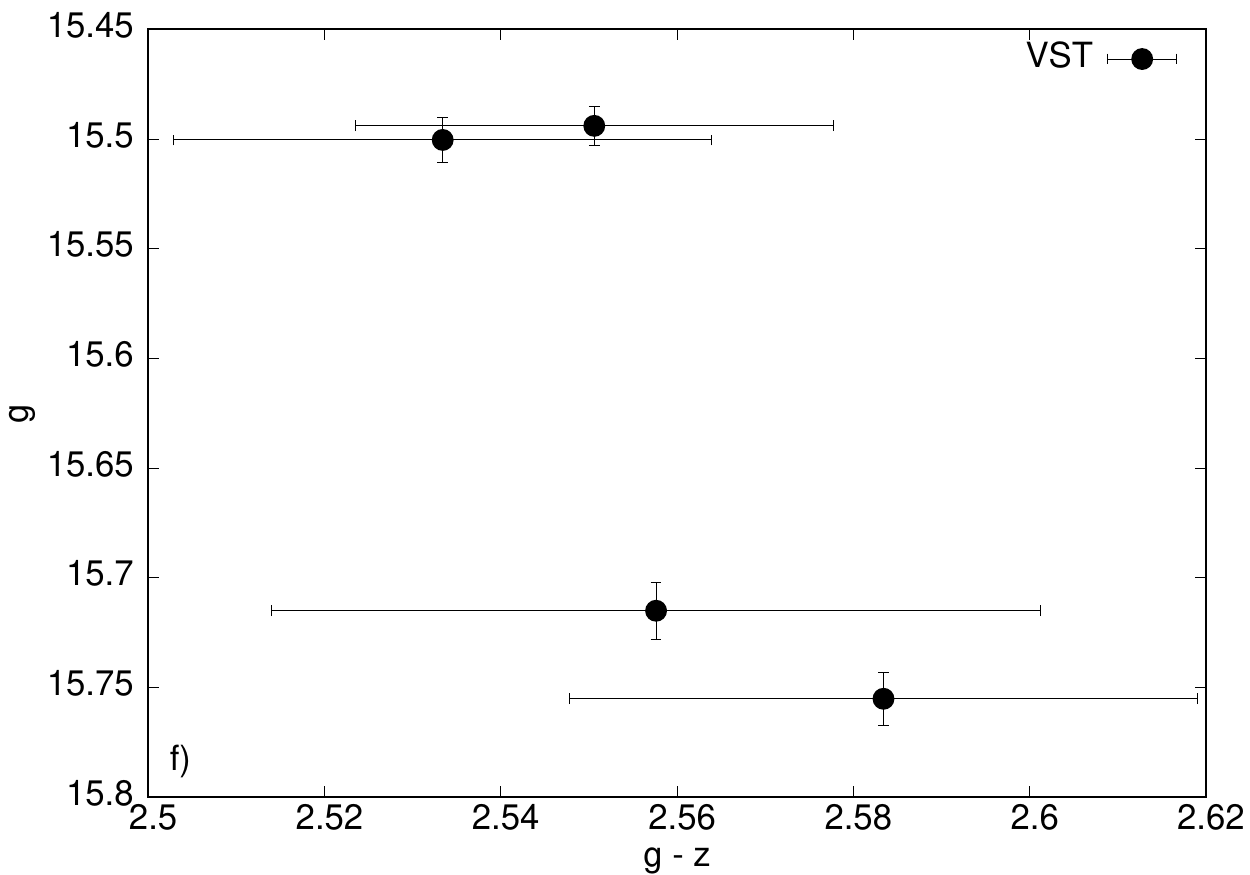}
    \includegraphics[width=0.45\textwidth]{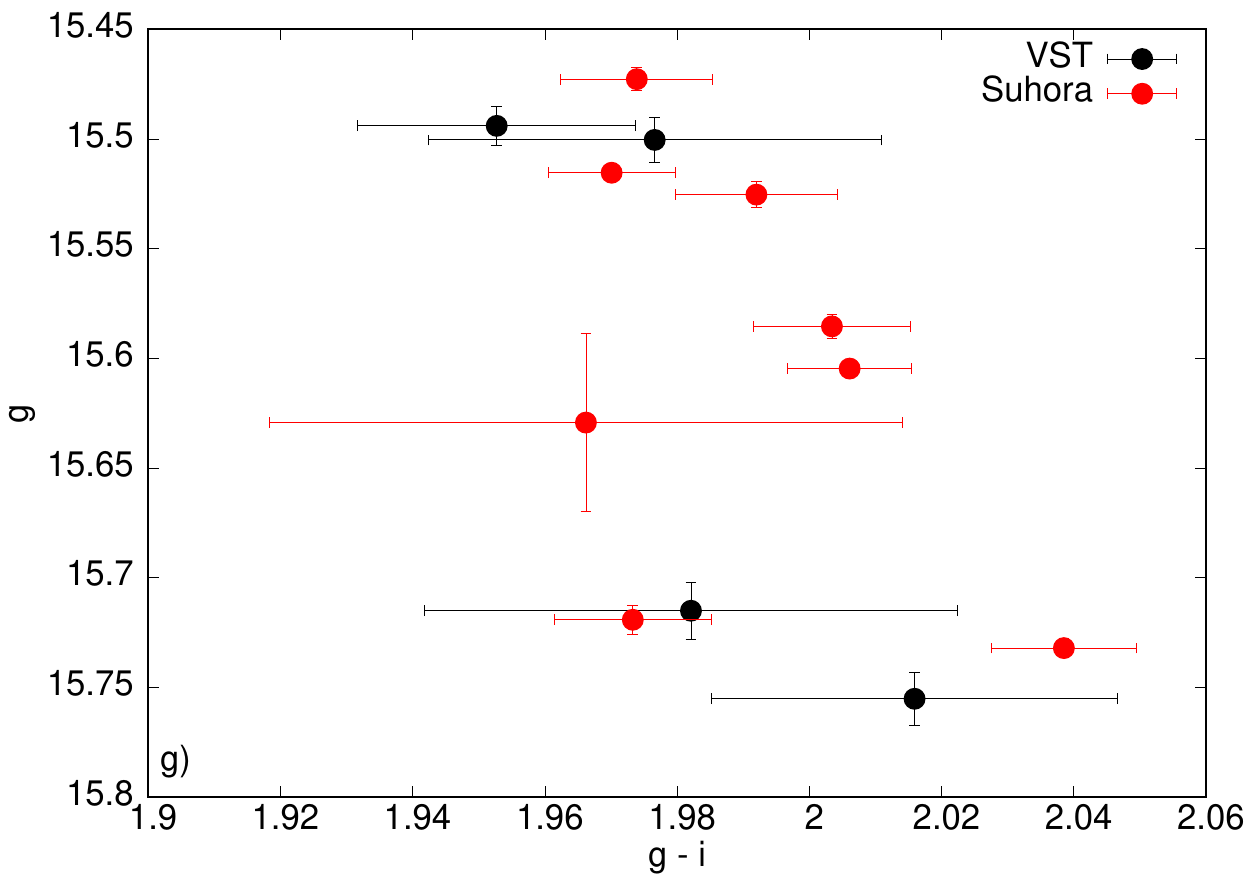}
    \includegraphics[width=0.45\textwidth]{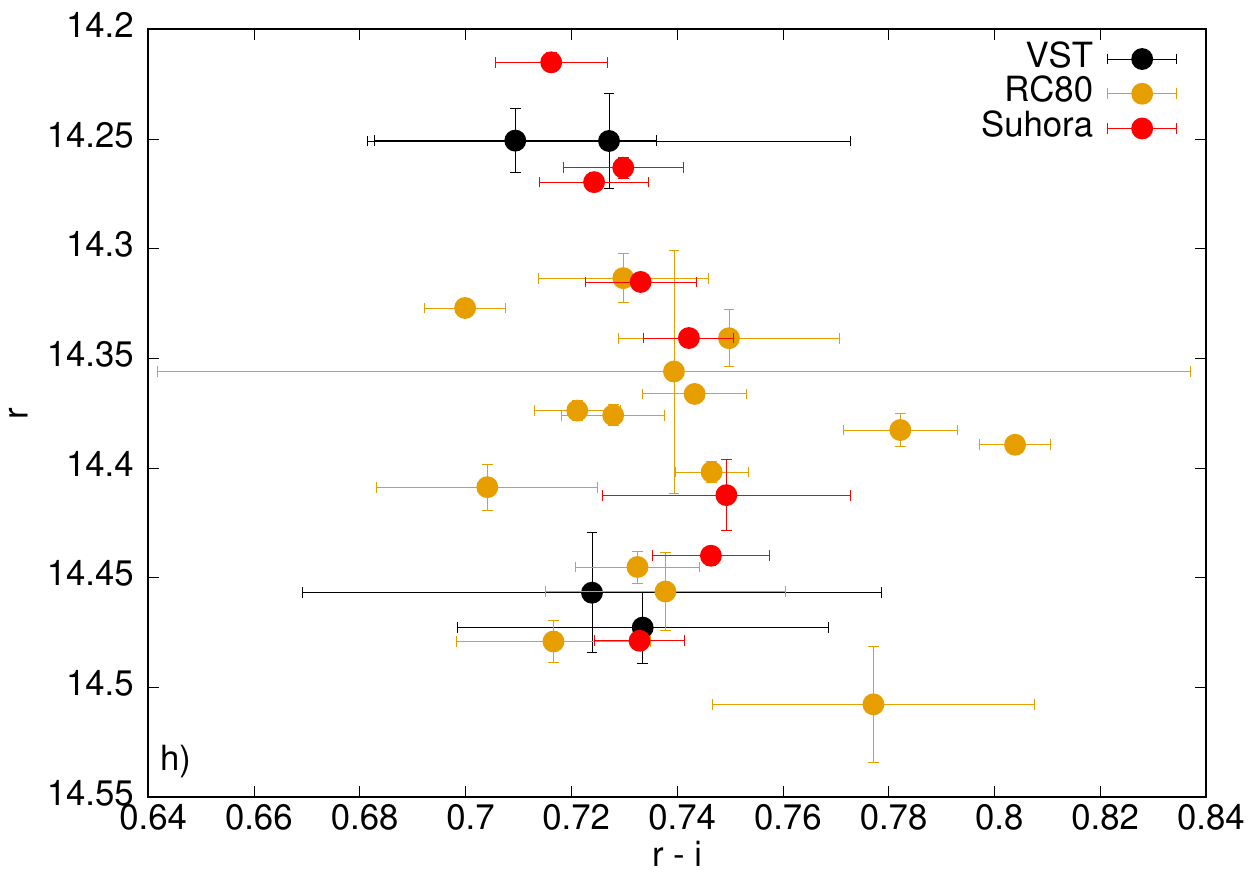}
\caption{Color-magnitude diagrams (CMD) of V899~Mon. Two upper rows (a--d): CMD from the archival and new Johnson filter data. Interstellar reddening vector is indicated by the continuous black line. Two bottom rows (e--h): CMD from the new Sloan filter data. \label{fig_CI_all}}
\end{figure*}

\subsubsection{Color relations}

The long-term color evolution has already been investigated by \citet{ninan2015}. The authors found that the object was reddest during the transition stages from outburst to quiescence and back. Their interpretation was that the outburst had to originate at some place in the disk, like in typical FUors.
We show these archival data together with new data points gathered in recent years in \autoref{fig_CI_all}~a--d. $V$ vs. $(V-R)$ and $V$ vs. $(V-I)$ color-magnitude diagrams (CMDs) constructed from archival and new data show that although the brightness in 2020/2021 (green symbols) was similar as during the transition stages from the first peak to the quiescence and back (black symbols), the colors of V899~Mon are currently bluer by 0.25 and 0.6~mag, respectively (\autoref{fig_CI_all}~c--d).

The major accretion burst (red symbols in \autoref{fig_light_events}), which dominated the variability in the 2017/2018 season, was observed only twice in $BVR_{C}I_{C}$ filters of the Schmidt telescope. The respective CMDs (\autoref{fig_CI_all}~b--d) show that all color indices were significantly bluer when the star was brighter. Although this behavior is similar to that observed in the first outburst in 2010, one should bear in mind that both are caused by different physical mechanisms: during the accretion burst, the color change is caused by the appearance of hot spots on the stellar photosphere, while in 2010 it was caused by the inner disk temperature increase solely through the enhanced accretion.

The same though much better established relation is observed during the 2020/2021 burst. The 2020/2021 data limited to the outburst duration show strong wavelength-dependency during the phenomena: the amplitude ($\sim$1~mag) observed in the $u$-filter is about 3--4 times larger than observed by TESS and in remaining filters (\autoref{fig_CI_all}~e--h). Unfortunately, the SAAO $U$-band data turned out to be too noisy to demonstrate this effect with similar precision as the ESO data. The accretion burst amplitude in $g$-band reached barely 0.25~mag, and is only slightly smaller in $z$-band (\autoref{fig_CI_all}~f). 

\subsection{Spectroscopy \label{sec_optical_spectroscopy}}
\begin{figure}[!htb]
    \centering
    \includegraphics[width=\columnwidth]{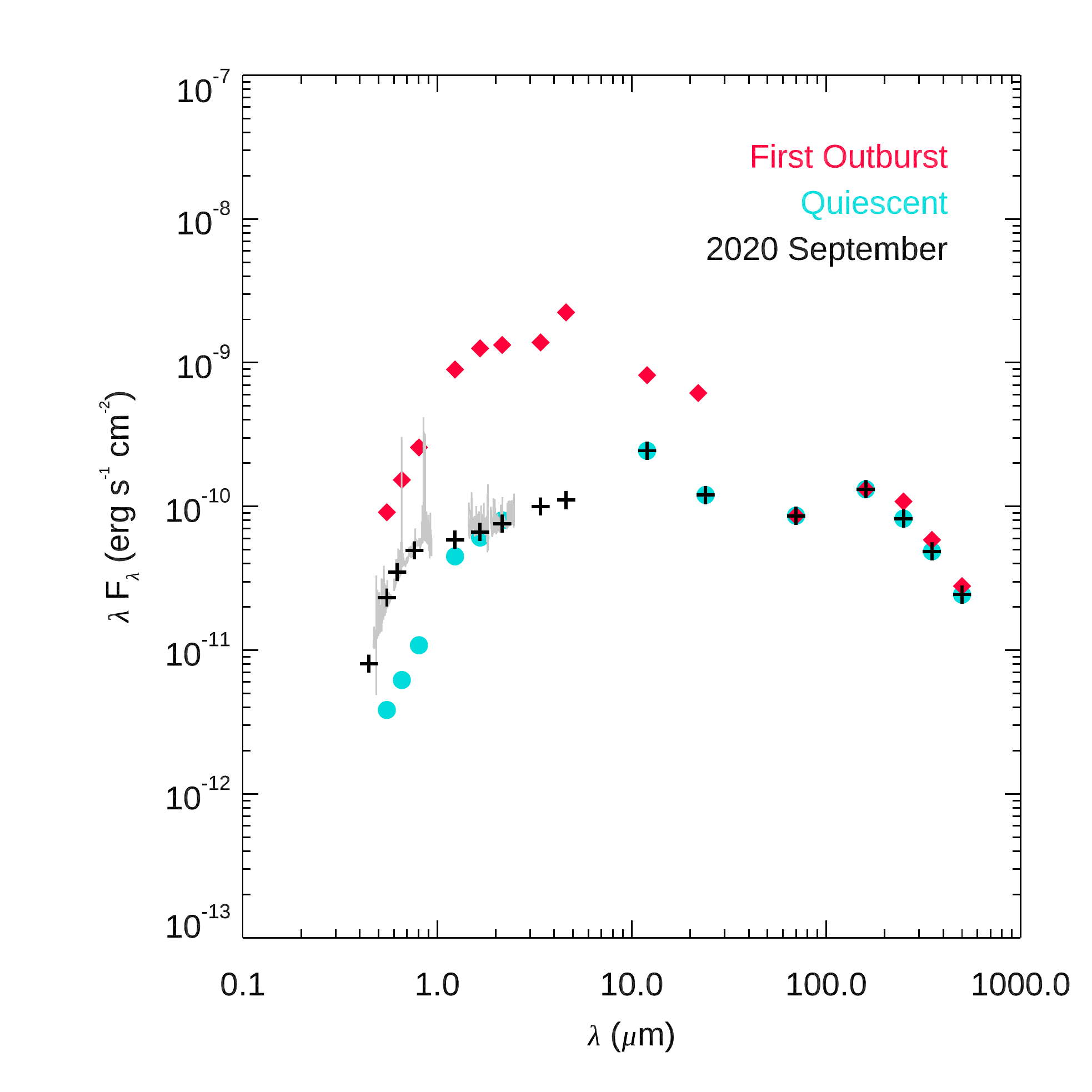}
    \caption{SED of V899 Mon. Red diamond and cyan circle symbols represent the first outburst and quiescent states from \citet{ninan2015}. Black cross symbol indicates our $BVr'i'$ observation and NEOWISE \citep[W1 and W2;][]{Mainzer2014} data observed in 2020 September (dashed gray line in the top panel in \autoref{fig_light}). $JHK$ data points are interpolated between optical and NEOWISE data. For data longer than 10~\um, the quiescent phase data \citep{ninan2015} are plotted. Our spectroscopic observations (gray line) were obtained at an intermediate brightness state between the outburst and quiescent phases. \label{fig_SED}}
\end{figure}

Optical and NIR spectroscopic observations were taken in 2021 January and 2020 November, respectively. To put our spectra into context, in \autoref{fig_SED} we show the SED of V899 Mon with our observations and those from \citet{ninan2015}. 
Our photometric (black crosses) and spectroscopic (gray line) observations were conducted in the brightness between the outburst and quiescent phases. 
We calculated the bolometric luminosity (${L_{\rm bol}}$) using our optical photometry and NEOWISE data obtained in September 2020 and JHK interpolated by optical and NEOWISE. For FIR data, quiescent data of \citet{ninan2015} were used. The obtained ${L_{\rm bol}}$ is 21\,${L_{\rm \odot}}$; the lower \Lbol indicates that V899~Mon is in the fading state compared to the outbursting stage \citep[$\sim$150\,${L_{\odot}}$;][]{ninan2015}. 

\begin{figure*}
    \centering
    \includegraphics[width=\textwidth]{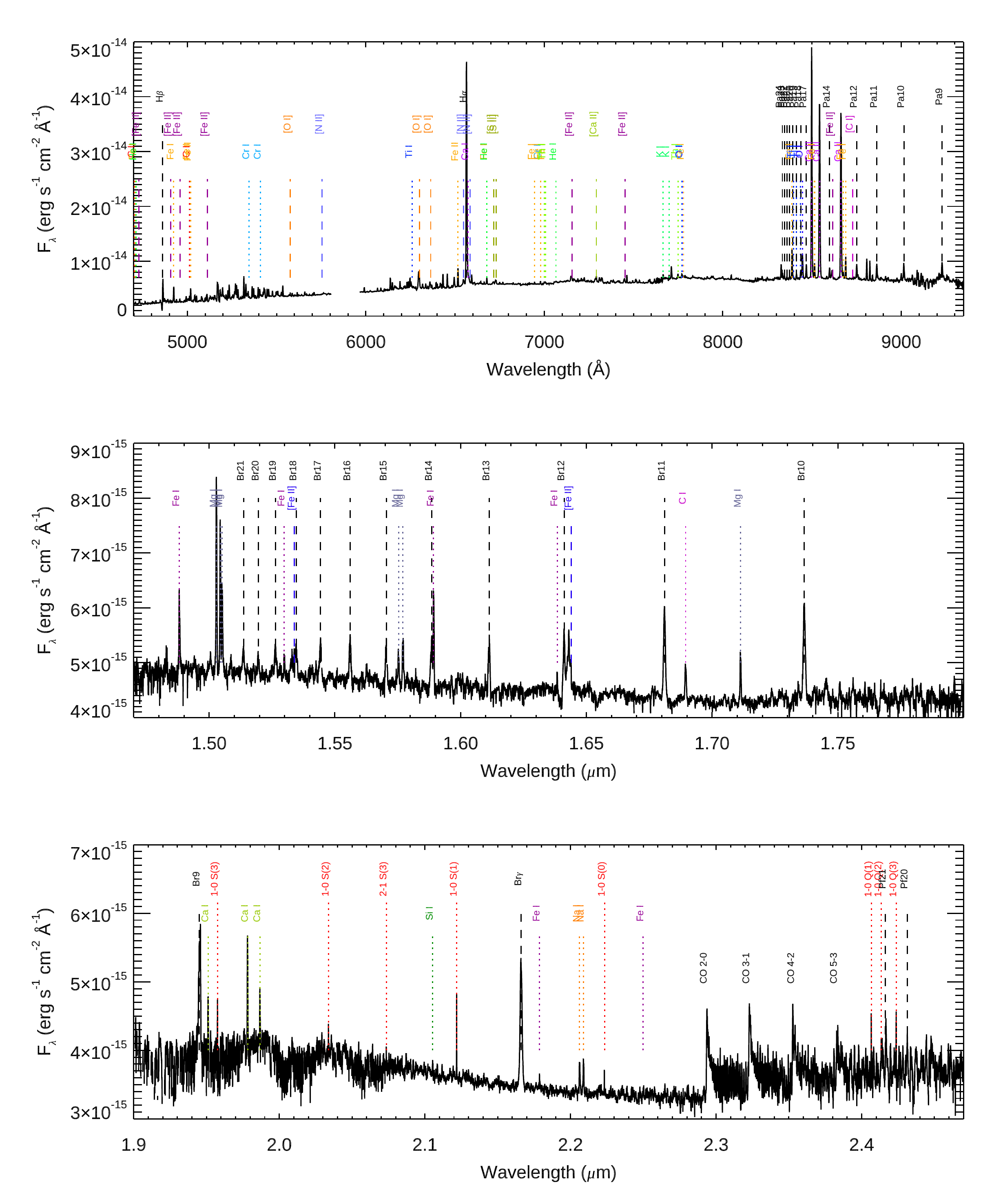} 
    \caption{Optical (top), H (middle), and K (bottom) band spectrum of V899 Mon. 
    Black dashed lines indicate hydrogen lines, and red dotted lines present molecular hydrogen lines. Different colors indicate different metallic species.    \label{fig_spec}}
\end{figure*}

\autoref{fig_spec} shows optical and NIR spectrum of V899~Mon with rich emission lines. In the following subsections, we discuss the various spectral features observed in V899~Mon.

\subsubsection{P Cygni profile} \label{sec_pcygni}
\begin{figure}
    \centering
    \includegraphics[width=\columnwidth]{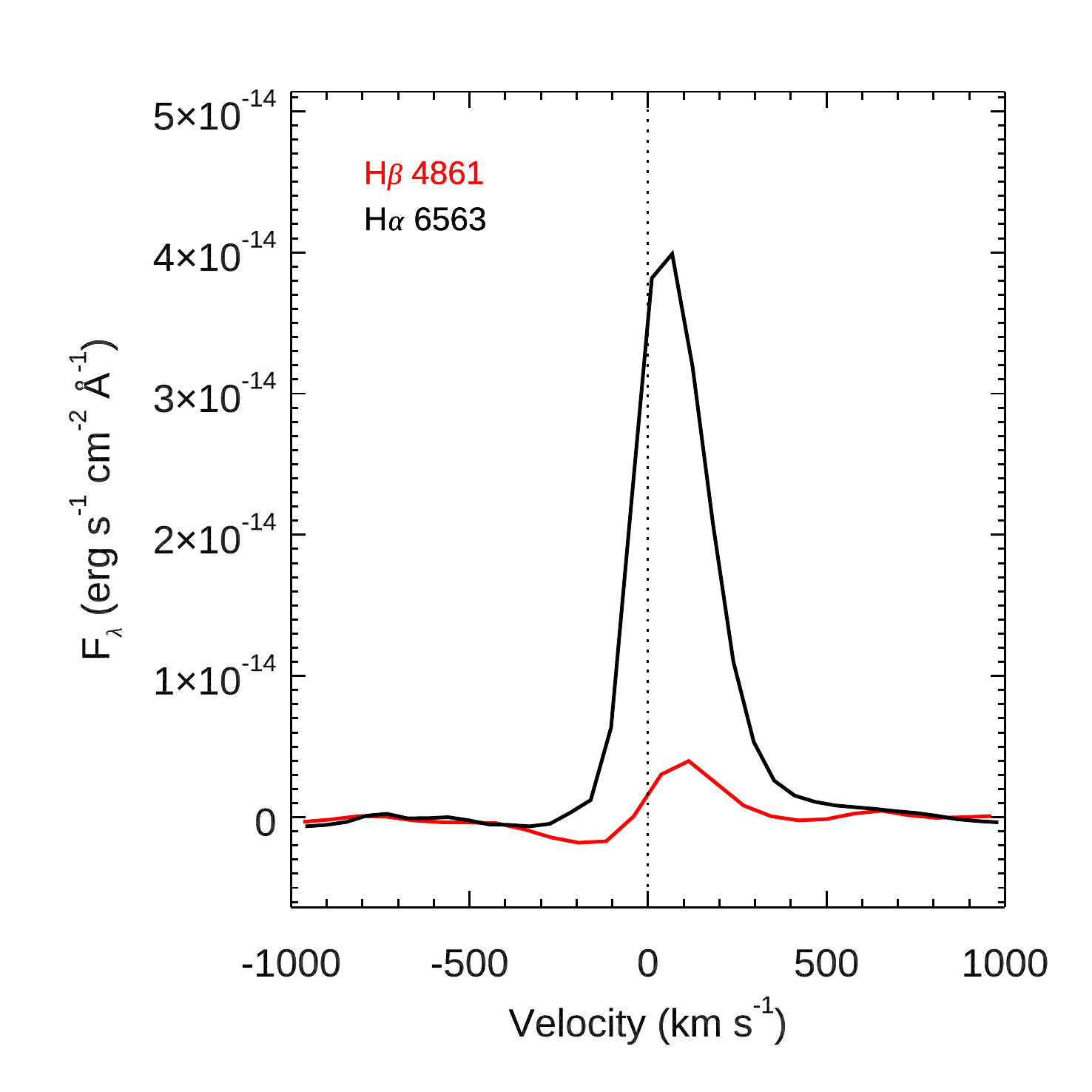} 
    \caption{P Cygni profile of \hbet\ 4861\,\AA{} (red) and \halp\ 6563\,\AA{} (black). \label{fig_pcyg}}
\end{figure}

In the optical spectrum of V899~Mon, \hbet~4861\,\AA{} and \halp~6563\,\AA{} lines have P~Cygni profiles (\autoref{fig_pcyg}).
The former line shows a relatively stronger blue-shifted absorption component to the latter.
\halp\ shows a shallow blue-shifted absorption component ($>$ --332\,km s$^{-1}$ where the intensity is minimum) and a strong emission profile ($\sim$ 60\,km s$^{-1}$) in our observation (black). The blue-shifted absorption component is produced by an outflowing wind \citep{hartmann1996, hartman2009}, and typically FUors show strongly blue-shifted absorption components.
Immediately after V899~Mon was discovered, \halp\ was observed as emission in 2009 November \citep{wils2009_ATel}. 
\halp\ showed a strong blue-shifted absorption component of the P~Cygni profile in the outbursting state in \citet{ninan2015} and was also observed as a P~Cygni profile in \citet{herczeg2016}. Then the blue-shifted absorption component became weaker as V899~Mon faded. 
A nearly symmetric profile of \halp\ was observed in the quiescent phase \citep[Figure~17 in][]{ninan2015}, and our observations recovered a similar profile.

According to previous studies \citep{ninan2015, ninan2016}, \CaII\ IRT lines also showed P~Cygni profiles during the outbursting stage and evolved similarly to \halp. In our observation, \CaII\ shows only symmetric emission profiles (\autoref{fig_CaII}). Combined with the \halp\ line profile, the weakening of the blue-shifted absorption component of the P~Cygni profiles indicates that the strength of the outflowing wind became weaker compared to the outbursting stage.

The line ratio of \halp/\hbet\ in our observation is about 18, which is 2 or 3 times higher than the ratio found by \citet{ninan2015}. 
The higher ratio in our observation also indicates the weakening of the outflowing wind based on the line profile of \halp. 

\subsubsection{\CaII\ IRT lines}

\begin{figure}
    \centering
    \includegraphics[width=\columnwidth]{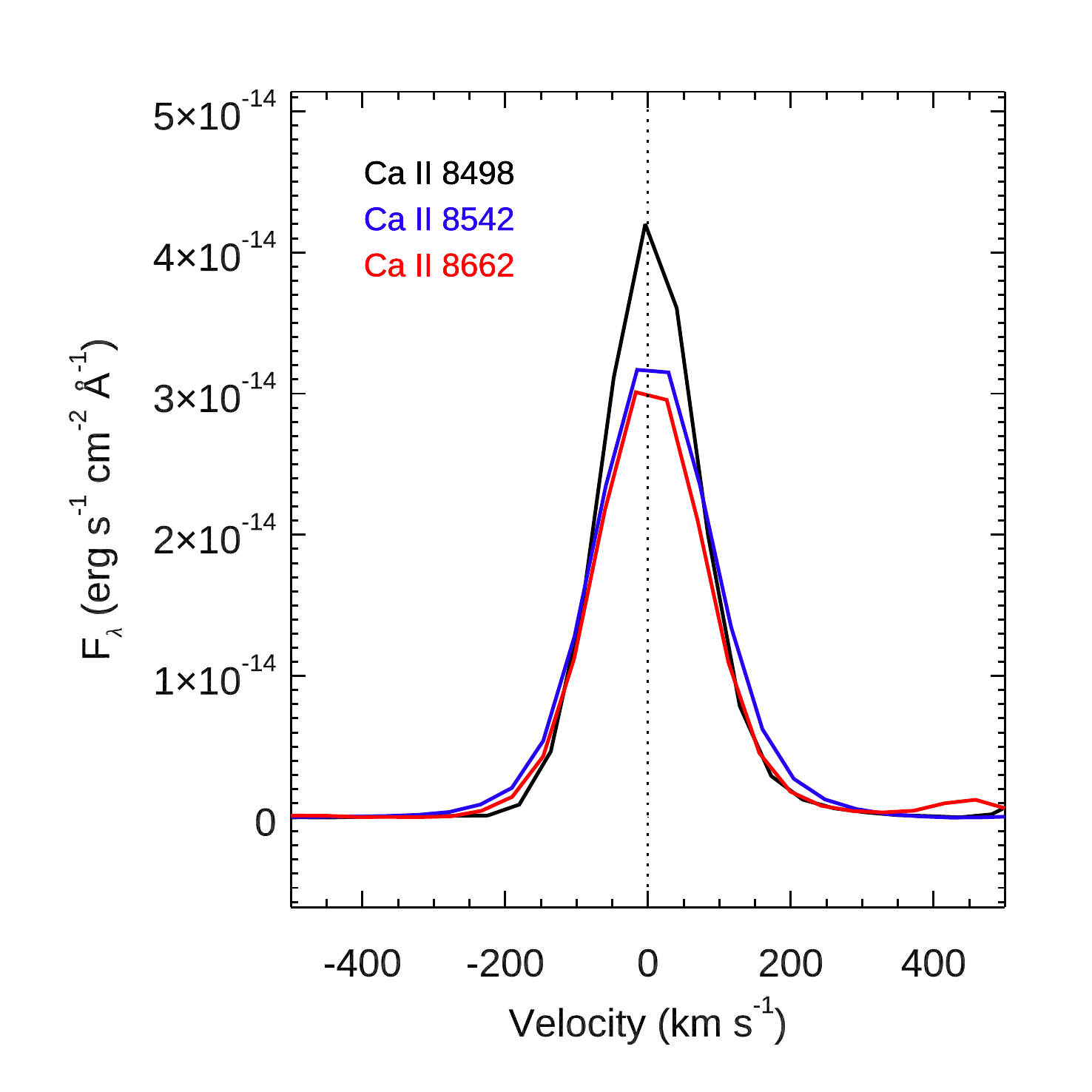} 
    \caption{\CaII\ IRT emission lines: 8498\,\AA{} (black), 8542\,\AA{} (blue), and 8662\,\AA{} (red). \label{fig_CaII}}
\end{figure}

As mentioned in Section~\ref{sec_pcygni}, \CaII\ IRT lines were observed in emission (FWHM $>$ 162~km s$^{-1}$, \autoref{fig_CaII}). This feature is often seen in CTTS and is interpreted as originating in the magnetospheric accretion \citep{muzerolle1998}, also seen in EXors \citep{herbig2008, sipos2009, sicilia-aguilar2012, hillenbrand2019, hodapp2019, hodapp2020}, and V1647~Ori \citep{aspin2010, aspin2011, ninan2013}. 

The intensity ratio between the \CaII\ IRT emission lines is nearly equal, 1.06:1.00:0.86, which suggests optically thick emission \citep{herbig1980_CaII, hamann1992, azevedo2006}. 
This ratio is similar to the previous study \citep[1:1.01:0.77;][]{ninan2015}, and also similar to CTTS \citep{herbig1980_CaII, hamann1992}, EXors \citep{hodapp2019, hodapp2020}, and V1647~Ori \citep{ninan2013}.
While the line ratio is similar to that of the outbursting stage, the relative intensity of the 8498\,\AA{} line became stronger.
The strength of the 8542\,\AA{} line should be the strongest according to the transition probability, but the opposite trend (8498\,\AA{} $>$ 8542\,\AA{}) is also observed in many CTTS \citep{hamann1992}.
In this case, the width of 8498\,\AA{} is smaller than that of 8542\,\AA{}, and the widths of 8542\,\AA{} and 8662\,\AA{} are similar.
The intensity ratios and line profiles of \CaII\ IRT of V899~Mon show similarity to CTTS.

\subsubsection{Permitted atomic lines}
\begin{figure}
    \centering
    \includegraphics[width=\columnwidth]{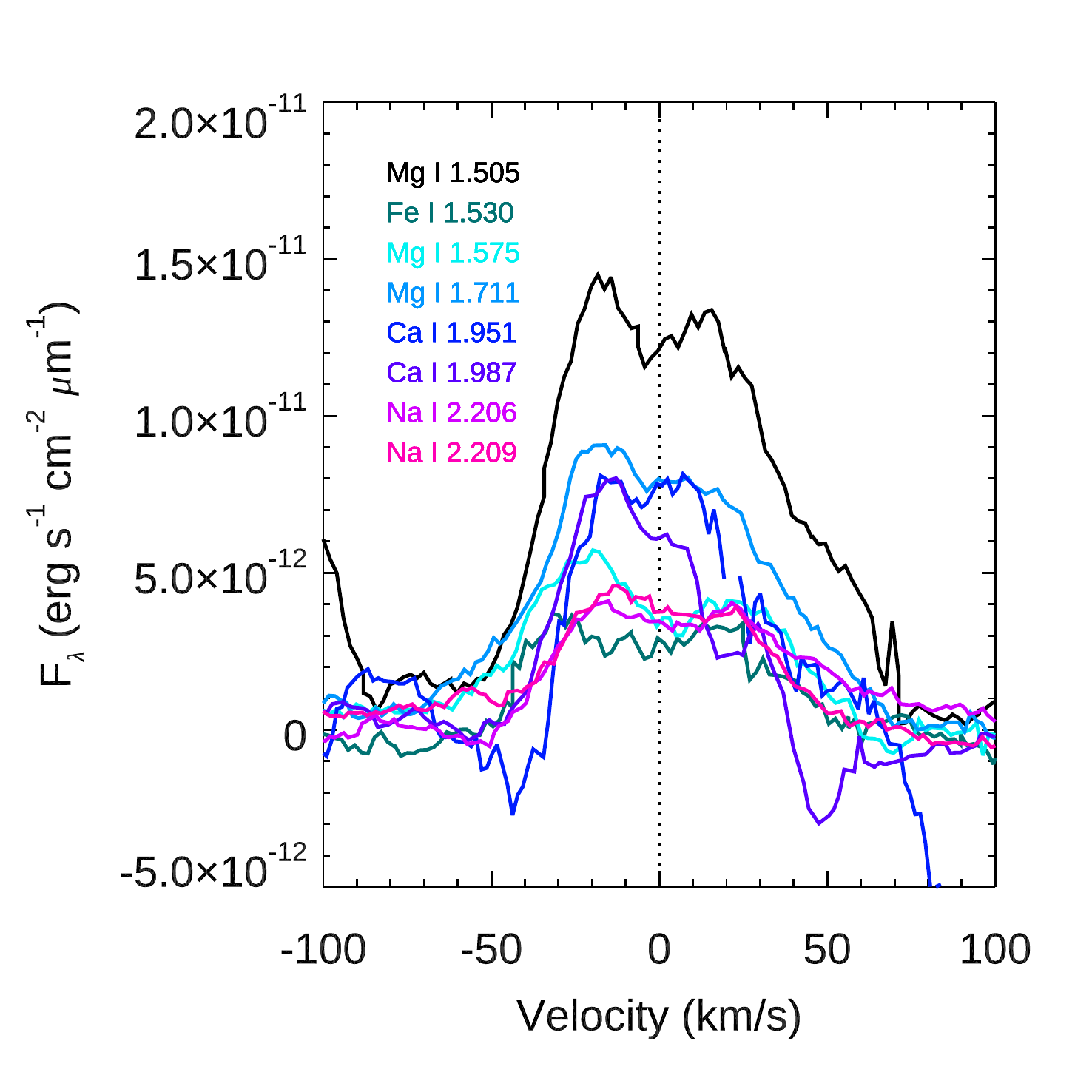}
    \caption{Permitted emission lines showing asymmetric double-peaked line profiles. Different colors indicate different lines. \label{fig_asymmetric}}
\end{figure}

\begin{figure}
    \centering
    \includegraphics[width=\columnwidth]{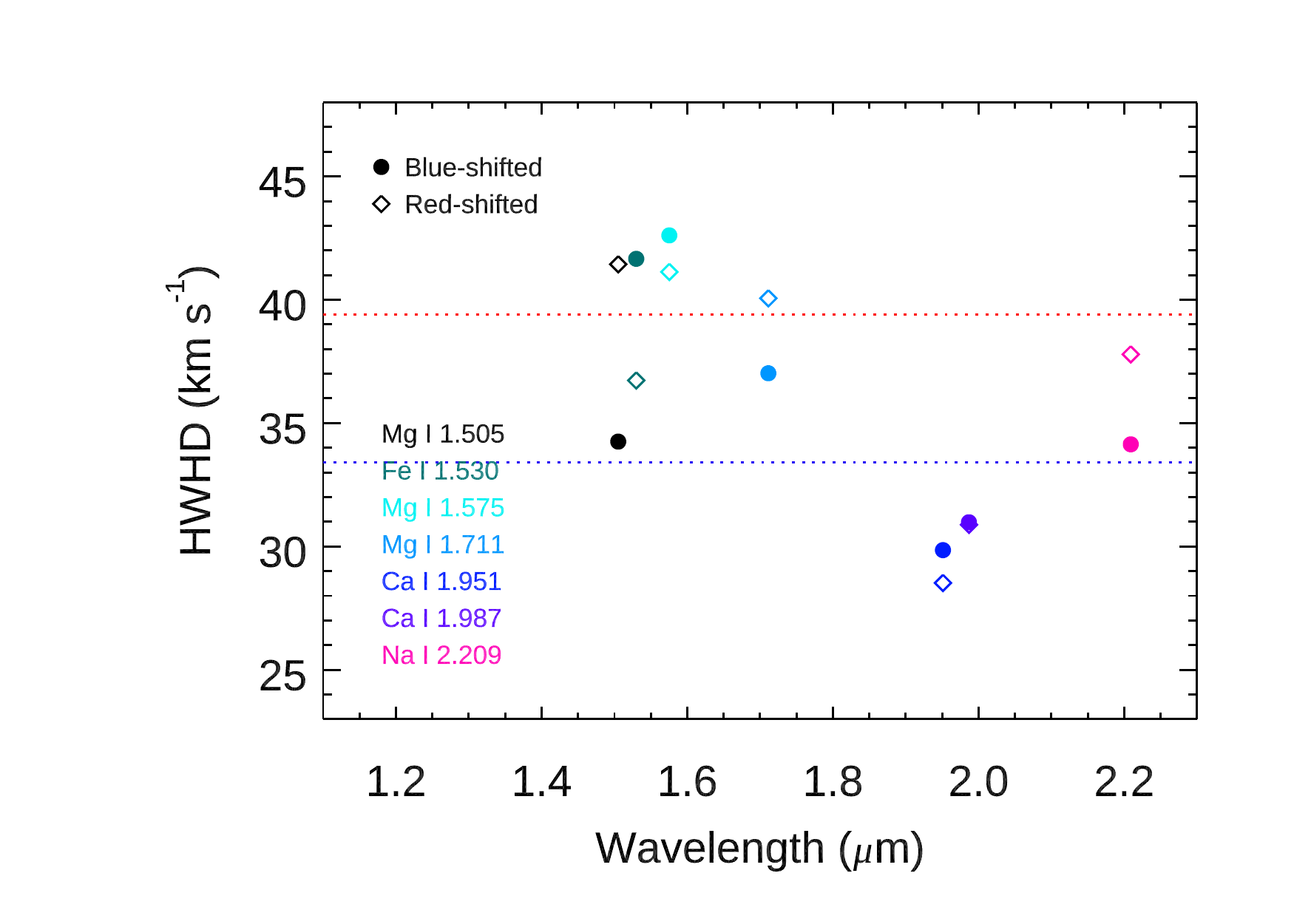}
    \caption{Half-width at half-depth (HWHD) as a function of wavelength. Circle and diamond symbols represent blue-shifted and red-shifted HWHD, respectively. Blue and red dotted lines show blue-shifted and red-shifted HWHD of the average permitted line profile (\autoref{fig_disk_metals}). Different colors indicate different lines. \label{fig_hwhd}}
\end{figure}

A number of permitted atomic emission lines are observed in the optical and NIR spectra of V899~Mon similar to CTTS \citep{muzerolle1998}, EXors \citep{lorenzetti2009, kospal2011, sicilia-aguilar2012, hodapp2019, hodapp2020}, and V1647~Ori \citep{fedele2007, aspin2011, ninan2013}. The observed lines are marked in \autoref{fig_spec}: {C\,{\footnotesize I}}, \FeI, \HeI, {Cr\,{\footnotesize I}}, {Ti\,{\footnotesize I}}, \CaI, {Th\,{\footnotesize I}}, \KI, {Cl\,{\footnotesize I}}, \OI, \MgI, {Si\,{\footnotesize I}}, and \NaI. 
The line flux of the relatively strong and isolated lines is measured and listed in Table~\ref{tbl_lines}.
Optical lines are observed with single Gaussian profiles, while NIR lines are resolved with double-peaked line profiles. \autoref{fig_asymmetric} shows NIR asymmetric double-peaked line profiles. 
We measured the half-width at half-depth \citep[HWHD;][]{petrov2008} of each blue-shifted and red-shifted wing because of the line asymmetry. \autoref{fig_hwhd} shows the measured HWHD as a function of wavelength. \NaI~2.2062\,\um\ line is blended by {Si\,{\footnotesize I}}~2.2069\,\um\ line; therefore, we did not measure the HWHD of this line. A general trend of HWHD decreases with increasing wavelength, suggesting that these atomic lines are formed at the disk \citep{zhu2009, lee2015, park2020}.

\subsubsection{Forbidden lines}
We detected numerous forbidden emission lines in V899~Mon. This makes it more similar to EXors or CTTS rather than FUors \citep{hamann1992, hamann1994, hodapp2019, hodapp2020}. These lines are: [\OI] 5577, 6300, 6364, [\NII] 6548, [\SII] 6716, 6731, [\FeII] 4728, 4907, 4959, 5112, 7155, 7453, 8617, [\CaII] 7291, [\CI] 8727~\AA, [\FeII] 1.533, and [\FeII] 1.644~\um. The central peak velocity of the detected forbidden emission line is blue-shifted with respect to the systemic velocity by about --200~km s$^{-1}$.

\begin{figure}
    \centering
    \includegraphics[width=\columnwidth]{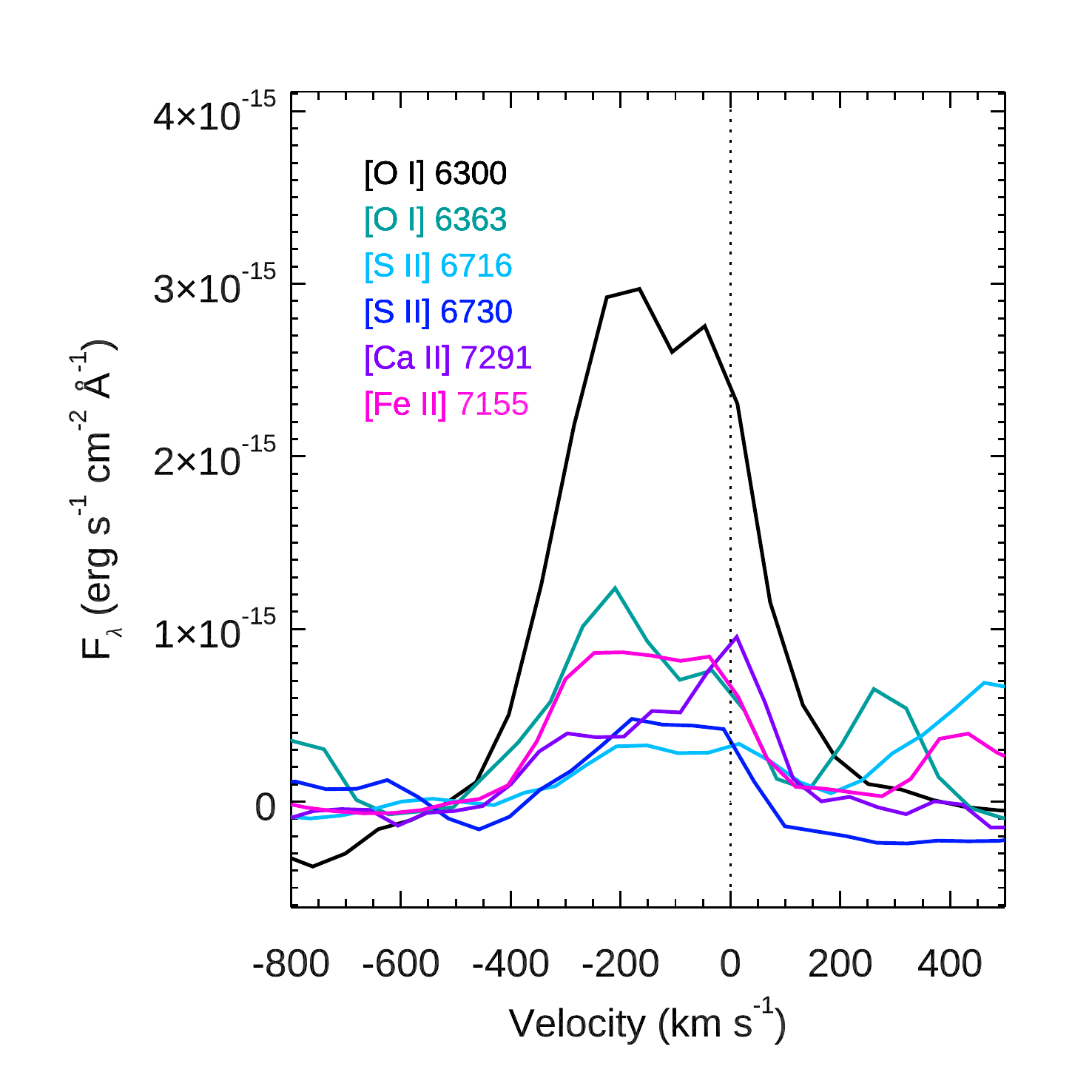}
    \caption{Example of strong forbidden emission lines detected in the optical spectrum. Most of the forbidden emission lines are blue-shifted about --200 km\,s$^{-1}$. \label{fig_forbidden}}
\end{figure}

Forbidden emission lines in CTTS are often interpreted as jet tracers and they are typically composed of two or more velocity components \citep{hartigan1995}.
The forbidden emission lines at optical wavelengths in V899~Mon also show two components similar to CTTS \citep{hartigan1995, hirth1997, pyo2003, Banzatti2019}: higher velocity components between --400 and --100 km\,s$^{-1}$ and relatively lower velocity components between --100 and 0 km\,s$^{-1}$. \autoref{fig_forbidden} shows examples of forbidden emission lines observed in the optical spectrum.

\begin{figure}
    \centering
    \includegraphics[width=\columnwidth]{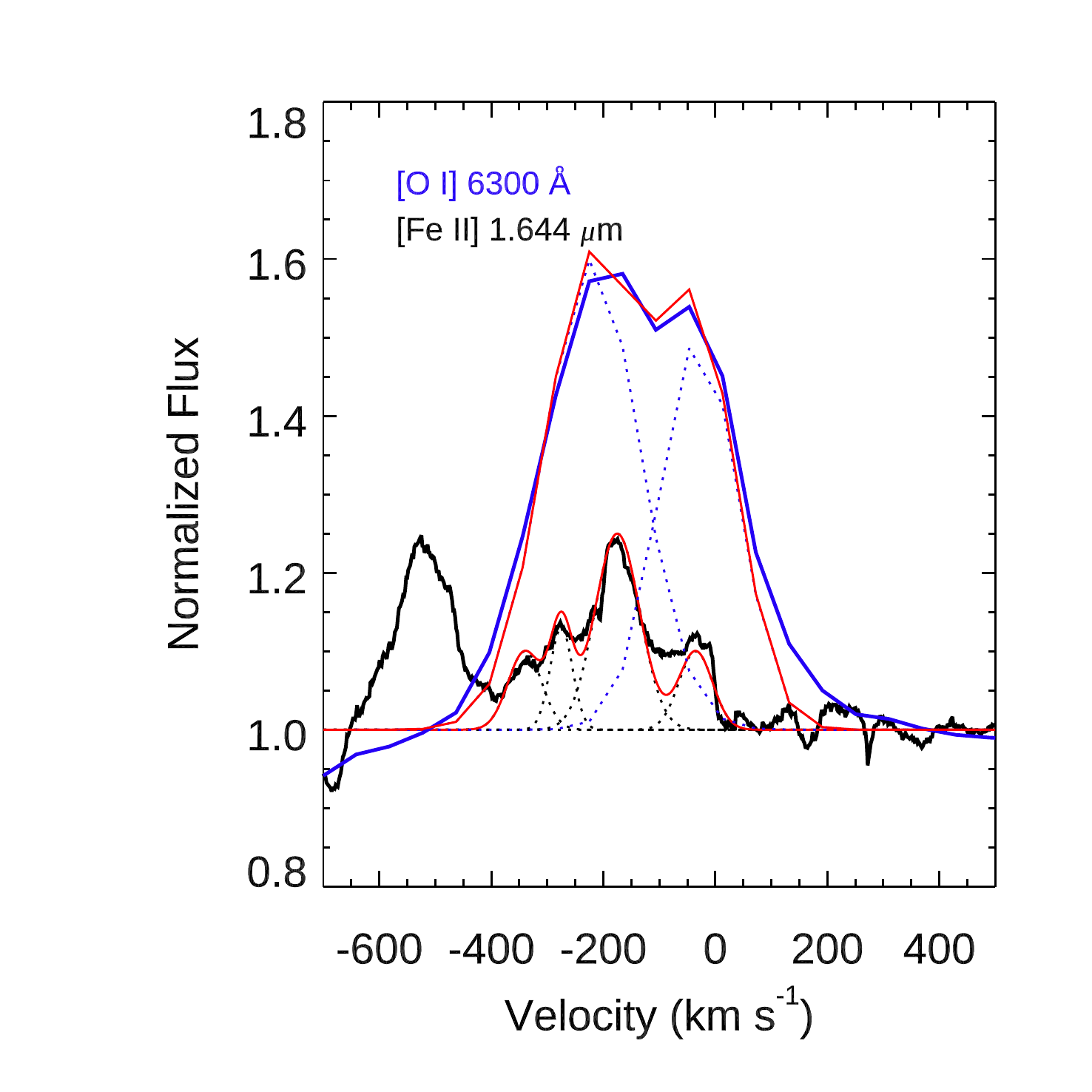}
    \caption{Different velocity components are detected at the [\OI] 6300\,\AA{} (blue) and [\FeII] 1.644\,\um\ (black) line. [\OI] 6300\,\AA{} shows two velocity components around --220 and --30\,km s$^{-1}$, and [\FeII] 1.644\,\um\ shows four velocity components around --340, --275, --175, and --35\,km s$^{-1}$. Solid and dotted lines indicate the observed spectrum and each Gaussian profile. Red line presents total Gaussian profile. \label{fig_forbidden_gauss}}
\end{figure}

\autoref{fig_forbidden_gauss} shows relatively strong forbidden emission lines in optical ([\OI] 6300\,\AA) and NIR ([\FeII] 1.644\,\um), with multiple Gaussian fittings.
Higher resolution NIR [\FeII] 1.533\,\um\ and [\FeII] 1.644\,\um\ lines can be decomposed into four components around --340, --275, --175, and --35 km\,s$^{-1}$. 
The resolved highest velocity component of the [\FeII]\,1.644\,\um\ line is consistent with the minimum intensity of the blue-shifted absorption component (--332\,km s$^{-1}$) of the \halp\ P~Cygni profile.

The temperature and density of the outflow region can be obtained using the forbidden emission line ratios. 
The electron density ($n_{e}$) can be estimated by using three flux ratios: [\SII]~6716/6731, [\OI]~5577/6300 and [\FeII]~1.644/1.533 \citep{hamann1994, nisini2002}.
The [\SII]~6716/6731 ratio is sensitive to the density while insensitive to the temperature, and based on Figure~7 from \citet{hamann1994}, its value (0.49) indicates the estimate $n_{e}$ is higher than 10$^4$\,\rm cm$^{-3}$. The [\OI]~5577/6300 ratio of 0.02 implies a temperature of 9000\,K, and an $n_{e}$ of 10$^{5}$--10$^{6}$\,\rm cm$^{-3}$, consistent with the [\SII] ratio. 
Finally, the log([\FeII] 1.644/1.533) ratio, 0.40, points to $n_{e}$ higher than 10$^{5}$\,\rm cm$^{-3}$ \citep[Figure~8 of][]{nisini2002}, in agreement with its two previous estimations.
Thus, the $n_{e}$ of the outflow is higher than 10$^{4}$\,\rm cm$^{-3}$, consistent with the outbursting stage.

Electron temperature ($T_{e}$) can be estimated by the line ratio of [\CaII] 7291/[\OI] 6300, [\SII] 6731/[\OI]6300, and [\CI] 8727/[\OI] 6300 \citep[Figure~5 in][]{hamann1994}, and the obtained ratios are 0.26, 0.11, and 0.05, respectively. 
The [\CaII]/[\OI] and [\CI]/[\OI] line ratios indicate $T_{e}$ is below 9000~K, while the [\SII]/[\OI] points towards higher temperatures than this. Therefore, we assume $T_{e}$ must be around 9000~K.
The obtained temperature is similar to that of the outbursting stage \citep[9000\,K;][]{ninan2015}, which suggests the outflow temperature and density of V899~Mon might not be directly affected by the accretion process.

\subsubsection{Mass accretion rates}
Several atomic emission lines formed by the accretion process are found in the optical and NIR spectrum. 
In order to estimate the mass accretion rate, the interstellar extinction of \Av=2.6 \citep{ninan2015} was corrected by adopting the extinction curve of \citet{cardelli1989}.

The line flux (\Fline) of emission lines was measured by fitting a Gaussian with a Monte Carlo method. The line flux of each line was measured 100 times with random Gaussian errors multiplied by the observation errors. The standard deviation derived from all 100 measurements was adopted as the uncertainty of line flux. The measured line fluxes and uncertainties are listed in Table~\ref{tbl_lines}.
Then, the line luminosity (\Lline), accretion luminosity (\Lacc), and mass accretion rate (\Macc) are estimated by the following Eqs.~\ref{eq_Lline}--\ref{eq_Macc}. 
Accretion luminosity was calculated by adopting the coefficients of \citet{alcala2017} for a and b in Eq.~\ref{eq_Lacc_a17}.

\begin{equation} \label{eq_Lline}
    L_{\rm line}=4\pi d^{2} \cdot F_{\rm line}
\end{equation}

\begin{equation} \label{eq_Lacc_a17}
    log \left(\frac{L_{\rm acc}} {L_{\odot}} \right)=a \cdot log\left(\frac{L_{\rm line}} {L_{\odot}}\right) + b
\end{equation}

\begin{equation} \label{eq_Macc}
    \dot{M}_{\rm acc}=\left(1- \frac{R_{*}} {R_{\rm in}}\right)^{-1}~\frac{L_{\rm acc} R_{*}} {G M_{*}}~\approx~1.25~\frac{L_{\rm acc} R_{*}} {G M_{*}},
\end{equation}
where $R_{\rm in}$ is the disk inner radius and assumed to be $5R_{*}$~\citep{gullbring1998}.

In order to compare \Macc\ in our observations with those of the outbursting phase, we adopted the same distance (905~pc), stellar mass (2.5~$M_{\odot}$), and stellar radius (4~$R_{\odot}$) as \citet{ninan2015} assumed.
The calculated \Macc\ is listed in Table~\ref{tbl_lines}.
The \Macc\ estimated in our observations are in the range of 6.0 $\times$ 10$^{-8}$ to 4.8 $\times$ 10$^{-7}$ $M_{\odot}$~yr$^{-1}$, which are about an order of magnitude lower than those of the outbursting phase \citep[see Figure~15 in][]{ninan2015}. 
The mean and standard deviation of \Macc\ is (2.2 $\pm$ 1.3) $\times$ 10$^{-7}$ $M_{\odot}$~yr$^{-1}$. 
If we use the 2$M_{\odot}$ obtained from our disk model (see Section~\ref{sec_disk_model}), the calculated \Macc\ is between 7.4$\times$ 10$^{-8}$ to 6.0$\times$ 10$^{-7}$ $M_{\odot}$~yr$^{-1}$. The mean and standard deviation of \Macc\ is (2.7 $\pm$ 1.6) $\times$ 10$^{-7}$ $M_{\odot}$~yr$^{-1}$, which are in agreement with those of \Macc\ obtained with 2.5 \Msun within the standard deviation.
The \Macc\ of V899~Mon is lower than FUors \citep[][and references therein]{audard2014} but similar to EX~Lup during its outburst \citep[(2 $\pm$ 0.5) $\times$ 10$^{-7}$ $M_{\odot}$~yr$^{-1}$;][]{aspin2010}.

\begin{deluxetable*}{lcccccc}
\tabletypesize{\scriptsize} 
\tablecaption{Line Information \label{tbl_lines}}
\tablewidth{0pt}
\tablehead{\colhead{Transition} & \colhead{Wavelength} & \colhead{$F_{\rm line}$ (Obs.)} & \colhead{$F_{\rm line}$ (\Av\ Corr.)} & \colhead{$L_{\rm line}$} & \colhead{$L_{\rm acc}$} & \colhead{$M_{\rm acc}$} \\ [-2mm]
\colhead{} & \colhead{(\AA{})} & \colhead{(erg s $^{-1}$ cm$^{-2}$)}  & \colhead{(erg s $^{-1}$ cm$^{-2}$)}  & \colhead{($L_{\odot}$)} & \colhead{($L_{\odot}$)} & \colhead{($M_{\odot}$ \yr)}}
\startdata
        H$\beta$ &    4861.33 &   1.19E-14 $\pm$   1.39E-15 &   1.97E-13 $\pm$   7.56E-14 &   5.02E-03 &   9.31E-01 &   5.95E-08 \\ \relax
       Fe\,{\scriptsize I} &    4921.93 &   9.15E-15 $\pm$   1.18E-15 &   1.57E-13 $\pm$   2.16E-14 &   4.00E-03 &   5.68E+00 &   3.63E-07 \\ \relax
       He\,{\scriptsize I} &    5015.68 &   3.73E-15 $\pm$   1.03E-15 &   5.51E-14 $\pm$   1.41E-14 &   1.40E-03 &   4.63E+00 &   2.96E-07 \\ \relax
        H$\alpha$ &    6562.80 &   2.17E-13 $\pm$   1.07E-14 &   1.54E-12 $\pm$   7.01E-14 &   3.92E-02 &   1.42E+00 &   9.05E-08 \\ \relax
       He\,{\scriptsize I} &    6678.15 &   3.92E-15 $\pm$   1.61E-15 &   2.58E-14 $\pm$   1.10E-14 &   6.58E-04 &   5.28E+00 &   3.37E-07 \\ \relax
       He\,{\scriptsize I} &    7065.19 &   1.26E-15 $\pm$   1.74E-15 &   8.38E-15 $\pm$   9.99E-15 &   2.14E-04 &   1.38E+00 &   8.80E-08 \\ \relax
        O\,{\scriptsize I} &    7773.05 &   2.32E-15 $\pm$   1.90E-15 &   1.07E-14 $\pm$   7.90E-15 &   2.73E-04 &   1.36E+00 &   8.69E-08 \\ \relax
      Ca\,{\scriptsize II} &    8498.02 &   2.00E-13 $\pm$   9.75E-15 &   7.13E-13 $\pm$   3.46E-14 &   1.82E-02 &   7.53E+00 &   4.81E-07 \\ \relax
      Ca\,{\scriptsize II} &    8542.09 &   1.92E-13 $\pm$   1.00E-14 &   6.75E-13 $\pm$   3.46E-14 &   1.72E-02 &   5.23E+00 &   3.34E-07 \\ \relax
      Ca\,{\scriptsize II} &    8662.14 &   1.71E-13 $\pm$   9.23E-15 &   5.83E-13 $\pm$   3.14E-14 &   1.49E-02 &   3.98E+00 &   2.54E-07 \\ \relax
        O\,{\scriptsize I} &    8446.36 &   2.11E-14 $\pm$   2.97E-15 &   7.81E-14 $\pm$   1.17E-14 &   1.99E-03 &   3.49E+00 &   2.23E-07 \\ \relax
      Pa10 &    9014.91 &   1.52E-14 $\pm$   2.55E-15 &   4.70E-14 $\pm$   8.59E-15 &   1.20E-03 &   1.74E+00 &   1.11E-07 \\ \relax
       Pa9 &    9229.01 &   1.53E-14 $\pm$   2.71E-15 &   4.67E-14 $\pm$   8.60E-15 &   1.19E-03 &   1.82E+00 &   1.16E-07 \\ \relax
       Br$\gamma$ &   21661.21 &   3.11E-14 $\pm$   2.95E-16 &   4.12E-14 $\pm$   2.90E-16 &   1.05E-03 &   2.99E+00 &   1.91E-07 \\
    \hline
        Fe\,{\scriptsize II} &    4958.82 &   1.87E-15 $\pm$   7.12E-16 &   2.95E-14 $\pm$   1.24E-14 &      \dots &      \dots &      \dots \\ \relax
        [O\,{\scriptsize I}] &    5577.34 &   6.22E-16 $\pm$   1.18E-15 &   5.55E-15 $\pm$   1.31E-14 &      \dots &      \dots &      \dots \\ \relax
      [O\,{\scriptsize I}] &    6300.30 &   2.52E-14 $\pm$   2.68E-15 &   1.97E-13 $\pm$   2.16E-14 &      \dots &      \dots &      \dots \\ \relax
      [O\,{\scriptsize I}] &    6363.78 &   8.38E-15 $\pm$   2.17E-15 &   6.58E-14 $\pm$   1.69E-14 &      \dots &      \dots &      \dots \\ \relax
      Fe\,{\scriptsize II} &    6516.08 &   1.08E-14 $\pm$   1.79E-15 &   8.50E-14 $\pm$   1.55E-14 &      \dots &      \dots &      \dots \\ \relax
     [N\,{\scriptsize II}] &    6548.05 &   5.74E-15 $\pm$   2.43E-15 &   4.20E-14 $\pm$   1.68E-14 &      \dots &      \dots &      \dots \\ \relax
     [S\,{\scriptsize II}] &    6716.44 &   2.86E-15 $\pm$   2.29E-15 &   1.97E-14 $\pm$   1.55E-14 &      \dots &      \dots &      \dots \\ \relax
     [S\,{\scriptsize II}] &    6730.82 &   5.85E-15 $\pm$   2.42E-15 &   4.34E-14 $\pm$   1.64E-14 &      \dots &      \dots &      \dots \\ \relax
    [Fe\,{\scriptsize II}] &    7155.17 &   7.74E-15 $\pm$   2.48E-15 &   4.59E-14 $\pm$   1.45E-14 &      \dots &      \dots &      \dots \\ \relax
    [Ca\,{\scriptsize II}] &    7291.47 &   6.64E-15 $\pm$   2.64E-15 &   3.42E-14 $\pm$   1.66E-14 &      \dots &      \dots &      \dots \\ \relax
    K\,{\scriptsize I}\ &    7664.90 &   2.36E-15 $\pm$   1.77E-15 &   1.16E-14 $\pm$   7.13E-15 &      \dots &      \dots &      \dots \\ \relax
    K\,{\scriptsize I} &    7698.96 &   1.41E-15 $\pm$   1.54E-15 &   6.35E-15 $\pm$   6.29E-15 &      \dots &      \dots &      \dots \\ \relax
    [C\,{\scriptsize I}] &    8727.13 &   1.19E-15 $\pm$   1.83E-15 &   3.51E-15 $\pm$   6.12E-15 &      \dots &      \dots &      \dots \\ \relax
    Fe\,{\scriptsize I} &    8387.77 &   1.52E-14 $\pm$   2.14E-15 &   6.65E-14 $\pm$   1.19E-14 &      \dots &      \dots &      \dots \\ 
    \hline \\[-3mm]
    Transition & Wavelength & Line Flux (Obs.) & Line Flux (Av Corr.) & & & \\
               & (\um)      & (erg s $^{-1}$ cm$^{-2}$)  & (erg s $^{-1}$ cm$^{-2}$)      & & & \\[1mm]
    \hline
      Br21 &     1.5137 &   3.47E-15 $\pm$   1.87E-16 &   5.60E-15 $\pm$   2.02E-16 &      \dots &      \dots &      \dots \\
      Br20 &     1.5196 &   2.23E-15 $\pm$   1.81E-16 &   3.63E-15 $\pm$   1.83E-16 &      \dots &      \dots &      \dots \\
      Br19 &     1.5265 &   5.14E-15 $\pm$   5.05E-16 &   8.41E-15 $\pm$   4.11E-16 &      \dots &      \dots &      \dots \\
      Br17 &     1.5443 &   4.99E-15 $\pm$   1.95E-16 &   8.12E-15 $\pm$   2.03E-16 &      \dots &      \dots &      \dots \\
      Br16 &     1.5561 &   5.75E-15 $\pm$   1.89E-16 &   9.20E-15 $\pm$   1.86E-16 &      \dots &      \dots &      \dots \\
      Br15 &     1.5705 &   4.55E-15 $\pm$   2.34E-16 &   7.15E-15 $\pm$   2.04E-16 &      \dots &      \dots &      \dots \\
      Br13 &     1.6114 &   6.46E-15 $\pm$   2.14E-16 &   1.02E-14 $\pm$   2.10E-16 &      \dots &      \dots &      \dots \\
      Br12 &     1.6412 &   7.68E-15 $\pm$   2.97E-15 &   1.19E-14 $\pm$   2.27E-15 &      \dots &      \dots &      \dots \\
      Br11 &     1.6811 &   1.45E-14 $\pm$   1.79E-16 &   2.19E-14 $\pm$   2.09E-16 &      \dots &      \dots &      \dots \\
      Br10 &     1.7367 &   1.69E-14 $\pm$   2.26E-16 &   2.52E-14 $\pm$   2.09E-16 &      \dots &      \dots &      \dots \\
       Br9 &     1.9451 &   2.44E-14 $\pm$   2.90E-16 &   3.39E-14 $\pm$   2.52E-16 &      \dots &      \dots &      \dots \\
      Pf21 &     2.4164 &   5.86E-15 $\pm$   1.95E-16 &   7.43E-15 $\pm$   1.95E-16 &      \dots &      \dots &      \dots \\
      Pf20 &     2.4314 &   3.29E-15 $\pm$   1.74E-16 &   4.15E-15 $\pm$   1.81E-16 &      \dots &      \dots &      \dots \\
\hline
   1-0S(3) &     1.9580 &   1.10E-15 $\pm$   8.65E-17 &   1.52E-15 $\pm$   8.78E-17 &      \dots &      \dots &      \dots \\
   1-0S(2) &     2.0340 &   6.26E-16 $\pm$   8.64E-17 &   8.48E-16 $\pm$   9.72E-17 &      \dots &      \dots &      \dots \\
   2-1S(3) &     2.0730 &   4.88E-16 $\pm$   8.44E-17 &   6.64E-16 $\pm$   7.71E-17 &      \dots &      \dots &      \dots \\
   1-0S(1) &     2.1220 &   1.66E-15 $\pm$   9.24E-17 &   2.21E-15 $\pm$   8.65E-17 &      \dots &      \dots &      \dots \\
   1-0S(0) &     2.2230 &   5.43E-16 $\pm$   1.15E-16 &   7.10E-16 $\pm$   9.85E-17 &      \dots &      \dots &      \dots \\
   1-0Q(1) &     2.4070 &   8.97E-16 $\pm$   6.92E-17 &   1.13E-15 $\pm$   6.88E-17 &      \dots &      \dots &      \dots \\
   1-0Q(3) &     2.4240 &   1.75E-15 $\pm$   2.68E-16 &   2.20E-15 $\pm$   2.94E-16 &      \dots &      \dots &      \dots \\
\enddata
\end{deluxetable*}

\subsubsection{Mass loss rates}
Mass loss rate was estimated by using the relation of log$\dot{M}$ = --8.6 + 0.7\,log$L_{\rm bol}$ \citep{nisini1995}. \Lbol~of 21~$L_{\odot}$ is used for this relation that is obtained from the SED of our observations. The calculated mass loss rate is about 2.1 $\times$ 10$^{-8}$~$M_{\odot}$~yr$^{-1}$, which is about one order of magnitude lower than that of the outbursting phase \citep[1 $\times$ 10$^{-7}$~$M_{\odot}$~yr$^{-1}$;][]{ninan2015}.
The estimated mass loss rate is lower than those of FUors, while similar to those of CTTS \citep[][and references therein]{hartmann1996}. 
In addition, the estimated mass loss rate is about 10\% of the mass accretion rate, which is consistent with those of CTTS \citep{hartmann1996, hartman2009, ellerbroek2013, bally2016}.


\subsubsection{Molecular Hydrogen Lines}
\hmol~emission lines are known as tracers of outflows in young stars \citep{davis2003, davis2010, davis2011, bally2007, greene2010, bally2016, nisini2002, vandenAncker1999, fernandes2000}. Several \hmol~rovibrational transitions were detected in our observation (red dotted lines in \autoref{fig_spec}), which is not common either in FUors or in EXors. Only V346~Nor \citep{kospal2020_v346nor} shows several \hmol\ emission lines similar to V899~Mon.
The \hmol~2.122\,\um\ emission line was detected in the FUor V960~Mon \citep{park2020} and in some EXors \citep{kospal2011, hodapp2019, hodapp2020}, however, none of the other \hmol\ transitions were detected.

\begin{figure}
    \centering
    \includegraphics[width=\columnwidth]{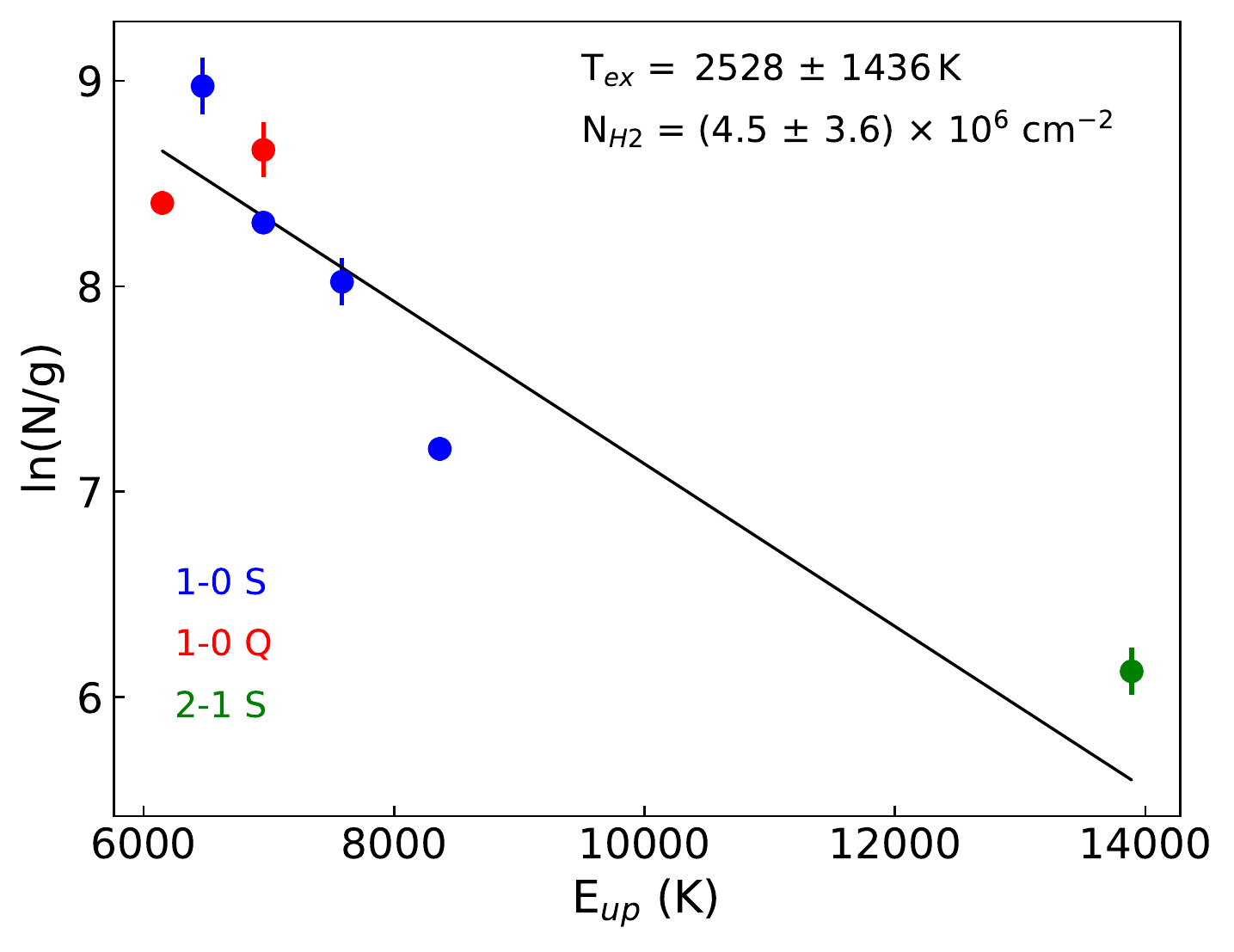}
    \caption{Excitational diagram of \hmol. \label{fig_H2}}
\end{figure}

The excitation diagram of \hmol~lines (\autoref{fig_H2}) is used to estimate the gas excitation temperature (\Tex) and the column density ($N_{\rm H2}$) \citep{vandenAncker1999, fernandes2000, nisini2002, davis2011}. 
The column density ($N_{v,J}$) was obtained using dereddened (\Av=2.6) line intensity ($I_{v,J}$), transition probability \citep[$A_{v,J}$;][]{turner1977}, and wavelength ($\lambda$) in micron  (Eq.~\ref{eq_Nobs}).
\begin{equation} \label{eq_Nobs}
        N_{v,J}=\frac{4 \pi \lambda I_{v,J}}{h c A_{v,J}}
\end{equation}
The column density ($N_{v,J}$) is proportional to the statistical weight ($g_{v,J}$) and upper energy level ($E_{v,J}$) temperature by assuming thermodynamic equilibrium.
\begin{equation}
    N_{v,J}=g_{v,J} {\rm exp}\Big(\frac{E_{v,J}}{kT_{\rm ex}}\Big)
\end{equation}
If the gas is thermalized, the excitation diagram will be fitted by a single straight line. The \Tex\ can be obtained from the inverse slope of the line, and the $N_{v,J}$ can be estimated from the y-intercept. 
The resulting \Tex\ and $N_{\rm H_{2}}$ are $2528 \pm 1436$~K and $4.5 \pm 3.6 \times 10^{6}~\rm cm^{-2}$, respectively.
The measured \Tex\ agrees well with V346~Nor \citep[\Tex\ = 2100 $\pm$ 100~K;][]{kospal2020_v346nor} within the uncertainty and is similar to the \Tex\ (2000--3000~K) of Class~I or Class~II, which is known to be caused by other shock-heated gases \citep{beck2008, davis2011}.

The intensity ratio of 1-0/2-1 S(1) is generally used for studying the shock mechanism. In our observation, only 1-0 S(1) line is detected, so we used the ratio of 1-0/2-1 S(3), which shows equivalent but better results than S(1) lines \citep{smith1995}. We determine a ratio value of 2.3, which is closer to the ratio of 4, indicative of J-shock, rather than a value of 20, indicative of cool-C shock \citep{smith1995}.   

\subsubsection{Hydrogen lines}

During the second outburst stage, several Balmer series lines were observed \citep[\halp\ to H8;][]{ninan2015} as P~Cygni, emission, or absorption profiles. 
However, our observations revealed only two Balmer lines (\halp\ and \hbet) to have the P~Cygni profile. In addition, we detected several Paschen series (Pa~9 to Pa~24) in emission. These line profiles are shown in \autoref{fig_hydrogen_optical} in \autoref{sec_appendix_spec}.

The \brgam\ 2.166\,\um\ line was not detected by \citet{ninan2015} during the second outburst observation period, consistent with the characteristics of FUors \citep{connelley2018}. On the other hand, during our observations, Brackett series lines from Br~7 to Br~21 and two Pfund series lines (Pf~20 and Pf~21) are observed in emission. Relatively strong Br lines are presented in \autoref{fig_Br}, and all the detected Br and Pf lines are presented in Appendix~\ref{sec_appendix_spec} (\autoref{fig_hydrogen_nir}). The detection of these Hydrogen emission lines is similar to EXors or CTTS \citep{aspin2011, kospal2011, hodapp2019, hodapp2020}.

In order to estimate the physical conditions of the emitting region, we used the excitation diagram of the Br series in \autoref{fig_br_ratio}. The observed line fluxes with respect to the \brgam\ were compared with the Case~B theory \citep{hummer1987}. Case~B theory assumes that the Lyman lines are optically thick and other lines are optically thin. The best fit model (red line) is for an electron temperature ($T_{e}$) of 5,000\,K and an electron density of ($N_{e}$) of 10$^{9}$\,cm$^{-3}$.
The best fit $T_{e}$ is lower than that of the outbursting stage of EX~Lup ($T_{e}$=10,000~K), while $N_{e}$ is higher than EX~Lup \citep[$N_{e}$=10$^{7}$\,cm$^{-3}$;][]{kospal2011}.
The lower temperature and higher density than the EX~Lup in the outbursting stage are may be because V899~Mon is in the fading stage.

\begin{figure}
    \centering
    \includegraphics[width=0.5\textwidth]{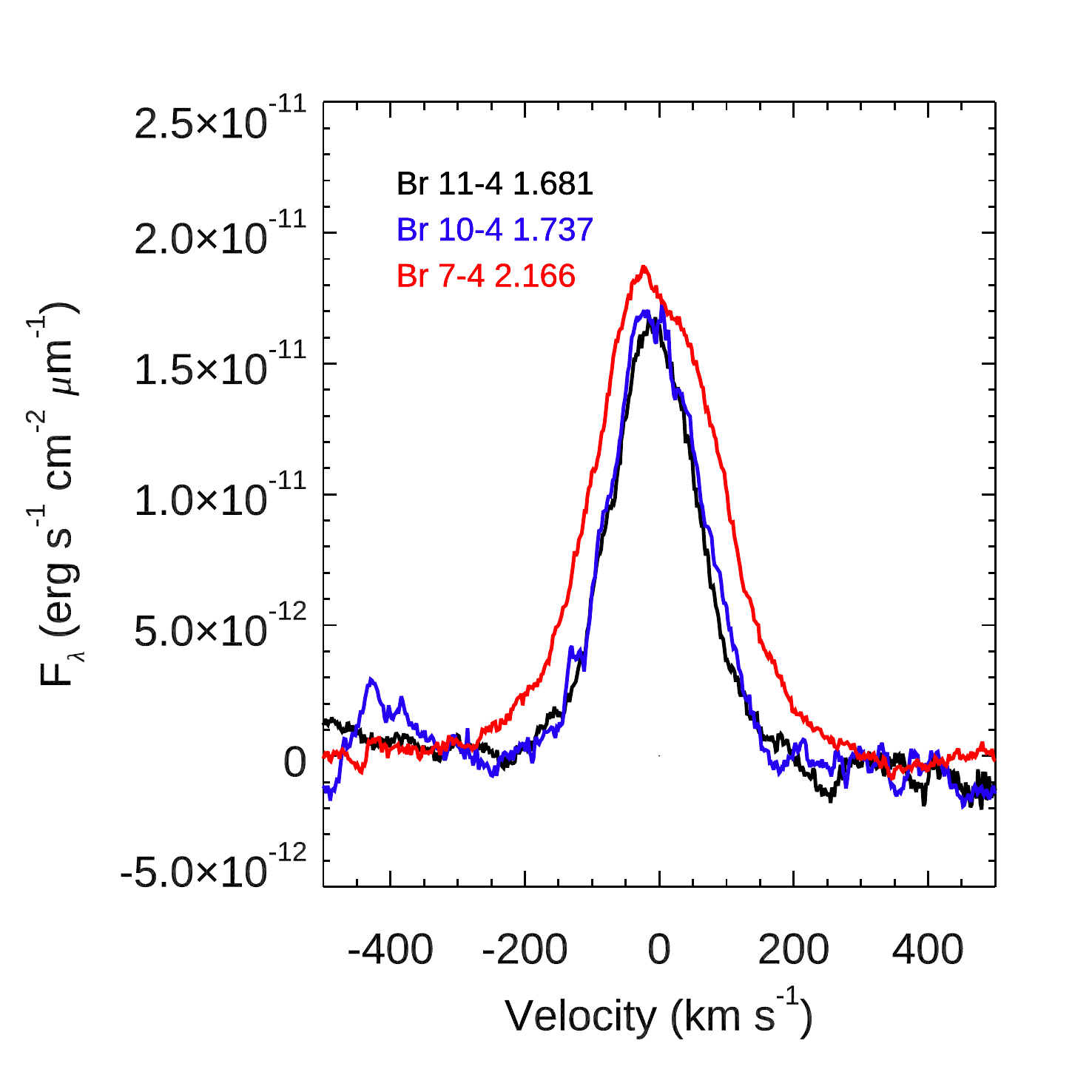}
    \caption{Relatively strong and clear Br series are presented: Br11 1.681\,\um\ (black), Br10 1.731\,\um\ (blue), and \brgam\ 2.166\,\um\ (red). \label{fig_Br}}
\end{figure}

\begin{figure}
    \centering
    \includegraphics[width=\columnwidth]{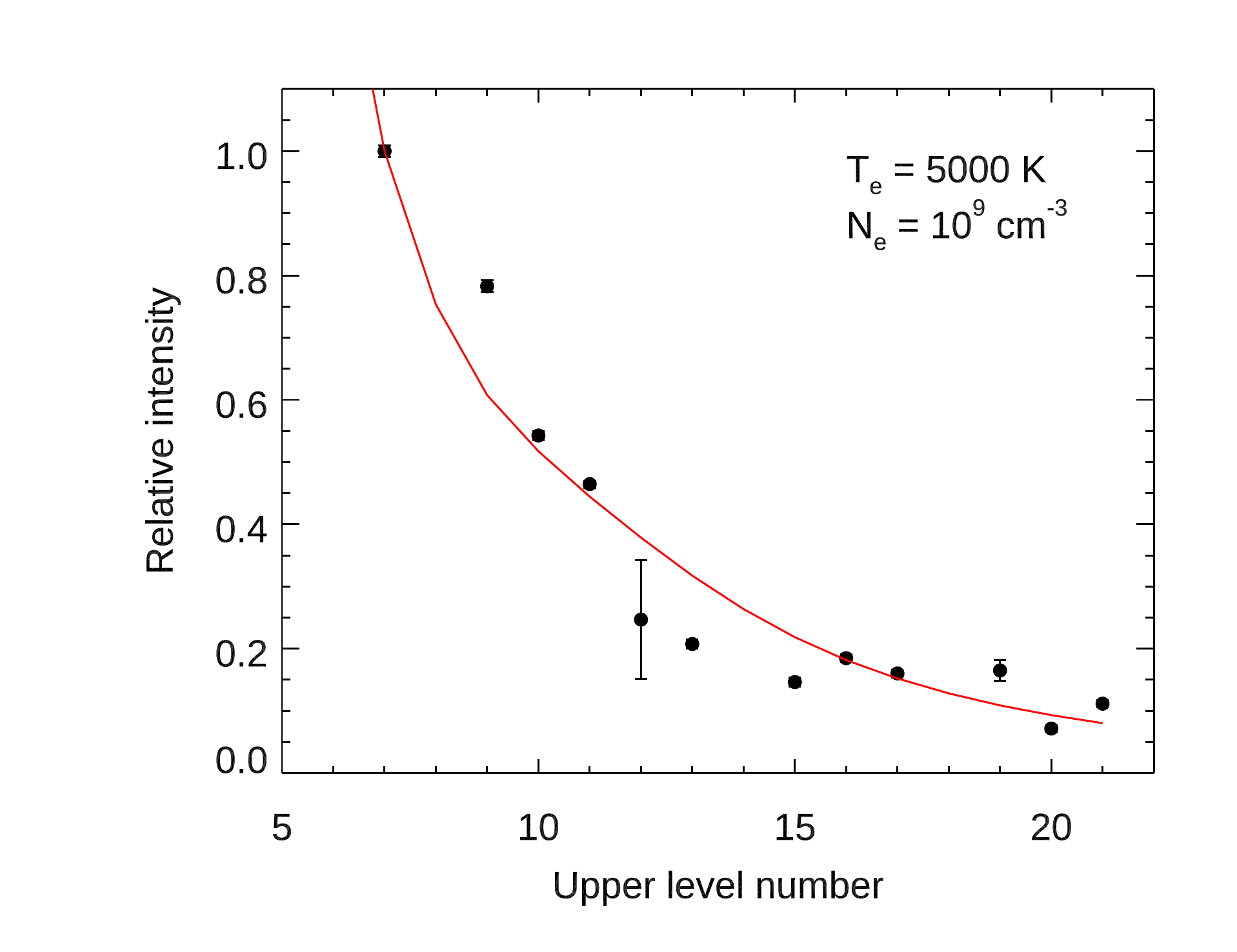}
    \caption{Excitation diagram of Brackett series. Black symbol indicates observed line flux divided by \brgam\ 2.166\,\um. Red solid line presents the Case B theory of $T_{e}$ = 5000\,K and $N_{e}$ = 10$^{9}$\,cm$^{-3}$ \citep{hummer1987}. \label{fig_br_ratio}}
\end{figure}

\subsubsection{Disk modeling of the CO and metallic features} \label{sec_disk_model}

\begin{figure*}
    \centering
    \includegraphics[angle=90,width=\textwidth, trim=170 0 150 0, clip]{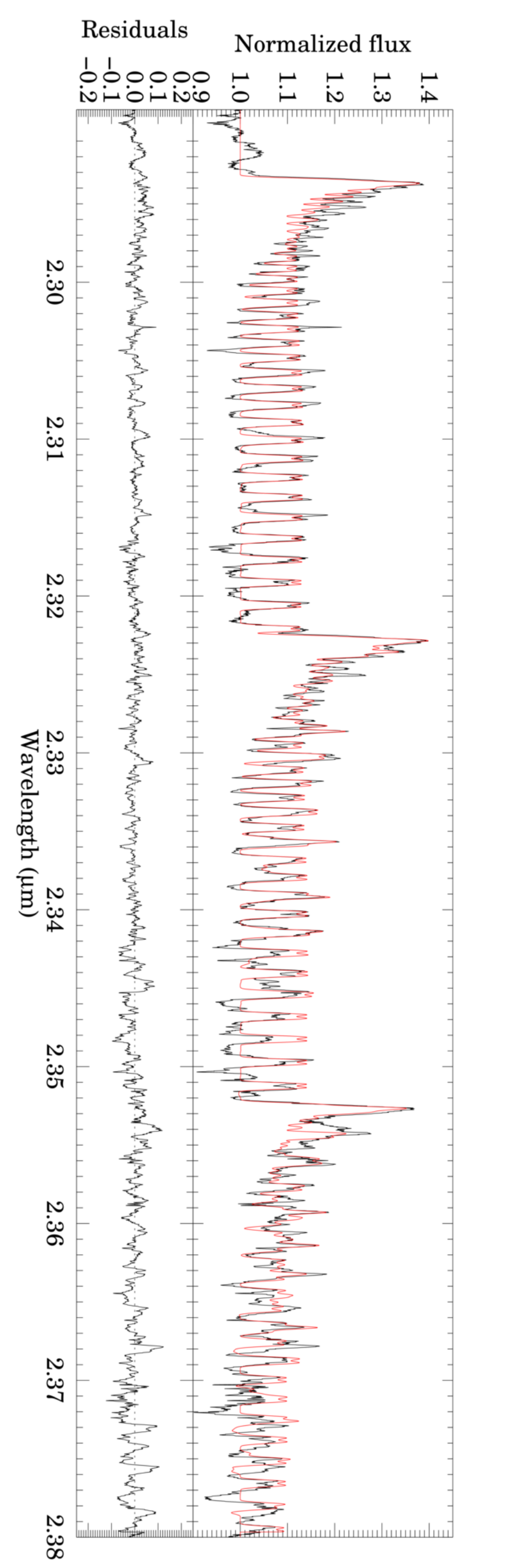}
    \caption{In the top panel with the black line, we show the CO overtone bandhead emission in the spectrum of V899 Mon. The best-fitting model with $T_{\rm ex} = 2482$\,K and $N_{\rm CO} = 10^{22.18}$\,cm$^{-2}$ is plotted in red, while the lower panel shows the residuals. \label{fig_co_bestfit}}
\end{figure*}

\begin{figure}
    \centering
    \includegraphics[angle=90,width=\columnwidth]{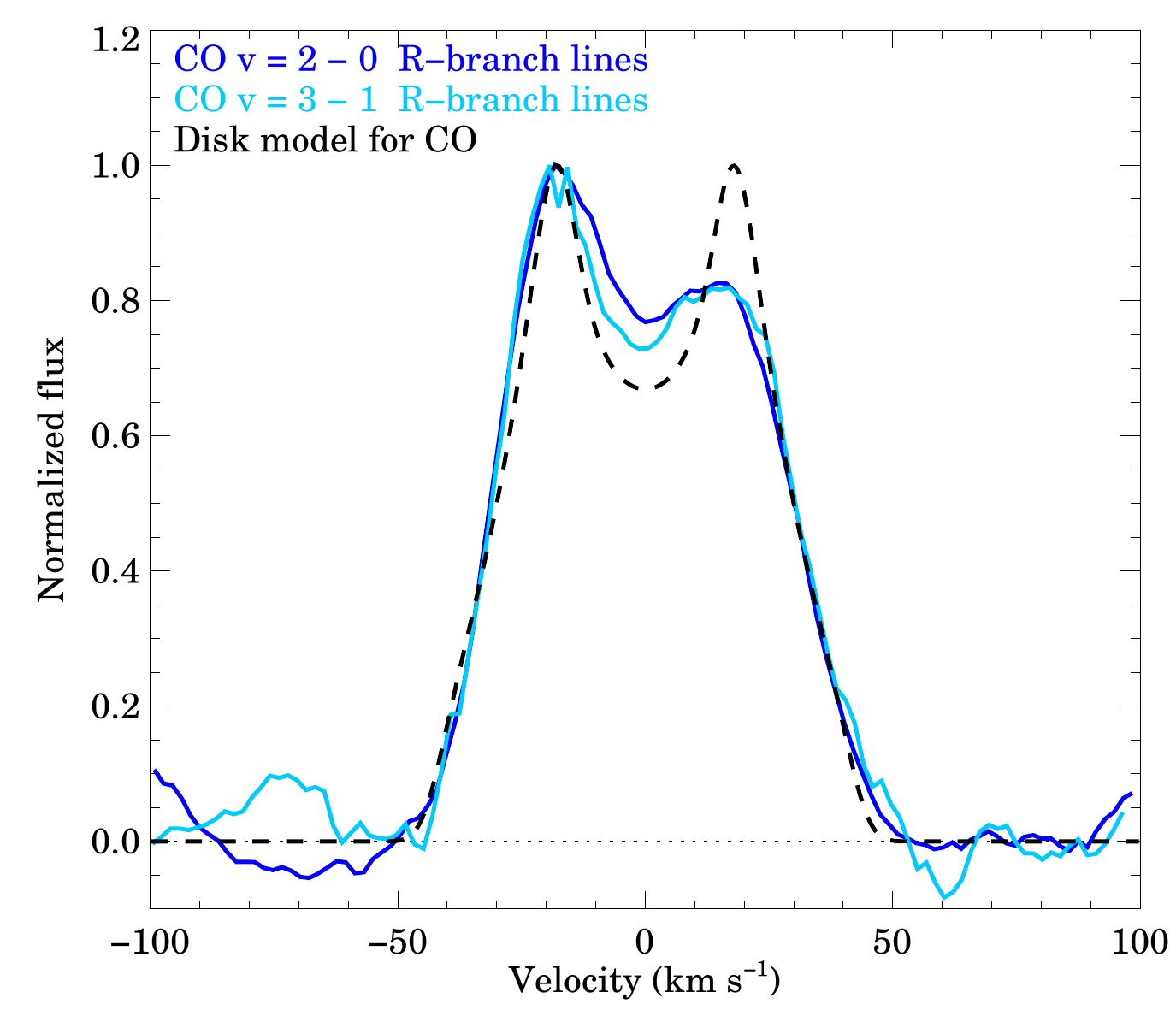}
    \caption{Average line profiles of the $v=2-0$ and $v=3-1$ rovibrational lines of V899~Mon. The disk model that fits the blue-shifted side of the line profiles is indicated with the dashed curve. The corresponding model parameters are in \autoref{tab_diskfit}. \label{fig_disk_co}}
\end{figure}

In order to study the CO bandhead emission features, we normalized the $K$ band spectrum by fitting a fifth order polynomial to the 2.10--2.29$\,\mu$m and 2.40--2.47$\,\mu$m wavelength ranges using a robust fitting method and divided the spectrum by this polynomial. The resulting spectrum is plotted with black in \autoref{fig_co_bestfit}. The spectral resolution of IGRINS is high enough to resolve some of the individual rovibrational lines that make up the CO bandhead. To obtain a high signal-to-noise ratio line profile, we averaged 14 R-branch lines from $J=14-15$ to $J=27-28$ for $v=2-0$ and 10 R-branch lines from $J=13-14$ to $J=22-23$ for $v=3-1$ after each line was normalized again by its peak value. These profiles can be seen in \autoref{fig_disk_co}. Both the $v=2-0$ and the $v=3-1$ profile show clear Keplerian double-peaked shape, but both are clearly asymmetric. The asymmetry is different from what is observed for the atomic lines. The atomic lines were broader towards the red and narrower towards the blue, while the CO profile is symmetric for velocities $|v|>25$\,km\,s$^{-1}$, and the asymmetry is only visible for the Keplerian peaks at $|v|<25$\,km\,s$^{-1}$.

To derive the geometrical parameters of the disk and the dynamical stellar mass, we calculated a simple disk model with inner and outer radii of $R_{\rm in}$ and $R_{\rm out}$, inclination of $i$, in Keplerian rotation around a star with a mass $M_*$, and having a power law brightness distribution with an exponent $\alpha$. We calculated the radial velocity for each point in the disk and shifted a Gaussian that corresponds to the finite instrumental spectral resolution of IGRINS (45,000) by this velocity. Then we computed the full disk's spectrum by integrating over the whole disk in each spectral channel. We considered $R_{\rm in}$ from 0.10\,au to 0.75\,au, with 0.05\,au increments, $R_{\rm out}$ from 2.5\,au to 5.5\,au, with 0.5\,au increments, $i$ from 41$^{\circ}$ to 59$^{\circ}$ with 3$^{\circ}$ increments, $M_*$ from 1.7\,$M_{\odot}$ to 2.3\,$M_{\odot}$ with 0.1\,$M_{\odot}$ increments, and a power law exponent of the brigthness profile, $\alpha$ from $-$1.85 to $-$2.15 in 0.05 increments. This resulted in grid of $14\times7\times7\times7\times7 = 33614$ models.

We compared the observed line profile with the models by calculating the $\chi^2$ taking into account only the blue-shifted channels between $-50$\,km\,s$^{-1}<v<0$\,km\,s$^{-1}$. We determined the best-fitting model parameters and their uncertainties by marginalizing in each parameter. We found that the results are not sensitive to the variation of the $\alpha$ parameter, so this was eventually fixed to $-$2. The best fitting parameters are listed in \autoref{tab_diskfit}. The parameters we obtained for the $v=2-0$ profile and for the $v=3-1$ profile are identical within the uncertainties, although there is a 1$\sigma$ hint that the outer radius is smaller for the $v=3-1$ lines than for the $v=2-0$ lines (3.6 au vs.~4.2 au). This is understandable if we consider that the $v=3-1$ lines are higher excitation lines, therefore there should be outer disk regions (between 3.6 au and 4.2 au) where they are not excited any more, while the lower excitation $v=2-0$ lines are still excited. In \autoref{fig_disk_co}, we plotted the model line profile that uses the average of the parameters obtained from the $v=2-0$ and $v=3-1$ fits. These parameters are also indicated in \autoref{tab_diskfit}.

To derive physical parameters such as the column density ($N_{\rm CO}$) and excitation temperature ($T_{\rm ex}$) of the CO gas that emits the bandhead feature, we modeled the CO lines of V899~Mon following our earlier approach in \citet{kospal2011} for EX~Lup. We used Eq.~10 from \citet{kraus2000} to calculate the absorption coefficient $\kappa$ of the CO molecules. We approximated the thermal and turbulent broadening using a simple Gaussian. We multiplied $\kappa$ with the $N_{\rm CO}$ to obtain the optical depth $\tau$. We assumed a simple slab, so that the $N_{\rm CO}$ and $T_{\rm ex}$ of the CO gas were identical everywhere in the disk. We calculated the intensity $I$ following:
\begin{equation}
I = BB(T_{\rm ex}) (1 - e^{-\tau/ \rm{cos}(i)}),
\end{equation}
where BB(T$_{\rm ex}$) is the Planck function corresponding to the $T_{\rm ex}$. We convolved the model spectrum with the disk's velocity profile determined in the previous step. We varied $T_{\rm ex}$ between 1500\,K and 6400\,K, in steps of 50\,K, and $N_{\rm CO}$ between $10^{19}$\,cm$^{-2}$ and $10^{24}$\,cm$^{-2}$ in logarithmic scale, in steps of 0.05 in the exponent. We used this model spectrum and the observed spectrum in \autoref{fig_co_bestfit} to calculate $\chi^2$, and after marginalization, obtained the following parameters: $T_{\rm ex} = 2482 \pm 326$\,K and $N_{\rm CO} = 10^{(22.18 \pm 0.30)}$\,cm$^{-2}$. The best fitting model is plotted with red in \autoref{fig_co_bestfit}.
In order to check the underestimated $v=2-0$ bandhead part, we also modeled only for the $v=2-0$. As a result, we obtained a similar temperature ($T_{\rm ex} = 2417 \pm 377$\,K) and column density ($N_{\rm CO} = 10^{(22.59 \pm 0.37)}$\,cm$^{-2}$) within the uncertainty. A similar temperature of best-fit results from the whole and only $v=2-0$ bandhead part suggests that the disk is thermalized \citep{lee2015}.


\begin{table*}
\begin{center}
\caption{Disk parameters and dynamical stellar mass for V899~Mon.\label{tab_diskfit}}
\begin{tabular}{lccccc}
\tableline\tableline
Parameter             & CO $v=2-0$ & CO $v=3-1$ & Average CO & Atomic (blue-shifted) & Atomic (red-shifted) \\
\tableline
$i$ ($^{\circ}$)      & 51.01 $\pm$ 10.82 & 52.54 $\pm$ 9.61 & 51.77 $\pm$ 10.22 & 51.77 (fixed)   & 51.77 (fixed)	  \\
$R_{\rm in}$ (au)     & 0.55 $\pm$ 0.20   & 0.62 $\pm$ 0.16  & 0.58 $\pm$ 0.18  & 0.41 $\pm$ 0.28  & 0.20 $\pm$ 0.15 \\
$R_{\rm out}$ (au)    & 4.17 $\pm$ 1.34   & 3.55 $\pm$ 1.04  & 3.86 $\pm$ 1.19  & 4.46 $\pm$ 1.94  & 3.73 $\pm$ 1.86 \\
$M_{*}$ (M$_{\odot}$) & 2.00 $\pm$ 0.67   & 2.08 $\pm$ 0.57  & 2.04 $\pm$ 0.42  & 2.04 (fixed)     & 2.04 (fixed)    \\
$\alpha$              & $-$2.00 (fixed)   & $-$2.00 (fixed)  & $-$2.00 (fixed) & $-$2.00 (fixed)   & $-$2.00 (fixed) \\
\tableline
\end{tabular}
\end{center}
\end{table*}

\begin{figure}
    \centering
    \includegraphics[angle=90,width=\columnwidth]{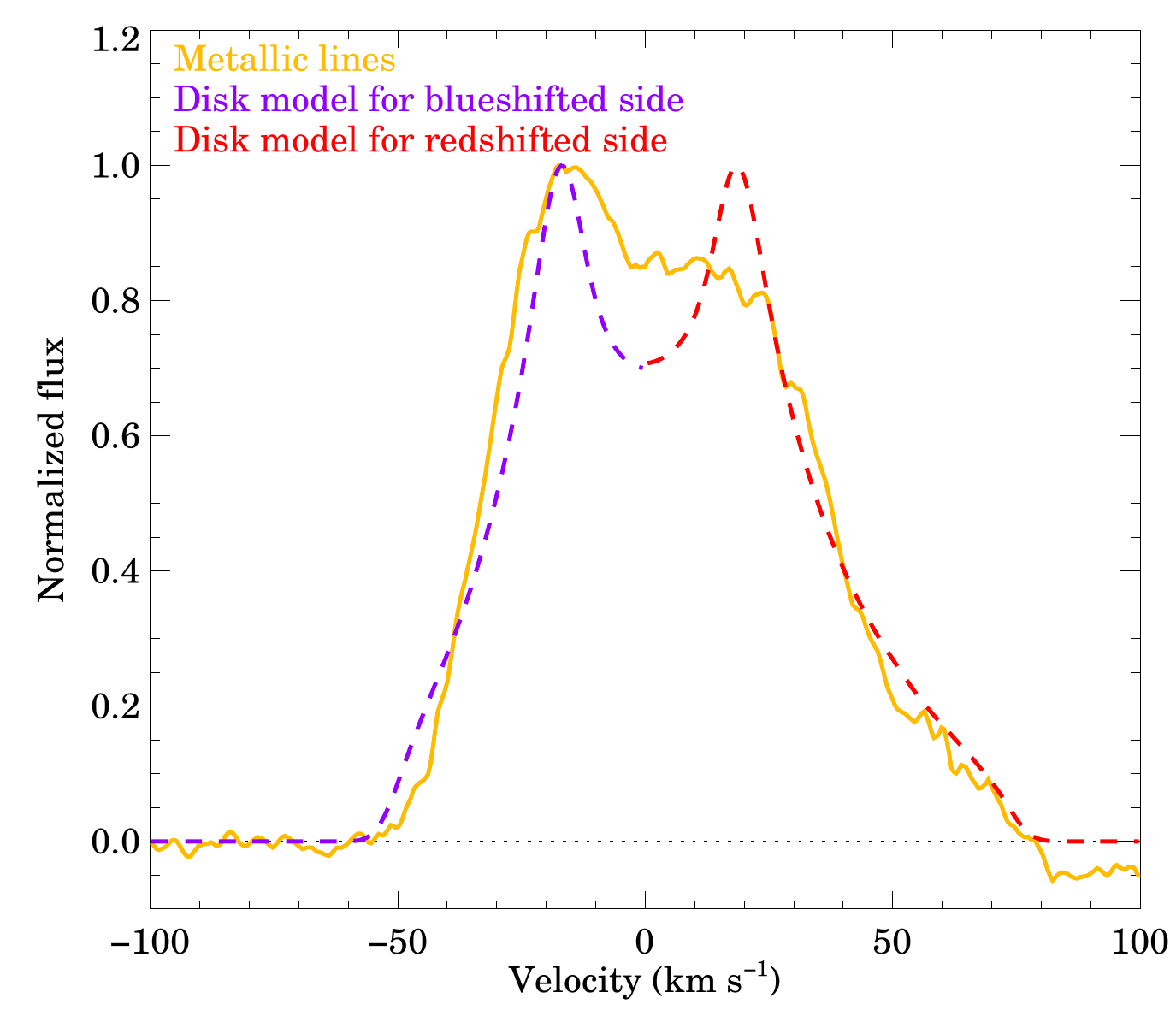}
    \caption{Average line profile of various atomic lines of V899~Mon. Two different disk models that fit the blue-shifted and red-shifted sides of the line profile are indicated with the dashed curves. \label{fig_disk_metals}}
\end{figure}

\autoref{fig_asymmetric} clearly shows that the atomic metal line profiles show the double-peaked Keplerian rotation. We may therefore assume that these lines are also emitted in a disk, just like the CO lines. Previously, we determined geometrical disk parameters, and in particular, we constrained the region of the disk that emits the CO bandhead feature in the K band. Here, we use the same disk model to find out the disk region that emits the metallic lines. Because the metallic lines are noisier than the CO line profile we obtained by averaging several R-branch lines, we did not attempt to fit the metallic lines separately, but we normalized them to their peak values and averaged them. This average metallic line profile is plotted in \autoref{fig_disk_metals}. This shows that not only the Keplerian peaks are asymmetric, but also the line is broader in the red side and narrower in the blue side. Therefore, we fitted the blue-shifted ($-60$\,km\,s$^{-1}<v<0$\,km\,s$^{-1}$) and red-shifted ($0$\,km\,s$^{-1}<v<80$\,km\,s$^{-1}$) sides of the profile separately. Because of the lower signal-to-noise ratio, we decided to fix not only the power law exponent of the brightness profile, but also the disk inclination and stellar mass. Therefore, the only free parameters were the inner and outer disk radii. The best-fitting parameters can be found in \autoref{tab_diskfit}.

Our results show that for the blue-shifted side, the emitting region of the metallic lines extends closer to the star than that of CO (0.41\,au instead of 0.58\,au), and it is also more extended outwards (4.46\,au instead of 3.86\,au). The smaller inner radius is apparent because the metallic line profile is broader than the CO line profile, while the larger outer radius is visible in the velocity channels around 0, where the metallic lines only show a small dip, while the CO profile has a deep dip between the two maxima. For the red-shifted side, which is even broader, we obtained an even smaller inner radius, 0.20\,au. This suggest that there is a significant asymmetry in the disk: on the red-shifted side, the hot gas (traced by the metals) extend much closer to the star than on the blue-shifted side. The CO profile does not show such an asymmetry, which is understandable if we consider that it traces cooler gas than the metals; there is not much CO emission inside of 0.6\,au, while the fit results for the metallic line profile suggest that the asymmetry happens somewhere between 0.2 and 0.4\,au.

\subsubsection{Spectro-astrometry}
We took advantage of MUSE's sensitivity and resolution to carry out a spectro-astrometric analysis of our target.
The presence of extended emission moving a different velocities with respect to the star would slightly shift the position of the star in each channel with respect to the continuum.
We began by fitting a 2D~Gaussian to the position of the star in each channel of the data cube.
Then for each line identified in the spectrum (top panel of \autoref{fig_spec}) we selected a velocity window of a few hundreds of km s$^{-1}$ and calculated the median (i.e. the position of the continuum).
Finally, for each channel in the velocity window, we calculated the difference between the best fitted position and the median.

\begin{figure*}
    \centering
    \includegraphics[width=\textwidth]{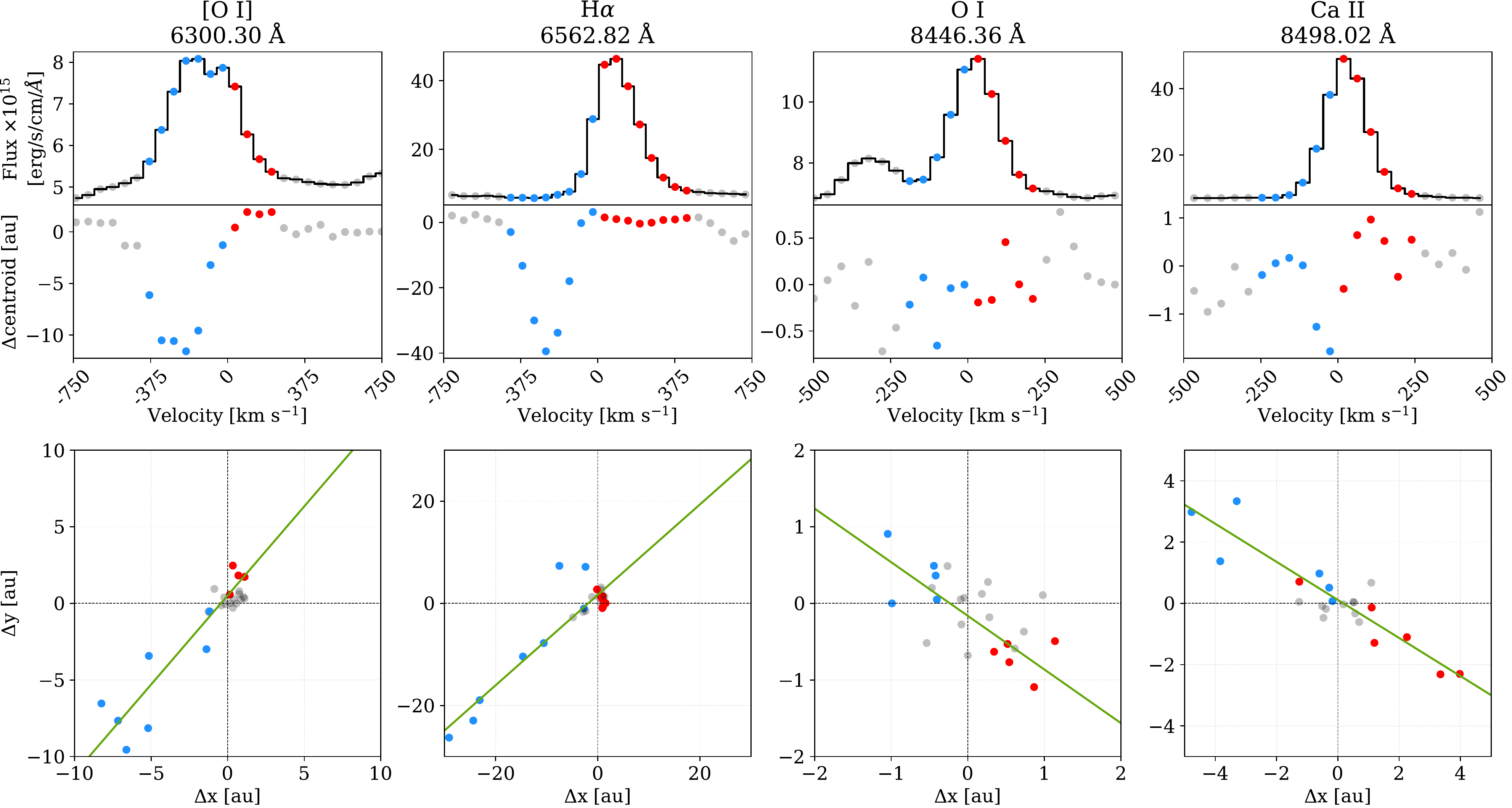}
    \caption{Representative spectro-astrometry data of [\OI] 6300\,\AA, \halp\ 6563\,\AA, \OI\ 8446\,\AA, and \CaII\ 8498\,\AA.
    Top panels show line profiles. 
    Middle panels show the offset of the line+continuum signal regarding the continuum emission centroid.
    Bottom panels show 2D plot of spectro-astrometric offsets at each position and the position angle (green line). The upper and lower directions are North and South, and the left and right directions are East and West. \label{fig_spectro_astrometry}}
\end{figure*}

In \autoref{fig_spectro_astrometry} we show the results of four lines: [\OI] 6300\,\AA{}, \halp\ 6563\,\AA, \OI{} 8446\,\AA{}, and \CaII\ 8498\,\AA.
The top row panels show the spectral line, the middle rows show the absolute differences in position, and the bottom panels show the position of the emission per channel relative to the center of the continuum.
In the case of the absolute difference of position, if the emission line traces infalling or outflowing material the absolute differences in position would be positive or negative, respectively.
However, if the line is tracing disk rotation then the differences would be positive on one side and negative on the other.
In the case of the relative positions, the infalling or outflowing material should be separated from the position of the star in one direction, and the disk rotation would be detected on opposite positions with respect to the star, thus indicating which side of the disk is blue-shifted or red-shifted.

Based on the middle row of \autoref{fig_spectro_astrometry}, [\OI]~6300\,\AA{} and \halp~6563\,\AA{} show outflow emission, and \OI~8446\,\AA{} and \CaII~8498\,\AA{} show disk rotation.
The former two lines show the blue-shifted emission extends towards the bottom left quadrant, i.e. the southeast of V899~Mon.
The latter two lines show blue-shifted and red-shifted disk emission in the upper left and lower right quadrants, respectively, indicating the disk has a Northeast-Southwest orientation.
We fitted a straight line to these points of extended emission, and found the [\OI]~6300\,\AA{} and H$\alpha$~6563\,\AA{} outflow emission have position angles of $\sim$140$^\circ$ and $\sim$130$^\circ$, respectively, and the point of \OI~8446\,\AA{} and \CaII~8498\,\AA{} have position angles of $\sim$55$^\circ$ and $\sim$58$^\circ$, respectively.
The disk position angles are within the uncertainties of those estimated from ALMA dust continuum observations \citep{kospal2021}, and are almost perpendicular ($\sim$80$^\circ$) to those of the high-velocity outflow emission. 
Spectro-astrometry of the \OI~8446\,\AA{} and \CaII~8498\,\AA{} lines (bottom panels in \autoref{fig_spectro_astrometry}) shows consistent results with permitted metallic lines with the disk size from 0.2 to 4.6~au (\autoref{tab_diskfit}).

The blue-shifted emission of [\OI]~6300\,\AA{} shows two peaks, at --24\,km s$^{-1}$ and --175\,km s$^{-1}$, indicative of a combination of low-velocity and high-velocity components, respectively, as has been detected on several T~Tauri stars \citep[e.g.][]{hartigan1995, hirth1997, pyo2003, Banzatti2019}.
The blue-shifted components of [\OI]~6300\,\AA{} and H$\alpha$~6563\,\AA{} reach maximum velocities of --500\,km s$^{-1}$.
We estimated the ejection dates for the outflows using an average outflow velocity of 250\,km s$^{-1}$, and the outflow separations are 10\,au for [\OI]~6300\,\AA{} and 40\,au for H$\alpha$~6563\,\AA{} (see bottom panels of \autoref{fig_spectro_astrometry}).
The deprojected outflow velocity, assuming an average system inclination of 50$^\circ$ (see \autoref{tbl_info}), is 385\,km s$^{-1}$.
Thus, under the assumption that outflow velocity remains constant after its ejection, we estimate ejection timelines of 45 and 180 days for [\OI]~6300\,\AA{} and H$\alpha$~6563\,\AA{}, respectively.
However, these time estimates must be taken with caution as the uncertainties in the astrometry and the spectral resolution of MUSE, and the large uncertainties in the disk inclination, can change this timelines by a factor of a few.
Nevertheless, the tentative estimated ejection date for [\OI]~6300\,\AA{} is early December 2020, which is in agreement with the 2020/2021 accretion burst (red symbols in \autoref{fig_light_events}).
And the estimated ejection date for H$\alpha$~6563\,\AA{} is late July 2020, which falls during a period when the photometric monitoring of V899~Mon is sparse.
We suggest this difference in time estimates can be partially explained by the different physical phenomena each line traces.
Both [\OI]~6300\,\AA{} and H$\alpha$~6563\,\AA{} trace outflow emission, however, the latter also traces accretion in the inner parts of the circumstellar disk.

\section{Discussion \label{sec_discussion}}

\subsection{Current status of V899~Mon}
The small-amplitude brightness variations observed in V899~Mon in recent years is common for CTTS. The accretion bursts that started to occur since 2016 (including the two unique, particularly well-characterized bursts marked by red symbols in \autoref{fig_light_events}) appear to be caused by the magnetospheric accretion process, similarly as in CTTS. The rich emission spectrum of V899~Mon is also similar to those of CTTS: hydrogen, permitted, and forbidden metallic emission lines \citep{muzerolle1998}. 
Calculated mass accretion rate and mass loss rate are about 10$^{-7}$~\Msunpyr{} and 10$^{-8}$~\Msunpyr{}, respectively. These values are lower than those of FUors but similar to those of CTTS \citep{muzerolle1998}, and the mass loss rate is about 10\% of the mass accretion rate that is also in agreement with CTTS \citep{hartmann1996, hartman2009, ellerbroek2013, bally2016}.
The intensity ratio of the \CaII\ IRT lines is nearly equal, and similar to those of CTTS \citep{herbig1980_CaII, hamann1992}.

From these photometric and spectroscopic observational results, we suggest that V899 Mon almost finished the outburst starting from 2012 and now returning to the CTTS stage where the mass accretion rate is lower than in the burst stage.

\subsection{Classification of V899~Mon}
V899~Mon shows spectral features of both FUors and EXors, similar to V1647~Ori \citep{briceno2004, reipurth2004, vacca2004, ojha2006, fedele2007, ninan2013}. Repetitive outbursts of relatively short duration are similar to those of V1647~Ori. In addition, the overall optical and NIR spectra of V899~Mon resembles V1647~Ori: \halp~and \hbet\ P~Cygni profiles, numerous metallic emission lines, forbidden emission lines, strong CO overtone bandhead emission, and \brgam\ emission. 

\begin{figure}
    \centering
    \includegraphics[width=\columnwidth, trim=40 10 0 10,clip]{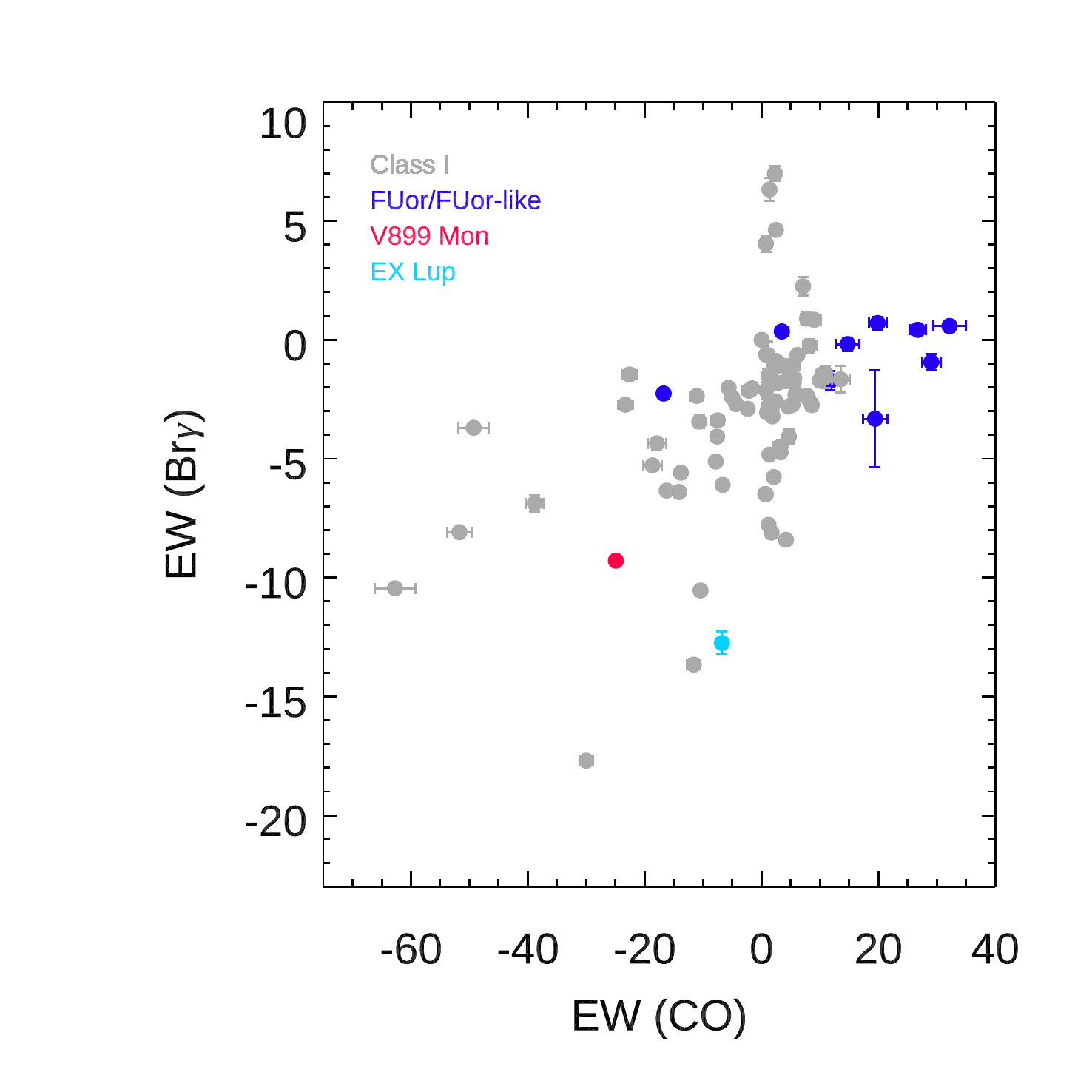}
    \caption{The EW of \brgam\ vs. CO. Gray and blue symbols indicate Class~I and FUor/FUor-like objects from \citet{connelley2010}. Red and sky blue symbols represent V899~Mon and EX~Lup, respectively. V899~Mon falls close to EX~Lup, compared to FUors/FUor-like objects. \label{fig_ew_brgam_co}}
\end{figure}

V1647 Ori was classified as a ``peculiar" object according to the NIR spectroscopic classification criteria \citep{connelley2018}. Peculiar objects have some spectroscopic or photometric characteristics similar to bonafide FUors. 
Based on their parameters to classify FUors, V899~Mon would be closer to a peculiar object than a bonafide FUor because of CO bandhead emission features, no water absorption features around the H band, and many metallic emission lines.
In order to compare V899~Mon with the classification of \citet{connelley2018}, we measured the equivalent width (EW) of CO bandhead, \NaI\ 2.202\,\um, \brgam\ 2.166\,\um, \hmol\ 2.122\,\um, and [\FeII] 1.644\,\um. The measured EWs are listed in Table~\ref{tbl_EW}. 
The EW of CO was obtained between 2.292 to 2.320\,\um\ because this feature is located at the edge of the order.
\NaI\ doublet lines are resolved, so we measured the EW of each line and used the total EW of the two lines. However, \CaI\ lines are indiscernible in our observation. 
V899~Mon shows rich emission spectrum similar to that of EX~Lup during the outburst stage \citep{kospal2011}, so the EWs of EX~Lup were also measured.
The measured EWs of \NaI\ + \CaI\ vs. CO are located closer to peculiar objects (including V1647~Ori) and EX~Lup rather than FUors and FUor-like objects \citep[see Figure~9 in][]{connelley2018}.
In addition, in the EW plot of \brgam\ vs. CO (\autoref{fig_ew_brgam_co}), V899~Mon shows similarity to EX~Lup rather than FUors and FUor-like objects.
Overall, our photometric and spectroscopic data in the current stage of V899~Mon shows more similarities to EXors.

\begin{deluxetable}{cccc}
\tabletypesize{\scriptsize} 
\tablecaption{EW of Spectral Lines \label{tbl_EW}}
\tablewidth{0pt}
\tablehead{\colhead{Transition} & \colhead{Wavelength} & \colhead{V899~Mon} & \colhead{EX~Lup$^{a}$} \\ [-2mm]
\colhead{} & \colhead{(\um)} & \colhead{(\AA)} & \colhead{(\AA)}}
\startdata
[Fe\,{\scriptsize II}] & 1.6440 &  $-$3.22 $\pm$   0.09 &  $-$3.14 $\pm$   0.45 \\ 
   Na\,{\scriptsize I} & 2.2062 &  $-$1.01 $\pm$   0.06 &  $-$1.02 $\pm$   0.25 \\
   Na\,{\scriptsize I} & 2.2090 &  $-$0.89 $\pm$   0.05 &  $-$1.18 $\pm$   0.22 \\ \relax
 \hmol\ 1-0 S(1) & 2.1218 &  $-$0.23 $\pm$   0.05 &  $-$0.22 $\pm$   0.23 \\
  \brgam\ & 2.1661 &  $-$9.29 $\pm$   0.12 & $-$12.75 $\pm$   0.49 \\
   CO 2-0 & 2.2935 & $-$24.94 $\pm$   0.16 &  $-$6.78 $\pm$   0.73 \\
\enddata
\tablenotetext{a}{Spectrum from \citet{kospal2011}}
\end{deluxetable}

\section{Conclusion \label{sec_conclusion}}
We conducted photometric monitoring and spectroscopic observations of V899~Mon. Using our observations and archival data, we were able to monitor the changes in the physical properties of V899~Mon as the outburst progressed. 
The primary results obtained in this study are as follows:

1. The peak of the first outburst occurred in 2010 and then the star went through a relatively quick quiescent phase in 2011. After that, in 2012, the second brightening started. The second peak of the outburst was reached in 2018. Since then, the brightness of V899~Mon is gradually decreasing with an approximate rate of 0.30 $\pm$ 0.02 mag~yr$^{-1}$.

2. We find that the amplitude of the in-seasonal brightness variations (superimposed on the major light plateau caused by viscous heating of the disk due to enhanced accretion) between 2012 and 2016 is about 0.1~mag. Then, the amplitude became higher, about 0.2~mag, and the variations started to be dominated by accretion bursts and dips (a few of them per year were detected). The most dramatic, $\Delta V \approx$~1~mag strong burst occurred in the 2017/2018 season, and it lasted for about 200~days. Then, again a shorter (about 30~days long) and weaker ($\Delta V \approx$~0.3~mag) accretion burst was observed in 2020/2021. The accretion flow originating from a disk warp formed at different distances from the star can be responsible for these short accretion bursts and associated central dips. In the case of other bursts, either too sparse data, or optically thin properties of the funnel flow, or different geometry of visibility prevented us from seeing the central brightness dips. The increasing number of accretion bursts (typical for ordinary CTTS and low massive Herbig~Ae stars) in parallel with the disk brightness decrease suggests that V899~Mon is gradually switching from the still not well understood accretion mechanism during the enhanced accretion (possibly a boundary layer mechanism) to the magnetospheric accretion mechanism.

3. When comparing the line profiles observed while in outburst to those compared with our observations, we found the blue-shifted absorption component of P~Cygni profiles became weaker in \halp\ and \CaII\ IRT lines, indicating that the strength of outflowing wind became weaker. 

4. Numerous forbidden emission lines are detected, and the estimated outflow temperature and density are about 9000\,K and $>$ 10$^{4}$\,cm$^{-3}$, respectively. In contrast to the weakening of the outflowing wind in P~Cygni profiles, the outflow temperature and density have not changed much.

5. The estimated mass accretion rate and the mass loss rate during the fading stage are about 10$^{-7}$\,\Msunpyr\ and 10$^{-8}$\,\Msunpyr, respectively. These values are about one order of magnitude lower than those of the outbursting stage \citep{ninan2015}, implying the weakening outburst, similar to those of CTTS. Moreover, our measured mass accretion rate is similar to that of the EX~Lup during the outburst \citep{aspin2010}.

6. Many metallic emission lines and individual CO rovibrational lines show double-peaked profiles characteristics of Keplerian rotation. In order to derive the geometrical and physical parameters of the disk, we calculated a simple disk model and estimated the disk parameters. 
The derived stellar mass ($M_{*}$ $\sim$ 2\,${M_{\odot}}$) is similar to that of \citet{ninan2015}, but the obtained inclination ($i$ $\sim$ 52$^{\circ}$) is higher than face-on \citep{ninan2016}, consistent with the ALMA observation \citep{kospal2021}.

7. Our spectro-astrometry data shows both outflow and disk signatures. The outflowing velocity is similar to that found from the blue-shifted absorption component of \halp\ P~Cygni profile and extended towards the southeast. Disk signatures that are almost perpendicular to the outflow direction were also observed and distributed in the Northeast-Southwest. The disk size is consistent with that of disk modeling for the permitted metallic lines.

8. The small amplitude of photometric variations, rich emission lines, measured mass accretion rate, and mass loss rate show similar properties to those of CTTS. In addition to these properties, the gradual decreasing of the brightness indicates that V899~Mon is almost finishing its second outburst, and now it is on the way back to the quiescent phase. 

9. Our recent photometric and spectroscopic observations of V899~Mon  in the current evolutionary stage show more features typical for EXors.



\clearpage
\acknowledgments
This work used the Immersion Grating Infrared Spectrometer (IGRINS) that was developed under a collaboration between the University of Texas at Austin and the Korea Astronomy and Space Science Institute (KASI) with the financial support of the Mt. Cuba Astronomical Foundation, of the US National Science Foundation under grants AST-1229522 and AST-1702267, of the McDonald Observatory of the University of Texas at Austin, of the Korean GMT Project of KASI, and Gemini Observatory.
This work was supported by K-GMT Science Program (PID: GS-2020B-Q-218) of Korea Astronomy and Space Science Institute (KASI).
We gratefully acknowledge the work of the observers, R\'obert Szak\'ats, Eda Sonba\c{s}, and H\"useyin Er. 
This project has received funding from the European Research Council (ERC) under the European Union's Horizon 2020 research and innovation programme under grant agreement No 716155 (SACCRED), and from the ``Transient Astrophysical Objects'' GINOP 2.3.2-15-2016-00033 project of the National Research, Development and Innovation Office (NKFIH), Hungary, funded by the European Union.
ZsMSz is supported by the \'UNKP-20-2 New National Excellence Program of the Ministry for Innovation and Technology from the source of the National Research, Development and Innovation Fund.
L. Kriskovics is supported by the Bolyai J\'anos Research Scholarship of the Hungarian Academy of Sciences. L. Kriskovics  acknowledges the financial support of the Hungarian National Research, Development and Innovation Office grant NKFIH PD-134784.
This paper uses observations made at the South African Astronomical Observatory (SAAO) and at the Mount Suhora Astronomical Observatory.
 




\vspace{5mm}
\facilities{Gemini:South,VLT:Yepun}
\software{IRAF \citep{tody1986,tody1993}}



\clearpage

\appendix

\section{Photometry of V899 Mon} \label{sec_appendix_photometry}

Table~\ref{tab:photometry} shows the ground-based optical photometry we obtained for V899 Mon. \\

\tabletypesize{\tiny}
\begin{deluxetable*}{ccccccccccc}
\tablecaption{Photometry of V899 Mon\label{tab:photometry}}
\tablewidth{0pt}
\tablehead{
\colhead{HJD} & 
\colhead{$U$} & \colhead{$B$} & \colhead{$g'$} & \colhead{$V$} &
\colhead{$r'$} & \colhead{$R_C$} & \colhead{$i'$} & \colhead{$I_C$} & \colhead{$z'$} & \colhead{Obs/Tel}
}
\startdata
2458082.531 & \dots            & 14.788$\pm$0.022 & \dots            & 13.392$\pm$0.009 & \dots            & 12.433$\pm$0.009 & \dots            & 11.445$\pm$0.012 & \dots & Konkoly/Schmidt \\
2458085.493 & \dots            & 14.010$\pm$0.011 & \dots            & 12.737$\pm$0.015 & \dots            & 11.880$\pm$0.006 & \dots            & 10.949$\pm$0.005 & \dots & Konkoly/Schmidt \\
2459114.624 & \dots            & 16.446$\pm$0.009 & \dots            & 14.998$\pm$0.002 & 14.336$\pm$0.005 & \dots            & 13.627$\pm$0.003 & \dots            & \dots & Konkoly/RC80 \\
2459115.590 & \dots            & 16.435$\pm$0.044 & \dots            & 14.935$\pm$0.014 & 14.280$\pm$0.012 & \dots            & 13.571$\pm$0.005 & \dots            & \dots & Konkoly/RC80 \\
2459116.599 & \dots            & 16.365$\pm$0.018 & \dots            & \dots            & \dots            & \dots            & \dots            & \dots            & \dots & Konkoly/RC80 \\
2459117.579 & \dots            & 16.383$\pm$0.050 & \dots            & 14.886$\pm$0.050 & 14.250$\pm$0.050 & \dots            & 13.653$\pm$0.050 & \dots            & \dots & Konkoly/RC80 \\
2459182.537 & \dots            & \dots            & 15.705$\pm$0.017 & \dots            & 14.373$\pm$0.008 & \dots            & 13.664$\pm$0.012 & \dots            & \dots & Adiyaman/ADYU60 \\
2459183.452 & \dots            & \dots            & 15.849$\pm$0.029 & \dots            & 14.415$\pm$0.013 & \dots            & 13.716$\pm$0.014 & \dots            & \dots & Adiyaman/ADYU60 \\
2459185.403 & \dots            & 16.527$\pm$0.033 & \dots            & 15.108$\pm$0.008 & 14.420$\pm$0.010 & \dots            & 13.732$\pm$0.009 & \dots            & \dots & Konkoly/RC80 \\
2459193.499 & \dots            & \dots            & 15.721$\pm$0.004 & \dots            & \dots            & \dots            & \dots            & \dots            & \dots & Mount Suhora \\
2459195.462 & \dots            & \dots            & 15.704$\pm$0.009 & \dots            & 14.432$\pm$0.009 & \dots            & 13.716$\pm$0.011 & \dots            & \dots & Mount Suhora \\
2459200.502 & 18.121$\pm$0.066 & 16.461$\pm$0.006 & \dots            & 14.981$\pm$0.005 & \dots            & \dots            & \dots            & 13.060$\pm$0.004 & \dots & SAAO/Lesedi \\
2459201.357 & 17.841$\pm$0.038 & 16.459$\pm$0.005 & \dots            & 14.992$\pm$0.006 & \dots            & 14.079$\pm$0.007 & \dots            & 13.081$\pm$0.006 & \dots & SAAO/Lesedi \\
2459202.342 & 17.871$\pm$0.047 & 16.428$\pm$0.005 & \dots            & 14.943$\pm$0.003 & \dots            & 14.024$\pm$0.008 & \dots            & 13.028$\pm$0.006 & \dots & SAAO/Lesedi \\
2459203.337 & 17.739$\pm$0.045 & 16.364$\pm$0.004 & \dots            & 14.908$\pm$0.006 & \dots            & 14.000$\pm$0.006 & \dots            & 12.995$\pm$0.005 & \dots & SAAO/Lesedi \\
2459204.488 & \dots            & \dots            & 15.463$\pm$0.005 & \dots            & 14.165$\pm$0.004 & \dots            & 13.461$\pm$0.008 & \dots            & \dots & Mount Suhora \\
2459206.324 & 17.646$\pm$0.050 & 16.299$\pm$0.006 & \dots            & 14.848$\pm$0.005 & \dots            & 13.949$\pm$0.007 & \dots            & 12.957$\pm$0.004 & \dots & SAAO/Lesedi \\
2459207.342 & 17.723$\pm$0.059 & 16.284$\pm$0.013 & \dots            & 14.838$\pm$0.008 & \dots            & 13.927$\pm$0.008 & \dots            & 12.938$\pm$0.005 & \dots & SAAO/Lesedi \\
2459208.432 & 17.863$\pm$0.072 & 16.326$\pm$0.007 & \dots            & 14.859$\pm$0.006 & \dots            & 13.949$\pm$0.008 & \dots            & 12.970$\pm$0.005 & \dots & SAAO/Lesedi \\
2459208.806 & 17.434$\pm$0.020 & \dots            & 15.511$\pm$0.007 & \dots            & 14.198$\pm$0.007 & \dots            & 13.496$\pm$0.016 & \dots            & 13.017$\pm$0.014 & VST/OMEGACAM \\
2459209.333 & 17.627$\pm$0.068 & 16.286$\pm$0.011 & \dots            & 14.831$\pm$0.005 & \dots            & 13.925$\pm$0.007 & \dots            & 12.943$\pm$0.006 & \dots & SAAO/Lesedi \\
2459210.335 & 17.638$\pm$0.096 & 16.381$\pm$0.156 & \dots            & 14.922$\pm$0.009 & \dots            & 14.021$\pm$0.006 & \dots            & 13.033$\pm$0.004 & \dots & SAAO/Lesedi \\
2459210.452 & \dots            & \dots            & 15.573$\pm$0.006 & \dots            & 14.268$\pm$0.005 & \dots            & 13.549$\pm$0.008 & \dots            & \dots & Mount Suhora \\
2459211.313 & 17.803$\pm$0.068 & 16.521$\pm$0.020 & \dots            & 15.031$\pm$0.006 & \dots            & 14.112$\pm$0.008 & \dots            & 13.106$\pm$0.011 & \dots & SAAO/Lesedi \\
2459212.304 & \dots            & 16.508$\pm$0.012 & \dots            & 15.030$\pm$0.006 & \dots            & 14.119$\pm$0.007 & \dots            & 13.133$\pm$0.006 & \dots & SAAO/Lesedi \\
2459212.658 & 18.138$\pm$0.086 & \dots            & 15.771$\pm$0.021 & \dots            & 14.442$\pm$0.021 & \dots            & 13.715$\pm$0.027 & \dots            & 13.218$\pm$0.020 & VST/OMEGACAM \\
2459213.312 & 18.039$\pm$0.137 & 16.628$\pm$0.051 & \dots            & 15.121$\pm$0.012 & \dots            & 14.194$\pm$0.016 & \dots            & 13.202$\pm$0.013 & \dots & SAAO/Lesedi \\
2459213.803 & 18.696$\pm$0.747 & \dots            & 15.825$\pm$0.023 & \dots            & 14.455$\pm$0.016 & \dots            & 13.739$\pm$0.025 & \dots            & 13.259$\pm$0.016 & VST/OMEGACAM \\
2459214.296 & \dots            & 16.438$\pm$0.012 & \dots            & 14.933$\pm$0.018 & \dots            & 14.031$\pm$0.010 & \dots            & 13.032$\pm$0.006 & \dots & SAAO/Lesedi \\
2459216.476 & \dots            & \dots            & 15.513$\pm$0.007 & \dots            & 14.214$\pm$0.004 & \dots            & 13.495$\pm$0.007 & \dots            & \dots & Mount Suhora \\
2459217.767 & 17.343$\pm$0.078 & \dots            & 15.564$\pm$0.026 & \dots            & 14.256$\pm$0.021 & \dots            & 13.549$\pm$0.029 & \dots            & 13.041$\pm$0.015 & VST/OMEGACAM \\
2459218.432 & \dots            & \dots            & 15.507$\pm$0.005 & \dots            & 14.219$\pm$0.004 & \dots            & 13.508$\pm$0.007 & \dots            & \dots & Mount Suhora \\
2459225.387 & \dots            & \dots            & 15.692$\pm$0.005 & \dots            & 14.408$\pm$0.005 & \dots            & 13.690$\pm$0.012 & \dots            & \dots & Adiyaman/ADYU60 \\
2459235.410 & 17.947$\pm$0.027 & 16.472$\pm$0.009 & \dots            & 14.986$\pm$0.007 & \dots            & 14.073$\pm$0.008 & \dots            & 13.080$\pm$0.007 & \dots & SAAO/Lesedi \\
2459237.463 & 17.880$\pm$0.101 & 16.458$\pm$0.008 & \dots            & 15.021$\pm$0.008 & \dots            & 14.128$\pm$0.012 & \dots            & 13.116$\pm$0.007 & \dots & SAAO/Lesedi \\
2459239.412 & 17.895$\pm$0.103 & 16.466$\pm$0.033 & \dots            & 15.030$\pm$0.052 & \dots            & 14.120$\pm$0.027 & \dots            & 13.096$\pm$0.044 & \dots & SAAO/Lesedi \\
2459241.388 & 17.920$\pm$0.107 & 16.530$\pm$0.034 & \dots            & 15.020$\pm$0.011 & \dots            & 14.108$\pm$0.014 & \dots            & 13.096$\pm$0.005 & \dots & SAAO/Lesedi \\
2459251.217 & \dots            & \dots            & 15.701$\pm$0.009 & \dots            & 14.383$\pm$0.006 & \dots            & 13.686$\pm$0.013 & \dots            & \dots & Adiyaman/ADYU60 \\
2459252.214 & \dots            & \dots            & 15.715$\pm$0.008 & \dots            & 14.397$\pm$0.007 & \dots            & 13.695$\pm$0.009 & \dots            & \dots & Adiyaman/ADYU60 \\
2459258.196 & \dots            & \dots            & 15.758$\pm$0.009 & \dots            & 14.399$\pm$0.009 & \dots            & 13.679$\pm$0.029 & \dots            & \dots & Adiyaman/ADYU60 \\
2459269.223 & \dots            & \dots            & 15.629$\pm$0.043 & \dots            & 14.363$\pm$0.013 & \dots            & 13.637$\pm$0.010 & \dots            & \dots & Mount Suhora \\
2459272.338 & \dots            & 16.413$\pm$0.076 & \dots            & 15.045$\pm$0.005 & 14.403$\pm$0.011 & \dots            & 13.687$\pm$0.006 & \dots            & \dots & RC80 \\
2459273.447 & \dots            & \dots            & \dots            & 14.907$\pm$0.010 & 14.338$\pm$0.010 & \dots            & 13.549$\pm$0.010 & \dots            & \dots & RC80 \\
2459274.305 & \dots            & 16.360$\pm$0.021 & \dots            & 14.952$\pm$0.008 & 14.322$\pm$0.003 & \dots            & 13.614$\pm$0.002 & \dots            & \dots & RC80 \\
2459275.373 & \dots            & 16.285$\pm$0.006 & \dots            & 14.894$\pm$0.014 & 14.273$\pm$0.004 & \dots            & 13.585$\pm$0.002 & \dots            & \dots & RC80 \\
2459276.328 & \dots            & 16.298$\pm$0.031 & \dots            & 14.921$\pm$0.016 & 14.317$\pm$0.004 & \dots            & 13.603$\pm$0.004 & \dots            & \dots & RC80 \\
2459276.343 & \dots            & \dots            & 15.600$\pm$0.006 & \dots            & 14.293$\pm$0.003 & \dots            & 13.566$\pm$0.007 & \dots            & \dots & Mount Suhora \\
2459278.211 & \dots            & \dots            & 15.659$\pm$0.005 & \dots            & 14.355$\pm$0.004 & \dots            & 13.654$\pm$0.010 & \dots            & \dots & Adiyaman/ADYU60 \\
2459279.207 & \dots            & \dots            & 15.680$\pm$0.006 & \dots            & 14.387$\pm$0.004 & \dots            & 13.691$\pm$0.010 & \dots            & \dots & Adiyaman/ADYU60 \\
2459280.335 & \dots            & 16.402$\pm$0.016 & \dots            & 15.005$\pm$0.014 & 14.378$\pm$0.007 & \dots            & 13.683$\pm$0.004 & \dots            & \dots & RC80 \\
2459281.288 & \dots            & \dots            & 15.625$\pm$0.014 & \dots            & 14.346$\pm$0.008 & \dots            & 13.643$\pm$0.013 & \dots            & \dots & Adiyaman/ADYU60 \\
2459281.373 & \dots            & 16.356$\pm$0.005 & \dots            & 14.986$\pm$0.014 & 14.348$\pm$0.009 & \dots            & 13.678$\pm$0.010 & \dots            & \dots & RC80 \\
2459282.223 & \dots            & \dots            & 15.666$\pm$0.008 & \dots            & 14.389$\pm$0.005 & \dots            & 13.685$\pm$0.009 & \dots            & \dots & Adiyaman/ADYU60 \\
2459284.337 & \dots            & 16.507$\pm$0.022 & \dots            & 15.081$\pm$0.039 & 14.448$\pm$0.019 & \dots            & 13.704$\pm$0.002 & \dots            & \dots & RC80 \\
2459298.284 & \dots            & 16.374$\pm$0.033 & \dots            & 14.974$\pm$0.030 & 14.320$\pm$0.003 & \dots            & 13.585$\pm$0.005 & \dots            & \dots & RC80 \\
2459299.279 & \dots            & 16.239$\pm$0.021 & \dots            & 14.863$\pm$0.013 & 14.274$\pm$0.012 & \dots            & 13.555$\pm$0.005 & \dots            & \dots & RC80 \\
\enddata
\tablecomments{More information about the telescopes and instruments can be found in Section~\ref{sec_photometric_observations}.}
\end{deluxetable*}

\section{Systemic velocity of V899~Mon from ALMA observations} \label{sec_Vsys}
V899~Mon was observed as part of ALMA project 2016.1.00209.S (PI: Takami).
The observations include the J=2--1 transitions of $^{12}$CO, $^{13}$CO and C$^{18}$O using two ALMA configurations plus the ACA, resulting in an angular resolution of $\sim$0.15$''$ and a maximum recoverable scale of 29$''$.
We manually calibrated the data, applied self-calibration using the two continuum spectral windows and generated data cube for the three isotopologues using the CASA package version 6.1 \citep{mcmullin2007}.
Exploring the three data cubes we found $^{12}$CO is optically thick while the other two isotopologues are optically and trace the gas surrounding our target.
We used a 2$''$ aperture centered on the position of V899~Mon to extract the line profile, we fitted a Gaussian function to it, and determined the systemic velocity is the velocity of the peak of the best-fit.
The resulting velocities are 9.74~km~s$^{-1}$ and 9.52~km~s$^{-1}$ for for $^{13}$CO and C$^{18}$O, respectively.
Thus we estimate the systemic velocity of V899~Mon is 9.63$\pm$0.11~km~s$^{-1}$.

\begin{figure}
\centering
    \includegraphics[width=\columnwidth]{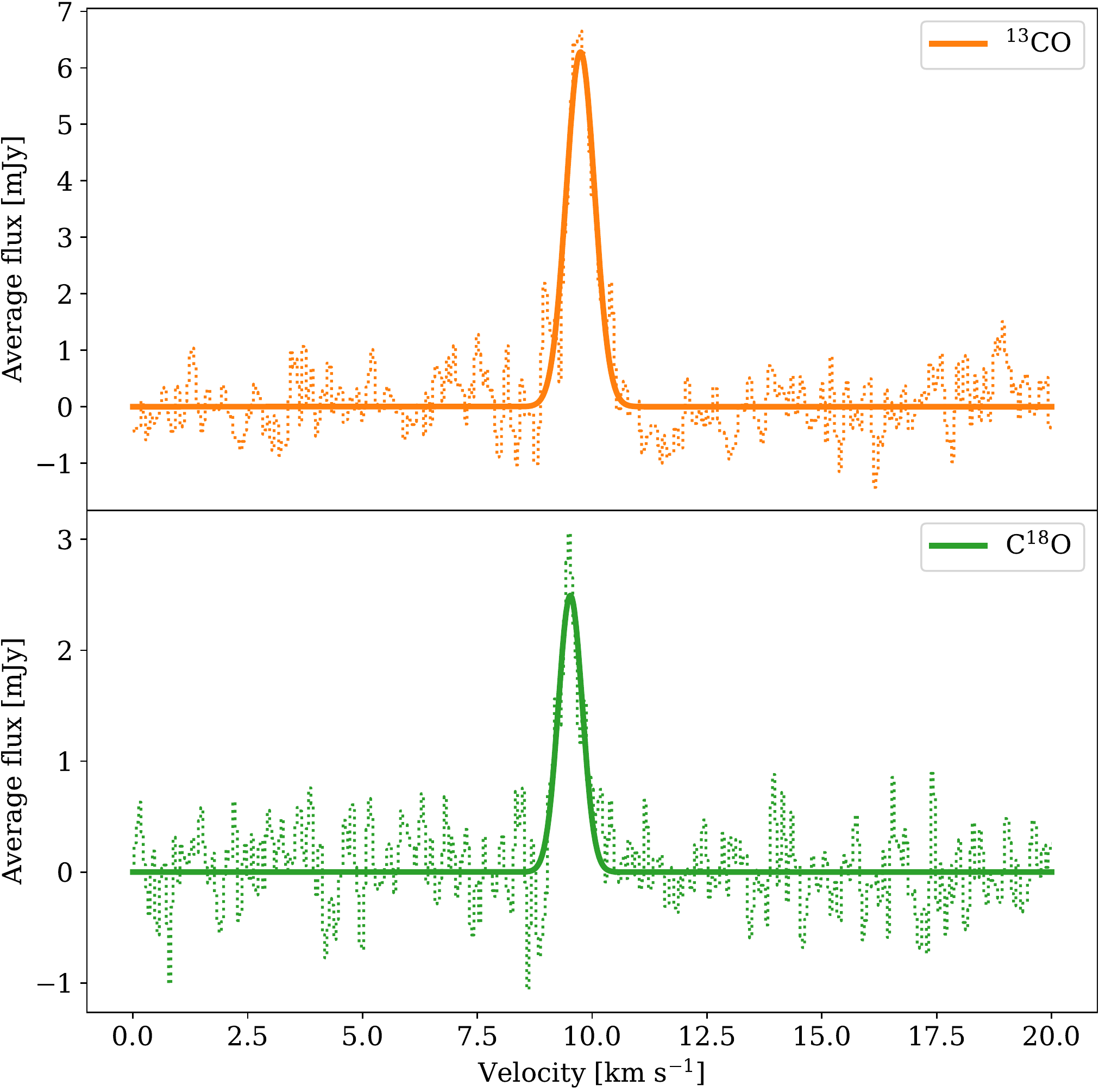}
    \caption{Line profiles of $^{13}$CO (top panel) and C$^{18}$O (bottom panel) observations with ALMA. The light dotted lines represent the flux averaged from a 2$''$ aperture centered on the position of the star. The thick continuum lines are the best fitted Gaussians used to determine the systemic velocity.  \label{fig_alma_co}}
\end{figure}

\section{Spectral Lines of V899 Mon} \label{sec_appendix_spec}

\begin{figure*}
    \centering
    \includegraphics[width=\textwidth]{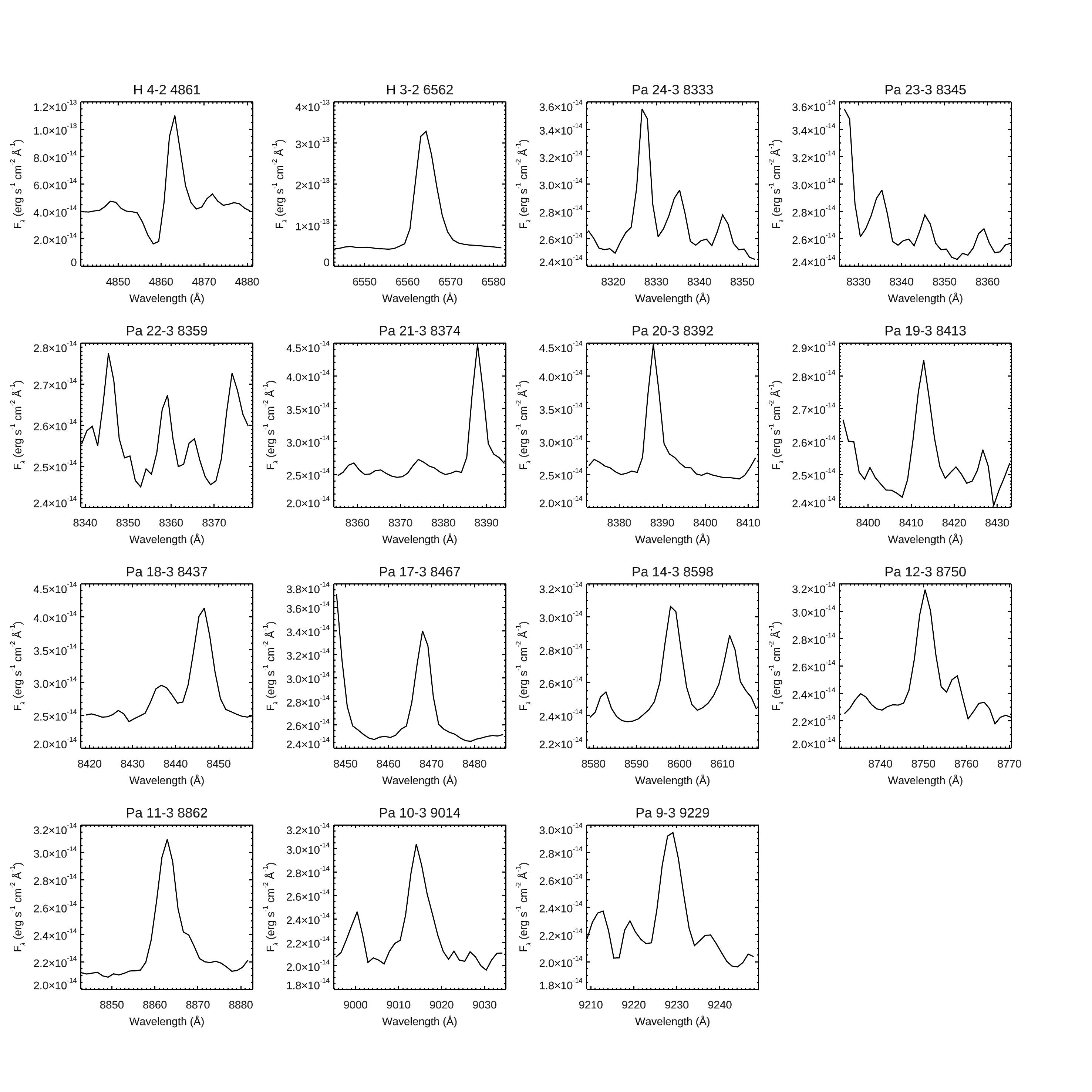}
    \caption{Hydrogen lines observed in the optical band. \label{fig_hydrogen_optical}}
\end{figure*}

\begin{figure*}
    \centering
    \includegraphics[width=\textwidth]{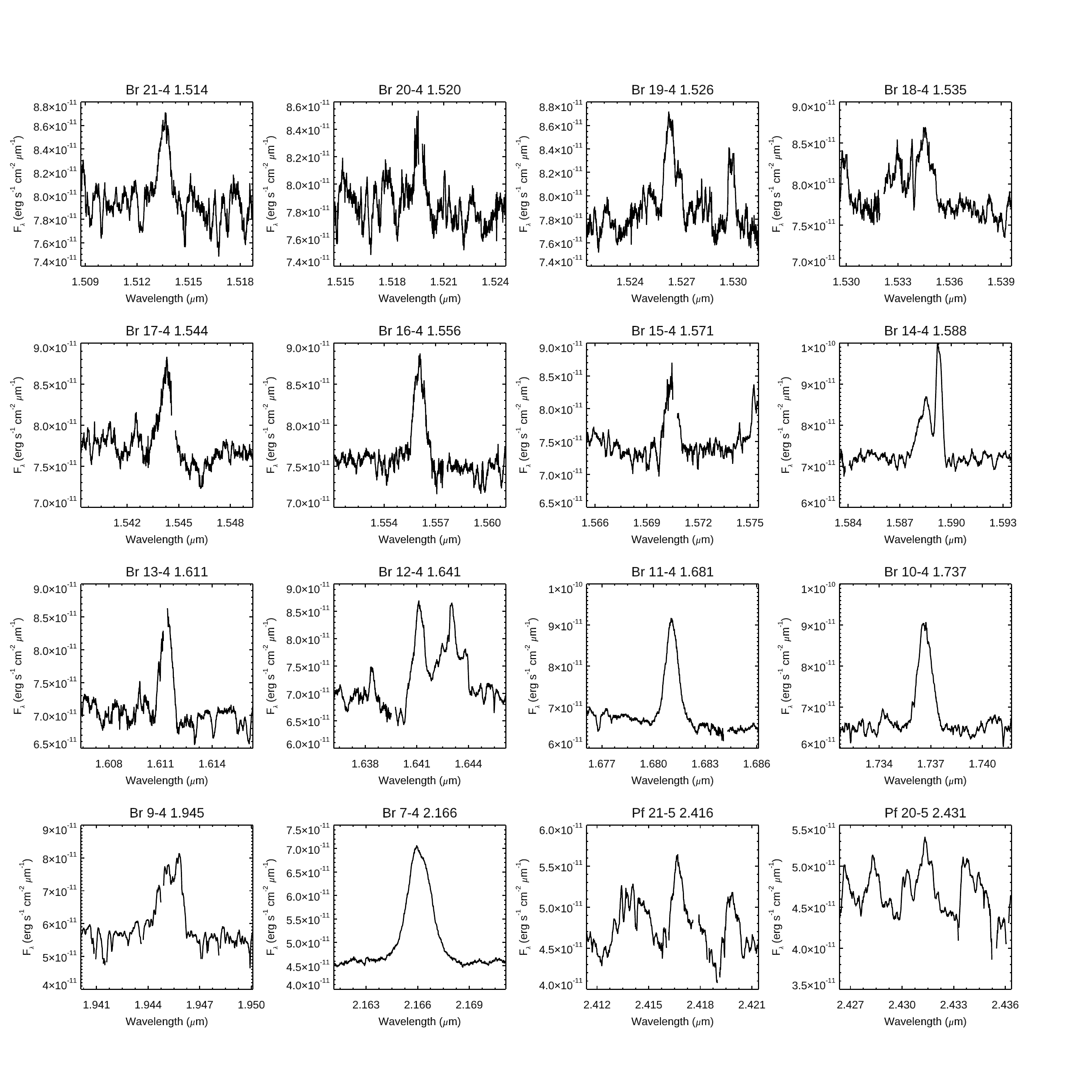}
    \caption{Hydrogen lines observed in the NIR. \label{fig_hydrogen_nir}}
\end{figure*}

\begin{figure*}
    \includegraphics[width=\textwidth]{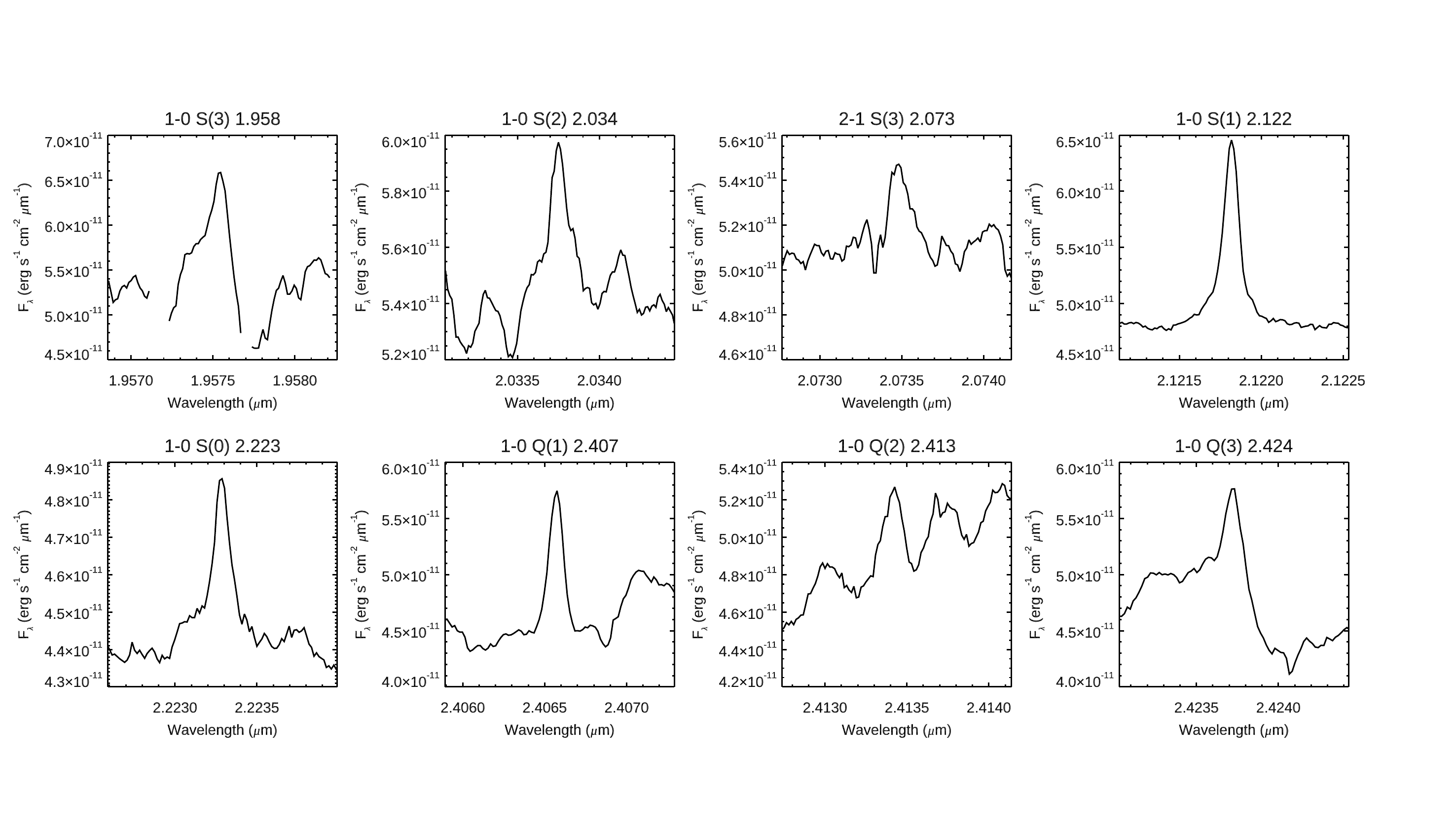}
    \caption{Molecular hydrogen lines. \label{fig_h2_all}}
\end{figure*}

In Figures~\ref{fig_hydrogen_optical}--\ref{fig_h2_all}, we provide the spectral lines of atomic hydrogen and \hmol\ lines. \\

\clearpage
\bibliography{V899Mon.bib}{}
\bibliographystyle{aasjournal}



\end{document}